\def\k{\kappa}

\def\m{\mu}
\def\n{\nu}

\newcommand{\kl}[3]{\mbox{$\rm #1$}^{\mu\nu , \alpha\beta}_{#2}(#3)}

\def\be{\begin{equation}}
\def\ee{\end{equation}}
\def\te{\end{equation}}
\def\bea{\begin{eqnarray}}
\def\nn{\nonumber}
\def\eea{\end{eqnarray}}
\def\tea{\end{eqnarray}}

\documentstyle[11pt]{article}

\makeatletter                    
\@addtoreset{equation}{section}  
\makeatother                     

\begin{document}

\title{Stochastic Gravity: Theory and Applications}

\author{B. L. Hu \\
         Department of Physics, University of Maryland,\\
             College Park, Maryland 20742-4111, U.S.A. \\
        e-mail:hub@physics.umd.edu\\
\\
E. Verdaguer \\
        Departament de Fisica Fonamental and C.E.R. in 
Astrophysics,\\Particles and Cosmology, 
Universitat de Barcelona,\\
             Av.~Diagonal 647, 08028 Barcelona, Spain\\
e-mail:verdague@ffn.ub.es \\
\\
}

\date{({\scriptsize To appear in {\it Living Reviews in Relativity}})}
\maketitle

\begin{abstract}
Stochastic semiclassical gravity of the 90's is a theory naturally
evolved  from semiclassical gravity of the 80's and quantum field
theory in curved spacetimes of the 70's.  Whereas semiclassical
gravity is based on the semiclassical Einstein equation with
sources given by the expectation value of the stress-energy
tensor of quantum fields, stochastic semiclassical gravity is
based on the Einstein-Langevin equation, which has in addition
sources due to the noise kernel. The noise kernel is the vacuum
expectation value of the  (operator-valued) stress-energy
bi-tensor which describes the fluctuations of quantum matter
fields in curved spacetimes. In the first part, we describe the
fundamentals of this new theory via two approaches: the axiomatic
and the functional. The axiomatic approach is useful to see the
structure of the theory from the framework of semiclassical
gravity, showing the link from the mean value of the stress-energy
tensor to their correlation functions. The functional
approach uses the Feynman-Vernon influence functional and the
Schwinger-Keldysh closed-time-path effective action methods which
are convenient for computations. It also brings out the open
systems concepts and the statistical and stochastic contents of
the theory such as dissipation, fluctuations, noise and
decoherence. We then focus on the properties of the stress energy
bi-tensor.
We obtain a general expression for the noise kernel of a quantum
field defined at two distinct points in an arbitrary curved
spacetime as products of covariant derivatives of the quantum
field's Green function, and show from this that the trace anomaly
of the noise kernel is zero for massless conformal scalar fields.
In the second part, we describe three applications of stochastic gravity
theory. First, we consider metric perturbations in a Minkowski
spacetime. We offer an analytical solution of
the Einstein-Langevin equation and compute
the two-point correlation functions
for the linearized Einstein tensor and
for the metric perturbations. Second, we discuss structure
formation from the stochastic gravity viewpoint, which can
go beyond the standard treatment by incorporating the full
quantum effect of the inflaton fluctuations. Third, we discuss the
backreaction of Hawking radiation in the gravitational background
of a quasi-static black hole (enclosed in a box). We derive a
fluctuation-dissipation relation between the fluctuations in the
radiation and the dissipative dynamics of metric fluctuations.
\end{abstract}



\section{Overview}
\label{introduction}
Stochastic semiclassical gravity\footnote{We
will often use the shortened term {\it stochastic gravity} as
there is no confusion as to the nature and source of
stochasticity in gravity being induced from the quantum fields
and not a priori from the classical spacetime.} is a theory
developed in the Nineties using semiclassical gravity
(quantum fields in classical spacetimes, solved self-consistently)
as the starting point and aiming at a theory of quantum gravity
as the goal. While semiclassical gravity is based on the
semiclassical Einstein equation with the source given by the
expectation value of the stress-energy tensor of quantum fields,
stochastic gravity includes also its fluctuations in a new
stochastic semiclassical or the Einstein-Langevin equation. If
the centerpiece in semiclassical gravity theory is the vacuum
expectation value of the stress-energy tensor of a quantum field,
and the central issues being how well the vacuum is defined and
how the divergences can be controlled by regularization and
renormalization, the centerpiece in stochastic semiclassical
gravity theory  is the stress-energy bi-tensor and its
expectation value known as the noise kernel. The mathematical
properties of this quantity and its physical content in relation
to the behavior of fluctuations of quantum fields in curved
spacetimes are the central issues of this new theory. How they
induce metric fluctuations and seed the structures of the
universe, how they affect the black hole horizons and the
backreaction of Hawking radiance in black hole dynamics,
including implications on trans-Planckian physics, are new
horizons to explore.  On the theoretical issues, stochastic
gravity is the necessary foundation to investigate the validity
of semiclassical gravity and the viability of inflationary
cosmology based on the appearance and sustenance of a vacuum
energy-dominated phase. It is also a useful beachhead supported
by well-established low energy (sub-Planckian) physics to explore
the connection with high energy (Planckian) physics in the realm
of quantum gravity.

In this review we summarize major work and results on this theory
since 1998. It is in the nature of a progress report rather than a
review. In fact we will have room only to discuss a handful of
topics of basic importance. A review of ideas leading to
stochastic gravity and further developments originating from it
can be found in Refs.~\cite{Physica,stogra}; a set of lectures
which include a discussion of the issue of the validity of
semiclassical gravity in Ref. ~\cite{HVErice}; a pedagogical
introduction of stochastic gravity theory with a more complete
treatment of backreaction problems in cosmology and black holes in
Ref. ~\cite{HuVer03a}. A comprehensive formal description of the
fundamentals is given in Refs.~\cite{MarVer99a,MarVer99} while
that of  the noise kernel in arbitrary spacetimes in
Refs.~\cite{MarVer99,PhiHu01,PhiHu03}. Here we will try to mention
all related work so the reader can at least trace out the parallel
and sequential developments. The references at the end of each
topic below are representative work where one can seek out further
treatments.

Stochastic gravity theory is built on three pillars:  general
relativity , quantum fields and nonequilibrium statistical
mechanics. The first two uphold semiclassical gravity, the last
two span statistical field theory.  Strictly speaking one can
understand a great deal without appealing to statistical
mechanics, and we will try to do so here. But concepts such as
quantum open systems \cite{Dav76,LinWes90,Wei93} and techniques
such as the influence functional \cite{FeyVer63,FeyHib65} (which
is related to the
closed-time-path effective action \cite{Sch61,BakMah63,Kel64,ChoEtal85,SuEtal88,CalHu89,CooEtal94,%
DeW86,Jor86,CalHu87,Jor87,Paz90}) were a great help in our
understanding of the physical meaning of issues involved toward
the construction of this new theory, foremost because quantum
fluctuations and correlation have become the focus. Quantum
statistical field theory and the statistical mechanics of quantum
field theory \cite{CalHu88,dch,cddn,CalHu00} also aided us in
searching for the connection with quantum gravity through the
retrieval of correlations and coherence. We show the scope of
stochastic gravity as follows:

\begin{description}
\item[I.] {\bf Ingredients} \item[A.] From General Relativity
\cite{MisThoWhe73,Wal84} to Semiclassical Gravity. \item[B.]
Quantum Field Theory in Curved Spacetimes
\cite{BirDav82,Ful89,Wal94,GriMamMos94}:
\begin{description}
\item[1.] Stress-energy tensor: Regularization and renormalization.
\item[2.] Self-consistent solution: Backreaction problems \cite{LukSta74,Gri76,HuPar77,HuPar78,Har81,And83,And84}.
\item[3.] Effective action: Closed time path, initial value
          formulation \cite{Sch61,BakMah63,Kel64,ChoEtal85,SuEtal88,%
CalHu89,CooEtal94,DeW86,Jor86,CalHu87,Jor87,Paz90}.
\item[4.] Equation of motion: Real and causal.
\end{description}
\item[C.] Nonequilibrium Statistical Mechanics:
\begin{description}
\item[1.] Open quantum systems \cite{Dav76,LinWes90,Wei93}.
\item[2.] Influence Functional: Stochastic equations
\cite{FeyVer63}. \item[3.] Noise and Decoherence: Quantum to
classical
  transition \cite{Zur81,Zur82,Zur86,Zur91,JooZeh85,CalLeg85,UnrZur89,%
Zur93,GiuEtal96,Gri84,Omn88a,Omn88b,Omn88c,Omn90,Omn92,Omn94,%
GelHar90,Har93,DowHal92,Hal93,Hal98,Bru93,PazZur93,Twa93,Ish94,%
IshLin94a,IshLin94b,Hal95,DowKen95,DowKen96,Ken96,Ken97,Ken98,IshLin98}.
\end{description}
\item[D.] Decoherence in Quantum Cosmology and
Emergence of Classical Spacetimes
\cite{Kie87,Hal89,Pad89,Hu90,Cal89,Cal91,HuPazSin93}.
\end{description}

\begin{description}
\item[II.] {\bf Theory}
\item[A.] Dissipation from  Particle Creation \cite{DeW86,Jor86,CalHu87,Jor87,Paz90,CamVer94}.
     Backreaction as Fluctuation-Dissipation Relation (FDR) \cite{HuSin95}.
\item[B.] Noise from Fluctuations of Quantum Fields
\cite{Physica,Banff,CalHu94}. \item[C.] Einstein-Langevin
Equations
\cite{CalHu94,HuMat95,HuSin95,CamVer96,CamVer97,CalCamVer97,LomMaz97,MarVer99a,MarVer99,MarVer99b}.
\item[D.] Metric Fluctuations in Minkowski spacetime
\cite{MarVer00}.
\end{description}

\begin{description}
\item[III.] {\bf Issues} \item[A.] Validity of Semiclassical
Gravity \cite{HuPhi00,PhiHu00}. \item[B.] Viability of Vacuum
Dominance and Inflationary Cosmology. \item[C.] Stress-Energy
Bitensor and Noise Kernel:
\begin{description}
\item[1.] Regularization Reassessed \cite{PhiHu01,PhiHu03}.
\end{description}
\end{description}

\begin{description}
\item[IV.] {\bf Applications:}  Early Universe and Black Holes.
\item[A.] Wave Propagation in Stochastic Geometry  \cite{HuShi98}.
\item[B.] Black Hole Horizon Fluctuations: Spontaneous/Active
versus Induced/Pas\-sive
\cite{ForSva97,WuFor99,Sor96,SorSud99,BarFroPar99,BarFroPar00,MasPar00,Par01,PhiHu03}.
\item[C.] Noise induced inflation \cite{CalVer99}. \item[D.]
Structure Formation
\cite{CalHu95,Mat97a,Mat97b,CalGon97,RouVer03a}.
     Trace anomaly-driven inflation
     \cite{Sta80,Vil85,HawHerRea01}.
\item[E.] Black Hole Backreaction as FDR
\cite{CanSci77,Sci79,SciCanDeu81,Mottola,Vishu,CamHu98,CamHu99,SinRavHu03}.
\end{description}

\begin{description}
\item[V.] {\bf Related Topics}
\item[A.]  Metric Fluctuations and Trans-Planckian Problem
  \cite{BarFroPar99,BarFroPar00,MasPar00,Par01,NiePar01}.
\item[B.] Spacetime Foam \cite{Car97,Car98,Gar98a,Gar98b,Gar99}.
\item[C.] Universal `Metric Conductance' Fluctuations  \cite{Shio}.
\end{description}

\begin{description}
\item[VI.] {\bf Ideas}
\item[A.] General Relativity as Geometro-Hydrodynamics \cite{grhydro}.
\item[B.] Semiclassical Gravity as Mesoscopic Physics \cite{meso}.
\item[C.] From Stochastic to Quantum Gravity:
\begin{description}
\item[1.] Via Correlation hierarchy of interacting quantum fields
  \cite{stogra,dch,CalHu00,kinQG}.
\item[2.] Possible relation to string theory and matrix theory.
\end{description}
\end{description}

{}We list only the latest work in the respective topics above
describing ongoing research. The reader should consult the
references therein for earlier work and the background material.
We do not seek a complete coverage here, but will discuss only the
selected topics in theory, issues and applications. We use the
$(+,+,+)$ sign conventions of Refs. \cite{MisThoWhe73,Wal84}, and
units in which $c=\hbar=1$.

\section{From Semiclassical to Stochastic Gravity }
\label{sec1}

There are three main steps that lead to the recent
development of stochastic gravity. The first step begins with
{\it{quantum field theory in curved spacetime}}
~\cite{DeW75,BirDav82,Ful89,Wal94,GriMamMos94}, which describes
the behavior of quantum matter fields propagating in a specified
(not dynamically determined by the quantum matter field as source)
background gravitational field. In this theory the gravitational
field is given by the classical spacetime metric determined from
classical sources by the classical Einstein equations, and the
quantum fields propagate as test fields in such a spacetime. An
important process described by quantum field theory in curved
spacetime is indeed particle creation from the vacuum, and effects
of vacuum fluctuations and polarizations, in the early universe
\cite{Par69,SexUrb69,Zel70,ZelSta71,Hu74,Ber74,Ber75a,Ber75b,DeW75,Fri89,CesVer90},
and Hawking radiation in black holes
\cite{Haw74,Haw75,Isr75,Par75,Wal75}.

The second step in the description of the interaction of gravity
with quantum fields is back-reaction, {\it i.e.}, the effect of
the quantum fields on the spacetime geometry. The source here is
the expectation value of the stress-energy operator for the matter
fields in some quantum state in the spacetime, a classical
observable. However, since this object is quadratic in the field
operators, which are only well defined as distributions on the
spacetime, it involves ill defined quantities. It contains
ultraviolet divergences the removal of which requires a
renormalization procedure \cite{DeW75,Chr76,Chr78}. The final
expectation value of the stress-energy operator using a reasonable
regularization technique is essentially unique, modulo some terms
which depend on the spacetime curvature and which are independent
of the quantum state. This uniqueness was proved by Wald
\cite{Wal77,Wal78} who investigated the criteria that a physically
meaningful expectation value of the  stress-energy tensor ought to
satisfy.

The theory obtained from a self-consistent solution of the
geometry of the spacetime and the quantum field is known as {\it
semiclassical gravity}. Incorporating the backreaction of the
quantum matter field on the spacetime is thus the central task in
semiclassical gravity. One assumes a general class of spacetime
where the quantum fields live in and act on, and seek a solution
which satisfies simultaneously the Einstein equation for the
spacetime and the field equations for the quantum fields. The
Einstein equation which has the expectation value of the
stress-energy operator of the quantum matter field as the source
is known as the {\it semiclassical Einstein equation}.
Semiclassical gravity was first investigated in cosmological
backreaction problems
\cite{LukSta74,Gri76,HuPar77,HuPar78,Har81,And83,And84,Har77,FisHarHu79,HarHu79},
an example is the
damping of anisotropy in Bianchi universes by the backreaction of
vacuum particle creation. Using the effect of quantum field
processes such as particle creation to explain why the universe is
so isotropic at the present was investigated in the context of
chaotic cosmology \cite{Mis69,BelKhaLif70,BelKhaLif82}
in the late seventies prior to
the inflationary cosmology proposal of the eighties
\cite{Gut81,AlbSte82,Lin82,Lin85},
which assumes the vacuum expectation value of an inflaton field as
the source, another, perhaps more well-known, example of
semiclassical gravity.

\subsection{The importance of quantum fluctuations}

For a free quantum field semiclassical gravity is fairly well
understood. The theory is in some sense unique, since the only
reasonable c-number stress-energy tensor that one may construct
\cite{Wal77,Wal78} with the stress-energy operator is a
renormalized expectation value. However the scope and limitations
of the theory are not so well understood. It is expected that the
semiclassical theory would  break down at the Planck scale. One
can conceivably assume that it would also break down when the
fluctuations of the stress-energy operator are large
\cite{For82,KuoFor93}. Calculations of the fluctuations of the
energy density for Minkowski, Casimir and hot flat spaces as well
as Einstein and de Sitter universes are available
\cite{KuoFor93,PhiHu97,HuPhi00,PhiHu00,PhiHu01,PhiHu03,PhiHu03b,MarVer99,MarVer00,%
RouVer99a,RouVer03a,OsbSho,CogGuiEli02}. It is less clear, however, how to
quantify what a large fluctuation is, and different criteria have
been proposed
\cite{KuoFor93,ForSCG,ForWu,HuPhi00,PhiHu00,AndMolMot02,AndMolMot03}.
The issue of the validity of the semiclassical gravity viewed in
the light of quantum fluctuations is summarized in our Erice
lectures \cite{HVErice}. One can see the essence of the problem by
the following example inspired by Ford \cite{For82}.

Let us assume a quantum state formed by an isolated system which
consists of a superposition with equal amplitude of  one
configuration of mass $M$ with the center of mass at $X_1$ and
another configuration of the same mass with the center of mass at
$X_2$. The semiclassical theory as described by the semiclassical
Einstein equation predicts that the center of mass of the
gravitational field of the system is centered at $(X_1+X_2)/2$.
However, one would expect that if we send a succession of test
particles to probe the gravitational field of the above system
half of the time they would react to a gravitational field of
mass $M$ centered at $X_1$ and half of the time to the field
centered at $X_2$. The two predictions are clearly different,
note that the fluctuation in the position of the center of masses
is of the order of $(X_1-X_2)^2$. Although this example raises
the issue of how to place the importance of fluctuations to the
mean, a word of caution should be added to the effect that it
should not be taken too literally. In fact, if the previous
masses are macroscopic the quantum system decoheres very quickly
\cite{Zur91,Zur93} and instead of being described by a pure quantum
state it is described by a density matrix which diagonalizes in a
certain pointer basis. For observables associated to such a
pointer basis the density matrix description is equivalent to
that provided by a statistical ensemble. The results will differ,
in any case, from the semiclassical prediction.

In other words, one would expect that a stochastic source that
describes the quantum fluctuations should enter into the
semiclassical equations. A significant step in this direction was
made in Ref.~\cite{Physica} where it was proposed to view the
back-reaction problem in the framework of an open quantum system:
the quantum fields seen as the ``environment" and the
gravitational field as the ``system". Following this proposal a
systematic study of the connection between semiclassical gravity
and open quantum systems resulted in the development of a new
conceptual and technical framework where (semiclassical)
Einstein-Langevin equations were derived
\cite{CalHu94,HuMat95,HuSin95,CamVer96,CamVer97,CalCamVer97,LomMaz97}.
The key technical factor to most of these results was the use of
the influence functional method of Feynman and Vernon
\cite{FeyVer63} when only the coarse-grained effect of the
environment on the system is of interest. Note that the word
semiclassical put in parentheses refers to the fact that the noise
source in the Einstein-Langevin equation arises from the quantum
field, while the background spacetime is classical; generally we
will not carry this word since there is no confusion that the
source which contributes to the stochastic features of this theory
comes from quantum fields.

In the language of the consistent histories formulation of quantum
mechanics \cite{Gri84,Omn88a,Omn88b,Omn88c,Omn90,Omn92,Omn94,GelHar90,%
Har93,DowHal92,Hal93,Hal98,Bru93,PazZur93,Twa93,Ish94,IshLin94a,%
IshLin94b,Hal95,DowKen95,DowKen96,Ken96,Ken97,Ken98,IshLin98}
for the existence of a semiclassical
regime for the dynamics of the system one needs two requirements:
The first is decoherence, which guarantees that probabilities can
be consistently assigned to histories describing the evolution of
the system, and the second is that these probabilities should peak
near histories which correspond to solutions of classical
equations of motion. The effect of the environment is crucial, on
the one hand, to provide decoherence and, on the other hand, to
produce both dissipation and noise to the system through
back-reaction, thus inducing a semiclassical stochastic dynamics
on the system. As shown by different authors
\cite{GelHar93,Zur81,Zur82,Zur86,Zur91,JooZeh85,%
CalLeg85,UnrZur89,Zur93,GiuEtal96}, indeed over a long history
predating the current revival of decoherence, stochastic
semiclassical equations are obtained in an open quantum system
after a coarse graining of the environmental degrees of freedom
and a further coarse graining in the system variables. It is
expected but has not yet been shown that this mechanism could also
work for decoherence and classicalization of the metric field.
Thus far, the analogy could be made formally \cite{MarVer99b} or
under certain assumptions, such as adopting the Born-Oppenheimer
approximation in quantum cosmology \cite{Paz91,PazSin92}.

An alternative axiomatic approach to the Einstein-Langevin
equation without invoking the open system paradigm was later
suggested based on the formulation of a self-consistent dynamical
equation for a perturbative extension of semiclassical gravity
able to account for the lowest order stress-energy fluctuations of
matter fields \cite{MarVer99a}. It was shown that the same
equation could be derived, in this general case, from the
influence functional of Feynman and Vernon \cite{MarVer99}. The
field equation is deduced via an effective action which is
computed assuming that the gravitational field is a c-number. The
important new element in the derivation of the Einstein-Langevin
equation, and of the stochastic gravity theory, is the physical
observable that measures the stress-energy fluctuations, namely,
the expectation value of the symmetrized bi-tensor constructed
with the stress-energy tensor operator: the {\it noise kernel}. It
is interesting to note that the Einstein-Langevin equation can
also be understood as a useful intermediary tool to compute
symmetrized two-point correlations of the quantum metric
perturbations on the semiclassical background, independent of a
suitable classicalization mechanism \cite{RouVer03b}.

\section{The Einstein-Langevin equation: Axiomatic approach}
\label{sec2}

In this section we introduce {\it stochastic semiclassical
gravity}, or {\it stochastic gravity} for short,
in an axiomatic way. It is introduced as an extension of
semiclassical gravity motivated by the search of self-consistent
equations which describe the back-reaction of the quantum
stress-energy fluctuations on the gravitational field
\cite{MarVer99a}.

\subsection{Semiclassical gravity}

Semiclassical gravity describes the interaction of a classical
gravitational field with quantum matter fields. This theory can
be formally derived as the leading $1/N$ approximation of quantum
gravity interacting with $N$ independent and identical free
quantum fields \cite{HorWal80,HorWal82,HarHor81,Tom77}
which interact with gravity only.  By keeping the value of $NG$
finite, where $G$ is Newton's gravitational constant, one arrives
at a theory in which formally the gravitational field can be
treated as a c-number field (i.e. quantized at tree level) while
matter fields are fully quantized.
The semiclassical theory may be summarized as follows.

Let $({\cal M},g_{ab})$ be a globally hyperbolic four-dimensional
spacetime manifold ${\cal M}$ with metric $g_{ab}$ and consider a real
scalar quantum field $\phi$ of mass $m$ propagating on that manifold;
we just assume a scalar field for
simplicity.
The classical action $S_m$ for this matter field is given by
the functional
\begin{equation}
S_m[g,\phi]=-{1\over2}\int d^4x\sqrt{-g}\left[g^{ab}
\nabla_a\phi\nabla_b\phi+\left(m^2+\xi R\right)\phi^2\right],
\label{2.1}
\end{equation}
where $\nabla_a$ is the covariant derivative associated to the
metric $g_{ab}$, $\xi$ is a  coupling parameter between the field
and the scalar curvature of the underlying spacetime $R$, and
$g={\rm det} g_{ab}$.

The field may be quantized in the manifold using the standard
canonical quantization formalism \cite{BirDav82,Ful89,Wal94}. The
field operator in the Heisenberg representation $\hat\phi$ is an
operator-valued distribution solution of the Klein-Gordon
equation, the field equation derived from  Eq. (\ref{2.1}),
\begin{equation}
(\Box-m^2 -\xi R)\hat\phi=0.
\label{2.2}
\end{equation}
We may write the field operator as $\hat\phi[g;x)$
to indicate that it is a functional of the metric $g_{ab}$ and a function
of the spacetime point $x$. This notation will be used also for other
operators and tensors.

The classical stress-energy tensor is obtained by functional derivation
of this action in the usual way
$T^{ab}(x)=(2/\sqrt{-g})\delta S_m/\delta g_{ab}$, leading to
\begin{eqnarray}
T^{ab}[g,\phi]&=&\nabla^a\phi\nabla^b\phi-{1\over2}g^{ab}
\left(\nabla^c\phi\nabla_c\phi+m^2\phi^2\right)
\nonumber \\
&&+\xi\left(g^{ab}\Box-\nabla^a\nabla^b+G^{ab}
\right)\phi^2,
\label{2.3}
\end{eqnarray}
where $\Box=\nabla_a\nabla^a$ and $G_{ab}$ is the Einstein tensor.
With the notation $T^{ab}[g,\phi]$ we explicitly
indicate that the stress-energy tensor is
a functional of the metric $g_{ab}$  and the field $\phi$.

The next step is to define a stress-energy tensor operator $\hat
T^{ab}[g;x)$. Naively one would replace the classical field
$\phi[g;x)$ in the above functional by the quantum operator
$\hat\phi[g;x)$, but this procedure involves taking the product of
two distributions at the same spacetime point. This is ill-defined
and we need a regularization procedure. There are several
regularization methods which one may use, one is the
point-splitting or point-separation regularization method
\cite{Chr76,Chr78} in which one introduces a point $y$ in a
neighborhood of the point $x$ and then uses as the regulator the
vector tangent at the point $x$ of the geodesic joining $x$ and
$y$; this method is discussed for instance in Refs.
\cite{PhiHu00,PhiHu01,PhiHu03} and in section \ref{sec4}. Another
well known method is dimensional regularization in which one works
in arbitrary $n$ dimensions, where $n$ is not necessarily an
integer, and then uses as the regulator the parameter
$\epsilon=n-4$; this method is implicitly used in this section.
The regularized stress-energy operator using the Weyl ordering
prescription, {\it i.e.} symmetrical ordering, can be written as
\be \hat{T}^{ab}[g] = {1\over 2} \{
     \nabla^{a}\hat{\phi}[g]\, , \,
     \nabla^{b}\hat{\phi}[g] \}
     + {\cal D}^{ab}[g]\, \hat{\phi}^2[g],
\label{regul s-t 2} \ee where ${\cal D}^{ab}[g]$ is the
differential operator:
\begin{equation}
{\cal D}^{ab} \equiv
\left(\xi-1/4\right) g^{ab} \Box+ \xi \left( R^{ab}- \nabla^{a}
\nabla^{b}\right).
\label{diff operator}
\end{equation}
Note that if dimensional regularization is
used, the field operator $\hat \phi[g;x)$ propagates in a
$n$-dimensional spacetime. Once the regularization prescription
has been introduced a regularized and renormalized stress-energy
operator $\hat T^R_{ab}[g;x)$ may be defined as
\begin{equation}
\hat T^R_{ab}[g;x)= \hat T_{ab}[g;x)+F^C_{ab}[g;x)\hat I,
\label{2.4}
\end{equation}
which differs from the regularized $\hat T_{ab}[g;x)$ by the
identity operator times some tensor counterterms $F^C_{ab}[g;x)$,
which depend on the regulator and are local functionals of the
metric, see Ref.~\cite{MarVer99} for details. The field states can
be chosen in such a way that for any pair of physically acceptable
states, {\it i.e.}, Hadamard states in the sense of
Ref.~\cite{Wal94}, $|\psi\rangle$, and $|\varphi\rangle$ the
matrix element $\langle\psi|T^R_{ab}|\varphi\rangle$, defined as
the limit when the regulator takes the physical value, is finite
and satisfies Wald's axioms \cite{Ful89,Wal77}. These counterterms
can be extracted from the singular part of a Schwinger-DeWitt
series \cite{Ful89,Chr76,Chr78,Bun79}. The choice of these
counterterms is not unique but this ambiguity can be absorbed into
the renormalized coupling constants which appear in the equations
of motion for the gravitational field.

The {\it semiclassical Einstein equation} for the metric $g_{ab}$
can then be written as
\begin{equation}
G_{ab}[g]+\Lambda g_{ab}
-2(\alpha A_{ab}+\beta B_{ab})[g]=
8\pi G \langle \hat T_{ab}^R[g]\rangle ,
\label{2.5}
\end{equation}
where $\langle \hat T_{ab}^R[g]\rangle $ is the expectation value
of the operator $\hat T_{ab}^R[g,x)$ after the regulator takes
the physical value in some physically acceptable state of the field on
$({\cal M},g_{ab})$. Note that both the stress tensor and the
quantum state are functionals of the metric, hence the notation.
The parameters $G$, $\Lambda$, $\alpha$ and $\beta$ are the
renormalized coupling constants, respectively, the gravitational
constant, the cosmological constant and two dimensionless  coupling
constants which are zero in the classical Einstein equation.
These constants must be understood as the result of  ``dressing''
the bare constants which appear in the classical action
before renormalization. The values of these constants must be
determined by experiment.
The left hand side of Eq. (\ref{2.5}) may be derived
from the gravitational action
\begin{equation}
S_g[g]= {1\over 8\pi G}\int d^4 x \sqrt{-g}\left[ {1\over 2}
R-\Lambda +\alpha C_{abcd}C^{abcd}
+\beta R^2\right],
\label{2.6}
\end{equation}
where $C_{abcd}$ is the Weyl tensor. The tensors
$A_{ab}$ and $B_{ab}$ come from the functional
derivatives with respect to the metric of the terms quadratic
in the curvature in Eq. (\ref{2.6}), they are explicitly given by
\begin{eqnarray}
A^{ab}&=&\frac{1}{\sqrt{-g}}\frac{\delta}{\delta g_{ab}}
\int d^4 \sqrt{-g} C_{cdef}C^{cdef}
\nonumber\\
&=&{1\over2}g^{ab}C_{cdef}
C^{cdef}-2R^{acde}
R^{b}_{\ cde}+4R^{ac}R_c^{\ b}
-{2\over3}RR^{ab}
\nonumber\\
&& -2\Box R^{ab}+{2\over3}\nabla^a\nabla^b R+
{1\over3}g^{ab}\Box R,
\label{2.7a}\\
B^{ab}&=&\frac{1}{\sqrt{-g}}\frac{\delta}{\delta g_{ab}}
\int d^4 \sqrt{-g} R^2
\nonumber\\
&=&{1\over2}g^{ab}R^2-2R R^{ab}
+2\nabla^a\nabla^b R-2g^{ab}\Box R,
\label{2.7b}
\end{eqnarray}
where $R_{abcd}$  and $R_{ab}$ are the Riemann and Ricci tensors,
respectively. These two tensors are, like the Einstein and metric
tensors, symmetric and divergenceless: $\nabla^a A_{ab}=0=\nabla^a
B_{ab}$.

A solution of semiclassical gravity consists of a spacetime
(${\cal M},g_{ab}$), a quantum field operator $\hat\phi[g]$
which satisfies the evolution equation (\ref{2.2}), and a physically
acceptable state  $|\psi[g]\rangle $ for this field, such that Eq.
(\ref{2.5}) is satisfied when the expectation value of the renormalized
stress-energy operator is evaluated in this state.

For a free quantum field  this theory is robust in the sense that
it is self-consistent and fairly well understood.
As long as the gravitational field is assumed to be described by a
classical metric, the above semiclassical Einstein
equations seems to be the only plausible dynamical equation
for this metric: the metric couples to matter fields via the
stress-energy tensor and for a given quantum state the only
physically observable c-number  stress-energy tensor that one
can construct is the above renormalized expectation value.
However, lacking a full quantum gravity theory the scope and
limits of the theory are not so well understood. It is assumed
that the semiclassical theory should break down at Planck scales,
which is when simple order of magnitude estimates suggest that
the quantum effects of gravity should not be ignored because the
energy of a quantum fluctuation in
a Planck size region, as determined by the Heisenberg uncertainty
principle, is comparable to the gravitational energy of
that fluctuation.

The theory is expected to break down when the fluctuations of the
stress-energy operator are large \cite{For82}. A criterion based
on the ratio of the fluctuations to the mean was proposed by Kuo
and Ford \cite{KuoFor93} (see also work via zeta-function methods
\cite{PhiHu97,CogGuiEli02}).  This proposal was questioned by Phillips and Hu
\cite{HuPhi00,PhiHu00,PhiHu01} because it does not contain a scale
at which the theory is probed or how accurately the theory can be
resolved. They suggested the use of a smearing scale or
point-separation distance, for integrating over the bi-tensor
quantities, equivalent to a stipulation of the resolution level of
measurements; see also the response by Ford \cite{ForSCG,ForWu}. A
different criterion is recently suggested by Anderson et al.
\cite{AndMolMot02,AndMolMot03} based on linear response theory. A
partial summary of this issue can be found in our Erice Lectures
\cite{HVErice}.

\subsection{Stochastic gravity}

The purpose of stochastic gravity is to extend the
semiclassical theory to account for these fluctuations in a
self-consistent way. A physical observable that describes these
fluctuations to lowest order is the {\it noise kernel} bi-tensor,
which is defined through the two point correlation of the
stress-energy operator as
\begin{equation}
N_{abcd}[g;x,y)={1\over2}\langle\{\hat t_{ab}[g;x),
\hat t_{cd}[g;y)\}\rangle,
\label{2.8}
\end{equation}
where the curly brackets mean anticommutator, and where
\begin{equation}
\hat t_{ab}[g;x)
\equiv \hat T_{ab}[g;x)-\langle \hat T_{ab}[g;x)\rangle.
\label{2.9}
\end{equation}
This bi-tensor can also be written $N_{ab,c^\prime
d^\prime}[g;x,y)$, or $N_{ab,c^\prime d^\prime}(x,y)$ as we do in
section \ref{sec4}, to emphasize that it is a tensor with respect
to the first two indices at the point $x$ and a tensor with
respect to the last two indices at the point $y$, but we shall not
follow this notation here. The noise kernel is defined in terms of
the unrenormalized stress-tensor operator $\hat T_{ab}[g;x)$ on a
given background metric $g_{ab}$, thus a regulator is implicitly
assumed on the right-hand side of Eq. (\ref{2.8}).  However, for a
linear quantum field the above kernel -- the expectation function
of a bi-tensor -- is free of ultraviolet divergences because the
regularized $T_{ab}[g;x)$ differs from the renormalized
$T_{ab}^R[g;x)$ by the identity operator times some tensor
counterterms, see Eq.~(\ref{2.4}), so that in the subtraction
(\ref{2.9}) the counterterms cancel. Consequently the ultraviolet
behavior of $\langle\hat T_{ab}(x)\hat T_{cd}(y)\rangle$ is the
same as that of $\langle\hat T_{ab}(x)\rangle \langle\hat
T_{cd}(y)\rangle$, and $\hat T_{ab}$ can be replaced by the
renormalized operator $\hat T_{ab}^R$ in Eq. (\ref{2.8}); an
alternative proof of this result is given in Ref.
\cite{PhiHu01,PhiHu03}. The noise kernel should be thought of as a
distribution function, the limit of coincidence points has meaning
only in the sense of distributions. The bi-tensor
$N_{abcd}[g;x,y)$, or $N_{abcd}(x,y)$ for short, is real and
positive semi-definite, as a consequence of $\hat T_{ab}^R$ being
self-adjoint. A simple proof is given in Ref. \cite{HuVer03a}.

Once the fluctuations of the stress-energy operator have been
characterized we can  perturbatively extend the semiclassical
theory to account for such fluctuations. Thus we will assume that
the background spacetime metric $g_{ab}$ is a solution of the
semiclassical Einstein Eqs.~(\ref{2.5}) and we will write the new
metric for the extended theory as $g_{ab}+h_{ab}$, where we will
assume that $h_{ab}$ is a perturbation to the background solution.
The renormalized stress-energy operator and the state of the
quantum field may now be denoted by $\hat T_{ab}^R[g+h]$ and
$|\psi[g+h]\rangle$, respectively, and $\langle\hat
T_{ab}^R[g+h]\rangle$ will be the corresponding expectation value.

Let us now introduce a Gaussian stochastic tensor field
$\xi_{ab}[g;x)$ defined by the following correlators:
\begin{equation}
\langle\xi_{ab}[g;x)\rangle_s=0,\ \ \
\langle\xi_{ab}[g;x)\xi_{cd}[g;y)\rangle_s=
N_{abcd}[g;x,y),
\label{2.10}
\end{equation}
where $\langle\dots\rangle_s$ means statistical average. The
symmetry and positive semi-definite property of the noise kernel
guarantees that the stochastic field tensor $\xi_{ab}[g,x)$, or
$\xi_{ab}(x)$ for short, just introduced is well defined. Note
that this stochastic tensor captures only partially the quantum
nature of the fluctuations of the stress-energy operator since it
assumes that cumulants of higher order are zero.

An important property of this stochastic tensor is that it is
covariantly conserved in the background spacetime
$\nabla^a\xi_{ab}[g;x)=0$. In fact, as a consequence of the
conservation of $\hat T_{ab}^R[g]$ one can see that $\nabla_x^a
N_{abcd}(x,y)=0$. Taking the divergence in Eq.~(\ref{2.10}) one
can then show that $\langle\nabla^a\xi_{ab}\rangle_s=0$ and
$\langle\nabla_x^a\xi_{ab}(x) \xi_{cd}(y)\rangle_s=0$ so that
$\nabla^a\xi_{ab}$ is deterministic and represents with certainty
the zero vector field in $\cal{M}$.

For a conformal field, {\it i.e.}, a field whose classical action
is conformally invariant, $\xi_{ab}$ is traceless:
$g^{ab}\xi_{ab}[g;x)=0$; so that, for a conformal matter field the
stochastic source gives no correction to the trace anomaly. In
fact, from the trace anomaly result which states that $g^{ab}\hat
T^R_{ab}[g]$ is, in this case, a local c-number functional of
$g_{ab}$ times the identity operator, we have that
$g^{ab}(x)N_{abcd}[g;x,y)=0$. It then follows from Eq.
(\ref{2.10}) that $\langle g^{ab}\xi_{ab}\rangle_s=0$ and $\langle
g^{ab}(x)\xi_{ab}(x) \xi_{cd}(y)\rangle_s=0$; an alternative proof
based on the point-separation method is given in Ref.
\cite{PhiHu01,PhiHu03}, see also section \ref{sec4}.

All these properties make it quite natural to incorporate into the
Einstein equations the stress-energy fluctuations by using the
stochastic tensor $\xi_{ab}[g;x)$ as the source
of the metric perturbations.
Thus we will write the following equation.
\begin{equation}
G_{ab}[g\!+\!h]\!+\! \Lambda (g_{ab}\!+\!h_{ab}) -2(\alpha
A_{ab}+\beta B_{ab})[g\!+\!h]\!=\!8\pi G\!\left( \!\langle \hat
T_{ab}^R[g\!+\!h]\rangle \!+ \!\xi_{ab}[g]\!\right). \label{2.11}
\end{equation}
This equation is in the form of a {\it (semiclassical)
Einstein-Langevin equation}, it is a dynamical equation for the
metric perturbation $h_{ab}$ to linear order. It describes the
back-reaction of the metric to the quantum fluctuations of the
stress-energy tensor of matter fields, and gives a first order
extension to semiclassical gravity as described by the
semiclassical Einstein equation (\ref{2.5}).

Note that we refer to the  Einstein-Langevin equation as a first
order extension to semiclassical Einstein equation of
semiclassical gravity and the lowest level representation of
stochastic gravity. However, stochastic gravity has a much broader
meaning, it refers to the range of theories based on second and
higher order correlation functions. Noise can be defined in
effectively open systems (e.g. correlation noise \cite{CalHu00} in
the Schwinger-Dyson equation hierarchy) to some degree but one
should not expect the Langevin form to prevail. In this sense we
say stochastic gravity is the intermediate theory between
semiclassical gravity (a mean field theory based on the
expectation values of the energy momentum tensor of quantum
fields) and quantum gravity (the full hierarchy of correlation
functions retaining complete quantum coherence
\cite{stogra,kinQG}.

The renormalization of the operator $\hat T_{ab}[g+h]$ is carried
out exactly as in the previous case, now in the perturbed metric
$g_{ab}+h_{ab}$. Note that the stochastic source $\xi_{ab}[g;x)$
is not dynamical, it is independent of $h_{ab}$ since it describes
the fluctuations of the stress tensor on the semiclassical
background $g_{ab}$.

An important property of the Einstein-Langevin equation is that it is
gauge invariant under the change of $h_{ab}$ by
$h_{ab}^\prime =h_{ab} +\nabla_a\zeta_b+\nabla_b\zeta_a$, where
$\zeta^a$ is a stochastic vector field on the background manifold ${\cal
M}$. Note that a tensor such as
$R_{ab}[g+h]$, transforms as
$R_{ab}[g+h^\prime]=R_{ab}[g+h]+{\cal L}_\zeta R_{ab}[g]$ to linear order
in the perturbations, where ${\cal L}_\zeta $ is the Lie derivative with
respect to $\zeta^a$. Now, let us write the source tensors in
Eqs.~(\ref{2.11}) and (\ref{2.5}) to the left-hand sides of these
equations. If we substitute  $h$ by $h^\prime$ in this new version of Eq.
(\ref{2.11}), we get the same expression, with $h$ instead of $h^\prime$,
plus the Lie derivative of the combination of tensors which appear on
the left-hand side of the new Eq. (\ref{2.5}). This last combination
vanishes when Eq. (\ref{2.5}) is satisfied, {\it i.e.}, when the
background metric $g_{ab}$ is a solution of semiclassical gravity.

A solution of Eq.~(\ref{2.11}) can be formally written as
$h_{ab}[\xi]$. This solution is characterized by the whole
family of its correlation functions. From the statistical average
of this  equation we have that $g_{ab}+\langle h_{ab}\rangle_s$
must be a solution of the semiclassical Einstein equation linearized
around the background  $g_{ab}$; this solution has been proposed as a test
for the validity of the semiclassical approximation
\cite{AndMolMot02,AndMolMot03}. The
fluctuation of the metric around this average are described by the moments
of the stochastic field $h_{ab}^s[\xi]=h_{ab}[\xi]-\langle
h_{ab}\rangle_s$. Thus the solutions of the Einstein-Langevin equation
will provide the two-point metric correlation functions $\langle h_{ab}^s(x)
h_{cd}^s(y)\rangle_s$.

We see that whereas the semiclassical theory depends on the
expectation value of the point-defined value of the stress-energy
operator, the stochastic theory carries information also on the
two point correlation of the stress-energy operator. We should
also emphasize that, even if the metric fluctuations appears
classical and stochastic, their origin is  quantum not only
because they are induced by the fluctuations of quantum matter,
but also because they are the suitably coarse-grained variables
left over from the quantum gravity fluctuations after some
mechanism for decoherence and classicalization of the metric field
\cite{GelHar93,hartle,DowHal92,Hal93,Hal98,Whe98}. One may, in
fact, derive the stochastic semiclassical theory from a full
quantum theory. This was done via the world-line influence
functional method for a moving charged particle in an
electromagnetic field in quantum electrodynamics \cite{JohHu02}.
From another viewpoint, quite independent of whether a
classicalization mechanism is mandatory or implementable,  the
Einstein-Langevin equation proves to be a useful tool to compute
the symmetrized two point correlations of the quantum metric
perturbations \cite{RouVer03b}. This is illustrated in the linear
toy model discussed in Ref.~\cite{HuVer03a}, which has features of
some quantum Brownian models
\cite{CalRouVer03,CalRouVer01,CalRouVer02}.

\section{The Einstein-Langevin equation: Functional approach}
\label{sec3}

The Einstein-Langevin equation (\ref{2.11}) may also be derived by
a method based on functional techniques  \cite{MarVer99}. Here
we will summarize these techniques starting with semiclassical gravity.

In
semiclassical gravity functional methods were used to study the
back-reaction of quantum fields in cosmological models
\cite{Har77,FisHarHu79,HarHu79}. The primary advantage of the effective action
approach is, in addition to the well-known fact that it is easy to
introduce perturbation schemes like loop expansion and nPI
formalisms, that it yields a {\it fully} self-consistent solution.
For a general discussion in the semiclassical context of these two
approaches, equation of motion versus effective action, see, e.g.,
the work of Hu and Parker (1978) versus Hartle and Hu (1979) in
\cite{LukSta74,Gri76,HuPar77,HuPar78,Har81,And83,And84}.
See also comments in Sec. 5.6 on the black hole
backreaction problem comparing the approach by York et al.
\cite{Yor83,Yor85,Yor86} versus that of Sinha, Raval and Hu \cite{SinRavHu03}.

The well known in-out effective action method treated in
textbooks, however, led to equations of motion which were not real
because they were tailored to compute transition elements of
quantum operators rather than expectation values. The correct
technique to use for the backreaction problem is the
Schwinger-Keldysh
\cite{Sch61,BakMah63,Kel64,ChoEtal85,SuEtal88,CalHu89,CooEtal94}
closed-time-path (CTP) or `in-in'
effective action. These techniques were adapted to the
gravitational context
\cite{DeW86,Jor86,CalHu87,Jor87,Paz90,CamVer94} and applied to
different problems in cosmology. One could deduce the
semiclassical Einstein equation from the CTP effective action for
the gravitational field (at tree level) with quantum matter
fields.

Furthermore, in this case the CTP functional formalism turns out
to be related
\cite{SuEtal88,CalHu94,CamVer96,LomMaz96,GreMul97,CamHu98,CamHu99,%
Mor86,LeeBoy93,MarVer99,MarVer99b} to the influence functional
formalism of Feynman and Vernon \cite{FeyVer63} since the full
quantum system may be understood as consisting of a distinguished
subsystem (the ``system'' of interest) interacting with the
remaining degrees of freedom (the environment). Integrating out
the environment variables in a CTP path integral yields the
influence functional, from which one can define an effective
action for the dynamics of the system
\cite{CalHu94,HuSin95,HuMat94,GreMul97}. This approach to
semiclassical gravity is motivated by the observation
\cite{Physica} that in some open quantum systems classicalization
and decoherence
\cite{Zur81,Zur82,Zur86,Zur91,JooZeh85,CalLeg85,UnrZur89,Zur93,GiuEtal96}
on the system may be brought about by interaction with an
environment, the environment being in this case the matter fields
and some ``high-momentum'' gravitational modes
\cite{Kie87,Hal89,Pad89,Hu90,Cal89,Cal91,HuPazSin93,Whe98}.
Unfortunately, since the form of a complete quantum theory of
gravity interacting with matter is unknown, we do not know what
these ``high-momentum'' gravitational modes are. Such a
fundamental quantum theory might not even be a field theory, in
which case the metric and scalar fields would not be fundamental
objects \cite{stogra}. Thus, in this case, we cannot attempt to
evaluate the influence action of Feynman and Vernon starting from
the fundamental quantum theory and performing the path
integrations in the environment variables. Instead, we introduce
the influence action for an effective quantum field theory of
gravity and matter
\cite{Don94a,Don94b,Don96a,Don96b,SinHu91,Paz91,PazSin92}, in
which such ``high-momentum'' gravitational modes are assumed to
have already been ``integrated out.''


\subsection{Influence action for semiclassical gravity}


Let us formulate semiclassical gravity in this functional
framework. Adopting the usual procedure of effective field
theories \cite{Wei95,Wei96,Don94a,Don94b,Don96a,Don96b,CalKan97}, one
has to take the effective action for the metric and the scalar
field of the most general local form compatible with general
covariance: $S[g,\phi] \equiv S_g[g]+S_m[g,\phi]+ \cdots ,$ where
$S_g[g]$ and $S_m[g,\phi]$ are given by Eqs. (\ref{2.6}) and
(\ref{2.1}), respectively, and the dots stand for terms of order
higher than two in the curvature and in the number of derivatives
of the scalar field. Here, we shall neglect the higher order terms
as well as self-interaction terms for the scalar field. The second
order terms are necessary to renormalize one-loop ultraviolet
divergences of the scalar field stress-energy tensor, as we have
already seen. Since ${\cal M}$ is a globally hyperbolic manifold,
we can foliate it by a family of $t\!=\! {\rm constant}$ Cauchy
hypersurfaces $\Sigma_{t}$, and we will indicate the initial and
final times by $t_i$ and $t_f$, respectively.

The {\it influence functional} corresponding to the action
(\ref{2.1}) describing a scalar field in a spacetime (coupled to
a metric field) may be introduced as a functional of two copies
of the metric, denoted by $g_{ab}^+$ and $g_{ab}^-$, which
coincide at some final time $t=t_f$. Let us assume that, in the
quantum effective theory, the state of the full system (the
scalar and the metric fields) in the Schr\"{o}dinger picture at
the initial time $t\! =\! t_{i}$ can be described by a density
operator which can be written as the tensor product of two
operators on the Hilbert spaces of the metric and of the scalar
field. Let $\rho_i(t_i)\equiv
\rho_i \left[\phi_+(t_i),\phi_-(t_i) \right] $ be the
matrix element of the density operator $\hat{\rho}^{\rm
\scriptscriptstyle S}(t_{i})$ describing the initial state of the
scalar field. The Feynman-Vernon influence functional is defined
as the following path integral over the two copies of the scalar
field:
\begin{equation}
{\cal F}_{\rm IF}[g^\pm] \equiv
\int\! {\cal D}\phi_+\;
{\cal D}\phi_- \;
\rho_i (t_i)
\delta\!\left[\phi_+(t_f)\!-\!\phi_-(t_f)  \right]\:
e^{i\left(S_m[g^+,\phi_+]-S_m[g^-,\phi_-]\right) }.
\label{path integral}
\end{equation}
Alternatively, the  above double path integral can be rewritten
as a closed time path (CTP) integral, namely, as a single path
integral in a complex time contour with two different time
branches, one going forward in time from $t_i$ to $t_f$, and the
other going backward in time from $t_f$ to $t_i$ (in practice one
usually takes $t_i\to -\infty$). {}From this influence functional,
the {\it influence action} $S_{\rm IF}[g^+,g^-]$, or $S_{\rm
IF}[g^\pm]$ for short,  defined  by \be {\cal F}_{\rm IF}[g^\pm]
\equiv e^{i S_{\rm IF}[g^\pm]}, \label{influence functional} \ee
carries all the information about the environment (the matter
fields) relevant to the system (the gravitational field). Then we
can define the CTP {\it effective action} for the gravitational
field, $S_{\rm eff}[g^\pm]$, as
\begin{equation}
S_{\rm eff}[g^\pm]\equiv S_{g}[g^+]-S_{g}[g^-] +S_{\rm
IF}[g^\pm]. \label{ctpif}
\end{equation}
This is the effective action for the classical gravitational
field in the CTP formalism. However, since the gravitational
field is treated only at the tree level, this is also the
effective classical action from which the classical equations of
motion can be derived.

Expression (\ref{path integral}) contains ultraviolet divergences
and must be regularized. We shall assume that dimensional
regularization can be applied, that is, it makes sense to
dimensionally continue all the quantities that appear in Eq.
(\ref{path integral}).  For this we need to work with the
$n$-dimensional actions corresponding to $S_m$ in (\ref{path
integral}) and $S_g$ in (\ref{2.6}). For example,  the parameters
$G$, $\Lambda$ $\alpha$ and $\beta$ of Eq. (\ref{2.6}) are the
bare parameters $G_B$, $\Lambda_B$, $\alpha_B$ and $\beta_B$, and
in $S_g[g]$, instead of the square of the Weyl tensor in Eq.
(\ref{2.6}),  one must use $(2/3)(R_{abcd}R^{abcd}- R_{ab}R^{ab})$
which by the Gauss-Bonnet theorem leads to the same equations of
motion as the action (\ref{2.6}) when $n \!=\! 4$. The form of
$S_g$ in $n$ dimensions is suggested by the Schwinger-DeWitt
analysis of the ultraviolet divergences in the matter
stress-energy tensor using dimensional regularization. One can
then write the Feynman-Vernon effective action $S_{\rm
eff}[g^\pm]$ in Eq. (\ref{ctpif}) in a form suitable for
dimensional regularization. Since both $S_m$ and $S_g$ contain
second order derivatives of the metric, one should also add some
boundary terms \cite{Wal84,HuSin95}. The effect of these terms is
to cancel out the boundary terms which appear when taking
variations of $S_{\rm eff}[g^\pm]$ keeping the value of $g^+_{ab}$
and $g^-_{ab}$ fixed at $\Sigma_{t_i}$ and $\Sigma_{t_f}$.
Alternatively, in order to obtain the equations of motion for the
metric in the semiclassical regime, we can work with the action
terms  without boundary terms and neglect all boundary terms when
taking variations with respect to $g^{\pm}_{ab}$. From now on, all
the functional derivatives with respect to the metric will be
understood in this sense.

The semiclassical Einstein equation (\ref{2.5}) can now be derived.
Using the definition of the stress-energy tensor
$T^{ab}(x)=(2/\sqrt{-g})\delta S_m/\delta g_{ab}$
and the definition
of the influence functional, Eqs.
(\ref{path integral}) and (\ref{influence functional}), we see that
\begin{equation}
\langle \hat{T}^{ab}[g;x) \rangle =
\left. {2\over\sqrt{- g(x)}} \,
 \frac{\delta S_{\rm IF}[g^\pm]}
{\delta g^+_{ab}(x)} \right|_{g^\pm=g},
\label{s-t expect value}
\end{equation}
where the expectation value is taken in the $n$-dimensional
spacetime generalization of the state described by
$\hat{\rho}^{\rm \scriptscriptstyle S}(t_i)$. Therefore,
differentiating $S_{\rm eff}[g^\pm]$ in Eq. (\ref{ctpif}) with
respect to $g^+_{ab}$, and then setting
$g^+_{ab}=g^-_{ab}=g_{ab}$, we get the semiclassical Einstein
equation in $n$ dimensions. This equation is then renormalized by
absorbing the divergences in the regularized $\langle\hat
T^{ab}[g]\rangle$ into the bare parameters. Taking the limit
$n\to 4$ we obtain the physical semiclassical Einstein equation
(\ref{2.5}).


\subsection{Influence action for stochastic gravity}


In the spirit of the previous derivation of the Einstein-Langevin
equation, we now seek a dynamical equation for a linear
perturbation $h_{ab}$ to the semiclassical metric $g_{ab}$,
solution of Eq. (\ref{2.5}). Strictly speaking if we use
dimensional regularization we must consider the $n$-dimensional
version of that equation. {}From
the results just described, if such an equation were simply a
linearized semiclassical Einstein equation, it could be obtained
from an expansion of the effective action $S_{\rm eff}[g+h^\pm]$.
In particular, since, from Eq. (\ref{s-t expect value}), we have
that
\begin{equation}
\langle \hat{T}^{ab}[g+h;x) \rangle =
\left. {2\over\sqrt{-\det (g\!+\!h)(x)}} \,
 \frac{\delta S_{\rm IF}
   [g\!+\!h^\pm]}{\delta h^+_{ab}(x)}
 \right|_{h^\pm=h},
\label{perturb s-t expect value}
\end{equation}
the expansion of $\langle \hat{T}^{ab}[g\!+\!h]\rangle $
to linear order in $h_{ab}$ can be obtained from an expansion of the
influence action $S_{\rm IF}[g+h^\pm]$ up to second order
in $h^{\pm}_{ab}$.

To perform the expansion of the influence action,
we have to compute the first and second order
functional derivatives of $S_{\rm IF}[g+h^\pm]$
and then set $h^+_{ab}\!=\!h^-_{ab}\!=\!h_{ab}$.
If we do so using the path integral representation
(\ref{path integral}), we can interpret these derivatives as
expectation values of operators.
The relevant second order derivatives are
\begin{eqnarray}
\left. {4\over\sqrt{\!- g(x)}\sqrt{\!- g(y)} }
 \frac{\delta^2 S_{\rm IF}[g+h^\pm]}
{\delta h^+_{ab}\!(x)\delta h^+_{cd}\!(y)}
 \right|_{h^\pm=h} \!\!\!\!\!\!
&=& \!\!\!\!\!-H_{\scriptscriptstyle \! {\rm S}}^{abcd}[g;x,y)
\!-\!K^{abcd}[g;x,y)
\nonumber\\
&&+i N^{abcd}[g;x,y),      \nonumber \\
\left. {4\over\sqrt{\!- g(x)}\sqrt{\!- g(y)} }
 \frac{\delta^2 S_{\rm IF}[g^\pm]}
{\delta h^+_{ab}\!(x)\delta h^-_{cd}\!(y)}
 \right|_{h^\pm=h} \!\!\!\!\!\!
&=& \!\!\!\!\!-H_{\scriptscriptstyle \! {\rm A}}^{abcd} [g;x,y)
\!-\! i N^{abcd}[g;x,y)\!, \label{derivatives}
\end{eqnarray}
where
$$
N^{abcd}[g;x,y) \equiv
{1\over 2} \left\langle  \bigl\{
 \hat{t}^{ab}[g;x) , \,
 \hat{t}^{cd}[g;y)
 \bigr\} \right\rangle ,
$$
$$
H_{\scriptscriptstyle \!
{\rm S}}^{abcd}
[g;x,y) \equiv
{\rm Im} \left\langle {\rm T}^*\!\!
\left( \hat{T}^{ab}[g;x) \hat{T}^{cd}[g;y)
\right) \right\rangle \!,
$$
$$
H_{\scriptscriptstyle \!
{\rm A}}^{abcd}
[g;x,y) \equiv
-{i\over 2} \left\langle
\bigl[ \hat{T}^{ab}[g;x), \, \hat{T}^{cd}[g;y)
\bigr] \right\rangle \!,\,
$$
$$
K^{abcd}[g;x,y) \equiv
\left. {-4\over\sqrt{- g(x)}\sqrt{- g(y)} } \, \left\langle
\frac{\delta^2 S_m[g+h,\phi]}
{\delta h_{ab}(x)\delta h_{cd}(y)}
\right|_{\phi=\hat{\phi}}\right\rangle \!,
$$
with $\hat{t}^{ab}$ defined in Eq. (\ref{2.9}), $[ \; , \: ]$
denotes the commutator and $\{ \; , \: \}$ the anti-commutator.
Here we use a Weyl ordering prescription for the operators.
The symbol ${\rm T}^*$ denotes the
following ordered operations: First, time order the field
operators $\hat{\phi}$ and then apply the derivative operators
which appear in each term of the product $T^{ab}(x) T^{cd}(y)$,
where $T^{ab}$ is the functional (\ref{2.3}). This ${\rm T}^{*}$
``time ordering'' arises because we have path integrals
containing products of derivatives of the field, which can be
expressed as derivatives of the path integrals which do not
contain such derivatives. Notice, from their definitions, that
all the kernels which appear in expressions (\ref{derivatives})
are real and also $H_{\scriptscriptstyle \!{\rm A}}^{abcd}$ is
free of ultraviolet divergences in the limit $n \to 4$.

{}From (\ref{s-t expect value}) and
(\ref{derivatives}), since
$S_{\rm IF}[g,g]=0$ and
$S_{\rm IF}[g^-,g^+]=
-S^{ {\displaystyle \ast}}_{\rm IF}[g^+,g^-]$, we can write the
expansion for the influence action
$S_{\rm IF}[g\!+\!h^\pm]$ around a background
metric $g_{ab}$ in terms of the previous kernels.
Taking into account that
these kernels satisfy the symmetry relations
\begin{equation}
H_{\scriptscriptstyle \!{\rm S}}^{abcd}(x,y)\!=\!
H_{\scriptscriptstyle \!{\rm S}}^{cdab}(y,x),
H_{\scriptscriptstyle \!{\rm A}}^{abcd}(x,y)\!=\!
-H_{\scriptscriptstyle \!{\rm A}}^{cdab}(y,x),
K^{abcd}(x,y) \!= \! K^{cdab}(y,x),
\label{symmetries}
\end{equation}
and introducing the new kernel
\begin{equation}
H^{abcd}(x,y)\equiv
H_{\scriptscriptstyle \!{\rm S}}^{abcd}(x,y)
+H_{\scriptscriptstyle \!{\rm A}}^{abcd}(x,y),
\label{H}
\end{equation}
the expansion of $S_{\rm IF}$ can be finally written as
\begin{eqnarray}
S_{\rm IF}[g\!+\!h^\pm]
&=& {1\over 2} \int\! d^4x\, \sqrt{- g(x)}\:
\langle \hat{T}^{ab}[g;x) \rangle  \,
\left[h_{ab}(x) \right] \nonumber\\
&&-{1\over 8} \int\! d^4x\, d^4y\, \sqrt{- g(x)}\sqrt{- g(y)}\,
\nonumber  \\
&& \ \ \ \ \times\left[h_{ab}(x)\right]
\left(H^{abcd}[g;x,y)\!
+\!K^{abcd}[g;x,y) \right)
\left\{ h_{cd}(y) \right\}  \nonumber  \\
&&
+{i\over 8} \int\! d^4x\, d^4y\, \sqrt{- g(x)}\sqrt{- g(y)}\,
\nonumber  \\
&& \ \ \ \ \times\left[h_{ab}(x) \right]
N^{abcd}[g;x,y)
\left[h_{cd}(y) \right]+0(h^3),
\label{expansion 2}
\end{eqnarray}
where we have used the notation
\begin{equation}
\left[h_{ab}\right] \equiv h^+_{ab}\!-\!h^-_{ab},
\hspace{5 ex}
\left\{ h_{ab}\right\} \equiv h^+_{ab}\!+\!h^-_{ab}.
\label{notation}
\end{equation}
{}From Eqs.~(\ref{expansion 2}) and
(\ref{perturb s-t expect value})
it is clear that the imaginary part of the
influence action does not contribute to the perturbed
semiclassical Einstein equation (the expectation value of the
stress-energy tensor is real), however, as it depends on the noise kernel,
it contains information on the fluctuations of the operator
$\hat{T}^{ab}[g]$.

We are now in a position to carry out the derivation of the
semiclassical Einstein-Langevin equation. The procedure is well
known
\cite{CalHu94,HuSin95,CamVer96,GleRam94,BoyEtal95,YamYok97,RamHuSty98}:
it consists of deriving a new ``stochastic'' effective action from
the observation that the effect of the imaginary part of the
influence action (\ref{expansion 2}) on the corresponding
influence functional is equivalent to the averaged effect of the
stochastic source $\xi^{ab}$ coupled linearly to the perturbations
$h_{ab}^{\pm}$. This observation follows from the identity first
invoked by Feynman and Vernon for such purpose:
\begin{eqnarray}
&&\exp\left(-{1\over 8} \!\int\! d^4x\, d^4y \, \sqrt{- g(x)}\sqrt{- g(y)}\,
\left[h_{ab}(x) \right]\,
N^{abcd}(x,y)\, \left[h_{cd}(y)\right] \right)
\nonumber  \\
&&\quad\quad =
\int\! {\cal D}\xi \: {\cal P}[\xi]\, \exp\left({i\over 2} \!\int\! d^4x \,
\sqrt{- g(x)}\,\xi^{ab}(x)\,\left[h_{ab}(x) \right] \right),
\label{Gaussian path integral}
\end{eqnarray}
where ${\cal P}[\xi]$ is the probability distribution
functional of a Gaussian stochastic tensor $\xi^{ab}$
characterized by the correlators (\ref{2.10})
with $N^{abcd}$ given by Eq.~(\ref{2.8}),
and where
the path integration measure is assumed to be a scalar under
diffeomorphisms of $({\cal M},g_{ab})$. The above identity follows
from the identification of the right-hand side of
(\ref{Gaussian path integral}) with the characteristic functional for
the stochastic field $\xi^{ab}$. The
probability distribution functional for $\xi^{ab}$ is explicitly
given by
\begin{equation}
{\cal P}[\xi]\!=\! {\rm det}(2\pi N)^{-1/2}\!\!
 \exp\!\left[\!-{1\over2}\!\int\!\! d^4x d^4y \!
\sqrt{\!-g(x)}\sqrt{\!-g(y)}\!
 \xi^{ab}\!(x) \! N^{-1}_{abcd}(x,y)\! \xi^{cd}\!(y)\!\right]\!.
\end{equation}

We may now introduce the {\it stochastic effective action} as
\begin{equation}
S^s_{\rm eff}[g+h^\pm,\xi] \equiv S_{g}[g+h^+]-S_{g}[g+h^-]+
S^s_{\rm IF}[g+h^\pm,\xi],
\label{stochastic eff action}
\end{equation}
where the ``stochastic'' influence action is defined as
\begin{equation}
S^s_{\rm IF}[g+h^\pm,\xi] \equiv {\rm Re}\, S_{\rm
IF}[g\!+\!h^\pm]+\! {1\over 2} \int\! d^4x \, \sqrt{-
g(x)}\,\xi^{ab}(x)\left[h_{ab}(x) \right]+ O(h^3). \label{eff
influence action}
\end{equation}
Note that, in fact, the influence functional can now be written as a
statistical average over $\xi^{ab}$:
$
{\cal F}_{\rm IF}[g+h^\pm]= \left\langle
\exp\left(i S^s_{\rm IF}[g+h^\pm,\xi]\right)
\right\rangle_{\! s}.
$
The stochastic equation of motion for $h_{ab}$ reads
\begin{equation}
\left.
\frac{\delta S^s_{\rm eff}[g+h^\pm,\xi]}{\delta h^+_{ab}(x)}
\right|_{h^\pm=h}=0,
\label{eq of motion}
\end{equation}
which is the Einstein-Langevin equation (\ref{2.11}); notice that only the
real part of $S_{IF}$ contributes to the expectation value
(\ref{perturb s-t expect value}).
To be precise we get
first the regularized
$n$-dimensional equations with the bare parameters,
and where instead of the tensor $A^{ab}$ we get
$(2/3)D^{ab}$ where
\begin{eqnarray}
D^{ab} &\equiv & {1\over\sqrt{- g}}   \frac{\delta}{\delta g_{ab}}
          \int \! d^n x \,\sqrt{- g}
\left(R_{cdef}R^{cdef}-
                         R_{cd}R^{cd}  \right)
\nonumber\\
   &=& {1\over2}\, g^{ab} \!
\left(  R_{cdef} R^{cdef}-
         R_{cd}R^{cd}+\Box  R \right)
      -2R^{acde}{R^b}_{cde}
\nonumber \\
&&
      -2 R^{acbd}R_{cd}
      +4R^{ac}{R_c}^b
      -3 \Box  R^{ab}
  +\nabla^{a}\nabla^{b} R.
\label{D}
\end{eqnarray}
Of course, when $n=4$ these tensors are related,
$A^{ab}=(2/3) D^{ab}$. After that
we renormalize and
take the limit $n\to 4$ to obtain the Einstein-Langevin
equations in the physical spacetime.


\subsection{Explicit form of the Einstein-Langevin equation}


We can write the Einstein-Langevin equation in a more explicit
form by working out the expansion of $\langle
\hat{T}^{ab}[g\!+\!h]\rangle $ up to linear order in the
perturbation $h_{ab}$. {}From Eq. (\ref{perturb s-t expect
value}), we see that this expansion can be easily obtained from
(\ref{expansion 2}). The result is
\begin{eqnarray}
\langle \hat{T}_n^{ab}[g\!+\!h;x) \rangle
\! &=&\!
\langle \hat{T}_n^{ab}[g,x) \rangle
 + \langle
\hat{T}_n^{{\scriptscriptstyle (1)}\hspace{0.1ex} ab} [g,h;x)
\hspace{-0.1ex} \rangle
\nonumber  \\
&&- \frac{1}{2} \!\int\! \hspace{-0.2ex}
d^ny \, \sqrt{- g(y)} \hspace{0.2ex}  H_n^{abcd}[g;x,y) h_{cd}(y)
+ 0(h^2). \label{s-t expect value expansion}
\end{eqnarray}
Here we use a subscript $n$ on a given tensor to indicate that we
are explicitly working in $n$-dimensions, as we use dimensional
regularization, and we also use the superindex ${\scriptstyle
(1)}$ to generally indicate that the tensor is the first order
correction, linear in $h_{ab}$, in a perturbative expansion around
the background $g_{ab}$.

Using the Klein-Gordon equation (\ref{2.2}), and expressions
(\ref{2.3})  for the stress-energy tensor and the corresponding
operator, we can write \be \hat{T}_n^{{\scriptscriptstyle
(1)}\hspace{0.1ex} ab} [g,h]=\left({1\over 2}\, g^{ab}h_{cd}-
\delta^a_c h^b_d- \delta^b_c h^a_d  \right) \hat{T}_{n}^{cd}[g]
+{\cal F}^{ab}[g,h]\, \hat{\phi}_{n}^2[g], \label{T(1) operator}
\ee where ${\cal F}^{ab}[g;h]$ is the differential operator \bea
{\cal F}^{ab}\!\!\!\!\!&\equiv&\!\!\!\! \!\left(\xi\!-\!{1\over
4}\right)\!\! \left(h^{ab}\!-\!{1\over 2}\, g^{ab} h^c_c \right)\!
\Box
\nonumber  \\
&&\!\!\!\!\!+\!{\xi \over 2} \left[ \nabla^{c} \nabla^{a} h^b_c+
\nabla^{c} \nabla^{b} h^a_c\!- \!\Box h^{ab}\!-\! \nabla^{a}
\nabla^{b}  h^c_c\!-\! g^{ab} \nabla^{c} \nabla^{d} h_{cd}+g^{ab}
\Box h^c_c
\right.   \nonumber \\
&&\left. \!\!\!\!\!+\!\left(\! \nabla^{a} h^b_c\!+\! \nabla^{b}
h^a_c\!-\!\nabla_{c} h^{ab}\!\!-\! 2 g^{ab} \nabla^{d} h_{cd} \!+
\!g^{ab} \nabla_{c}  h^d_d \right)\! \nabla^{c} \!\!-\!g^{ab}
h_{cd} \nabla^{c} \nabla^{d} \right]\!\!. \label{diff operator F}
\eea It is understood that indices are raised with the background
inverse metric $g^{ab}$ and that all the covariant derivatives are
associated to the metric $g_{ab}$.

Substituting (\ref{s-t expect value expansion}) into
the $n$-dimensional version of the Einstein-Langevin
Eq. (\ref{2.11}),
taking into account that
$g_{ab}$ satisfies the semiclassical Einstein equation
(\ref{2.5}), and substituting expression (\ref{T(1) operator})
we can write the Einstein-Langevin
equation in dimensional regularization as
\bea
&&{1\over 8 \pi G_{B}}\Biggl[
G^{{\scriptscriptstyle (1)}\hspace{0.1ex} ab}\!-\!
{1\over 2}\, g^{ab} G^{cd}
h_{cd}+ G^{ac} h^b_c+G^{bc} h^a_c+
\Lambda_{B} \left( h^{ab}\!-\!{1\over 2}\,
g^{ab} h^c_c \right)
\Biggr]
   \nn \\
&&
- \,
{4\alpha_{B}\over 3}  \left( D^{{\scriptscriptstyle
(1)}ab}
-{1\over 2} g^{ab} D^{cd} h_{cd}+
D^{ac} h^b_c+D^{bc} h^a_c
\right)\!
\nonumber  \\
&&-2\beta_{B}\!\left( B^{{\scriptscriptstyle (1)}ab}\!-\!
{1\over 2} g^{ab} B^{cd}
h_{cd}+ B^{ac} h^b_c+B^{bc} h^a_c
\right)   \nn \\
&&- \, \mu^{-(n-4)}\, {\cal F}^{ab}_x \langle
\hat{\phi}_{n}^2[g;x) \rangle +{1\over 2} \!\int\! d^ny  \sqrt{-
g(y)}\, \mu^{-(n-4)} H_n^{abcd}[g;x,y) h_{cd}(y)
\nonumber  \\
 &&= \mu^{-(n-4)}
\xi^{ab}_n, \label{Einstein-Langevin eq 3} \eea where the tensors
$G^{ab}$, $D^{ab}$ and $B^{ab}$ are computed from the
semiclassical metric $g_{ab}$, and where we have omitted the
functional dependence on $g_{ab}$ and $h_{ab}$ in
$G^{{\scriptscriptstyle (1)}ab}$, $D^{{\scriptscriptstyle
(1)}ab}$, $B^{{\scriptscriptstyle (1)}ab}$ and ${\cal F}^{ab}$ to
simplify the notation. The parameter $\mu$ is a mass scale which
relates the dimensions of the physical field $\phi$ with the
dimensions of the corresponding field in $n$-dimensions,
$\phi_n=\mu^{(n-4)/2}\phi$. Notice that, in Eq.
(\ref{Einstein-Langevin eq 3}), all the ultraviolet divergences
in the limit $n\to 4$, which must be removed by renormalization of
the coupling constants, are in $\langle \hat{\phi}_{n}^2(x)
\rangle$ and the symmetric part $H_{\scriptscriptstyle \! {\rm
S}_{\scriptstyle n}}^{abcd}(x,y)$ of the kernel $H_n^{abcd}(x,y)$,
whereas the kernels $N_n^{abcd}(x,y)$ and $H_{\scriptscriptstyle
\! {\rm A}_{\scriptstyle n}}^{abcd}(x,y)$ are free of ultraviolet
divergences. If we introduce the bi-tensor $F_{n}^{abcd}[g;x,y)$
defined by
\begin{equation}
F_{n}^{abcd}[g;x,y) \equiv
\left\langle \hat{t}_n^{ab}[g;x)\,
\hat{t}_n^{\rho\sigma}[g;y)
  \right\rangle
\label{bitensor F}
\end{equation}
where $\hat t^{ab}$ is defined by Eq. (\ref{2.9}), then the
kernels $N$ and $H_A$ can be written as \be N_n^{abcd}[g;x,y)=
{\rm Re} \, F_{n}^{abcd}[g;x,y), \hspace{7ex}
H_{\scriptscriptstyle \! {\rm A}_{\scriptstyle n}}^{abcd}[g;x,y)=
{\rm Im} \, F_{n}^{abcd}[g;x,y), \label{finite kernels} \ee where
we have used that $2 \langle \hat{t}^{ab}(x)\, \hat{t}^{cd}(y)
\rangle= \langle \{ \hat{t}^{ab}(x), \, \hat{t}^{cd}(y) \}\rangle
+ \langle [ \hat{t}^{ab}(x), \, \hat{t}^{cd}(y)]\rangle$, and the
fact that the first term on the right hand side of this identity
is real, whereas the second one is pure imaginary. Once we perform
the renormalization procedure in Eq.~(\ref{Einstein-Langevin eq
3}), setting $n = 4$ will yield the physical Einstein-Langevin
equation. Due to the presence of the kernel $H_n^{abcd}(x,y)$,
this equation will be usually non-local in the metric
perturbation. In section \ref{sec:flucminspa} we will carry out an
explicit evaluation of the physical Einstein-Langevin equation
which will illustrate the procedure.


\subsubsection{The kernels for the vacuum state}


When the expectation values in the Einstein-Langevin equation are
taken in a vacuum state $|0 \rangle$, such as, for instance, an
``in'' vacuum, we can be more explicit, since we can write the
expectation values in terms of the Wightman and Feynman functions,
defined as \be G_n^+[g;x,y)\! \equiv\! \langle 0| \!
   \hat{\phi}_{n}[g;x)  \hat{\phi}_{n}[g;y) \,
   \!|0 \rangle ,
i G\!_{\scriptscriptstyle F_{\scriptstyle \hspace{0.1ex}  n}}
 \hspace{-0.2ex}[g;x,y)
  \!\equiv\! \langle 0| \!
  {\rm T}\! \left(\!\hat{\phi}_{n}[g;x)  \hat{\phi}_{n}[g;y)\! \right)
  \!
  |0 \rangle.
\label{Wightman and Feynman functions}
\ee
These expressions for the kernels in the Einstein-Langevin
equation will be very useful for explicit
calculations.
To simplify the notation, we omit the functional
dependence on the semiclassical metric $g_{ab}$, which will be
understood in all the expressions below.

{}From Eqs. (\ref{finite kernels}), we see that the kernels
$N_n^{abcd}(x,y)$ and
$H_{\scriptscriptstyle \!
{\rm A}_{\scriptstyle n}}^{abcd}(x,y)$
are the real and imaginary parts,
respectively, of the bi-tensor
$F_{n}^{abcd}(x,y)$.
{}From the expression (\ref{regul s-t 2})
we see that
the stress-energy operator $\hat{T}_n^{ab}$
can be written as a
sum of terms of the form $\left\{ {\cal A}_x \hat{\phi}_{n}(x),
\,{\cal B}_x \hat{\phi}_{n}(x)\right\}$, where ${\cal A}_x$ and
${\cal B}_x$ are some differential operators. It  then follows
that we can express the bi-tensor
$F_{n}^{abcd}(x,y)$ in
terms of the Wightman function as
\bea
F_{n}^{abcd}(x,y)
\!\!\!\!&=&\!\!\!\!\nabla^{a}_x
 \nabla^{c}_y G_n^+(x,y)
 \nabla^{b}_x
 \nabla^{d}_y G_n^+(x,y)
+\nabla^{a}_x
 \nabla^{d}_y G_n^+(x,y)
 \nabla^{b}_x
 \nabla^{c}_y G_n^+(x,y)
   \nn \\
&&
+\, 2 {\cal D}^{ab}_{x}  \bigl(
  \nabla^{c}_y G_n^+(x,y)
  \nabla^{d}_y G_n^+(x,y) \bigr)
\nonumber  \\
&&
+2 {\cal D}^{cd}_{y} \bigl(
  \nabla^{a}_x G_n^+(x,y)
  \nabla^{b}_x G_n^+(x,y) \bigr)
+2 {\cal D}^{ab}_{x}
   {\cal D}^{cd}_{y}  \bigl(
 G_n^{+ 2}(x,y)  \bigr),
\label{Wightman expression 2}
\eea
where ${\cal D}^{ab}_{x}$ is the differential
operator (\ref{diff operator}).
{}From this expression and the relations
(\ref{finite kernels}), we get expressions for the kernels
$N_n$ and
$H_{\scriptscriptstyle \!{\rm A}_{\scriptstyle n}}$ in
terms of the Wightman function $G_n^+(x,y)$.

Similarly the kernel $H_{\scriptscriptstyle \! {\rm
S}_{\scriptstyle n}}^{abcd}(x,y)$, can be written in terms of the
Feynman function as \bea H_{\scriptscriptstyle \! {\rm
S}_{\scriptstyle n}}^{abcd}(x,y)&\!\!\!\!\!\!=\!\!\!\!\!\!& -
{\rm Im} \Bigl[
 \nabla^{a}_{{x}}
 \nabla^{c}_{{y}}
     G\!_{\scriptscriptstyle F_{\scriptstyle \hspace{0.1ex}  n}}
 \hspace{-0.2ex}(x,y)
 \nabla^{b}_{{x}}
 \nabla^{d}_{{y}}
     G\!_{\scriptscriptstyle F_{\scriptstyle \hspace{0.1ex}  n}}
 \hspace{-0.2ex}(x,y)
\nonumber  \\
&&
+\nabla^{a}_{{x}}
 \nabla^{d}_{{y}}
     G\!_{\scriptscriptstyle F_{\scriptstyle \hspace{0.1ex}  n}}
 \hspace{-0.2ex}(x,y)
 \nabla^{b}_{{x}}
 \nabla^{c}_{{y}}
     G\!_{\scriptscriptstyle F_{\scriptstyle \hspace{0.1ex}  n}}
 \hspace{-0.2ex}(x,y)   \nn \\
&&
-\,g^{ab}(x) \nabla^{e}_{{x}}
 \nabla^{c}_{{y}}
     G\!_{\scriptscriptstyle F_{\scriptstyle \hspace{0.1ex}  n}}
 \hspace{-0.2ex}(x,y)
 \nabla_{\!\!e}^{{x}}
 \nabla^{d}_{{y}}
     G\!_{\scriptscriptstyle F_{\scriptstyle \hspace{0.1ex}  n}}
 \hspace{-0.2ex}(x,y)
\nonumber  \\
&&
-g^{cd}(y) \nabla^{a}_{{x}}
 \nabla^{e}_{{y}}
     G\!_{\scriptscriptstyle F_{\scriptstyle \hspace{0.1ex}  n}}
 \hspace{-0.2ex}(x,y)
 \nabla^{b}_{{x}}
 \nabla_{\!\!e}^{{y}}
     G\!_{\scriptscriptstyle F_{\scriptstyle \hspace{0.1ex}  n}}
 \hspace{-0.2ex}(x,y)    \nn  \\
&&
+\,{1 \over 2}\, g^{ab}(x) g^{cd}(y)
 \nabla^{e}_{{x}}\!
 \nabla^{f}_{{y}}
     G\!_{\scriptscriptstyle F_{\scriptstyle \hspace{0.1ex}  n}}
 \hspace{-0.2ex}(x,y)
 \nabla_{\!\!e}^{{x} }
 \nabla_{\!\!f}^{{y} }
     G\!_{\scriptscriptstyle F_{\scriptstyle \hspace{0.1ex}  n}}
 \hspace{-0.2ex}(x,y)
\nonumber  \\
&&
+{\cal K}^{ab}_{ x}  \bigl(
 2 \hspace{-0.2ex} \nabla^{c}_{{y}}
   G\!_{\scriptscriptstyle F_{\scriptstyle \hspace{0.1ex}  n}}
   \hspace{-0.2ex}(x,y)
 \nabla^{d}_{{y}}
   G\!_{\scriptscriptstyle F_{\scriptstyle \hspace{0.1ex}  n}}
   \hspace{-0.2ex}(x,y)
     \nn   \\
&&
  -\, g^{cd}(y) \nabla^{e}_{{y}}
   G\!_{\scriptscriptstyle F_{\scriptstyle \hspace{0.1ex}  n}}
   \hspace{-0.2ex}(x,y)
\nabla_{\!\!e}^{{y} }
     G\!_{\scriptscriptstyle F_{\scriptstyle \hspace{0.1ex}  n}}
     \hspace{-0.2ex}(x,y) \bigr)
\nonumber  \\
&&
+{\cal K}^{cd}_{y}  \bigl(
 2 \hspace{-0.2ex} \nabla^{a}_{{x}}
   G\!_{\scriptscriptstyle F_{\scriptstyle \hspace{0.1ex}  n}}
   \hspace{-0.2ex}(x,y)
 \nabla^{b}_{{x}}
   G\!_{\scriptscriptstyle F_{\scriptstyle \hspace{0.1ex}  n}}
   \hspace{-0.2ex}(x,y)
     \nn   \\
&&
-\! g^{ab}(x) \nabla^{e}_{{x}}
   G\!_{\scriptscriptstyle F_{\scriptstyle \hspace{0.1ex}  n}}
   \hspace{-0.2ex}(x,y)
\nabla_{\!\!e}^{{x} }
     G\!_{\scriptscriptstyle F_{\scriptstyle \hspace{0.1ex}  n}}
     \hspace{-0.2ex}(x,y) \bigr)
\!+\!2 {\cal K}^{ab}_{x}
   {\cal K}^{cd}_{y} \! \bigl(
   G\!_{\scriptscriptstyle F_{\scriptstyle \hspace{0.1ex}  n}}^{\;\: 2}
   \hspace{-0.2ex}(x,y)  \bigr) \Bigr],
\label{Feynman expression 2}
\eea
where ${\cal K}^{ab}_{x}$ is the differential
operator
\be
{\cal K}^{ab}_{x} \equiv
\xi \left( g^{ab}(x) \Box_{x}
  -\nabla^{a}_{{x}}
   \nabla^{b}_{{x}}+
G^{ab}(x) \right)
-{1 \over 2}\, m^2 g^{ab}(x).
\label{diff operator K}
\ee
Note that, in the vacuum state
$|0 \rangle$, the term
$\langle \hat{\phi}_{n}^2 (x) \rangle$ in
equation (\ref{Einstein-Langevin eq 3}) can also be written as
$\langle \hat{\phi}_{n}^2(x) \rangle=
i G\!_{\scriptscriptstyle F_{\scriptstyle \hspace{0.1ex}  n}}
      \hspace{-0.2ex}(x,x)=G_n^+(x,x)$.

Finally, the causality of the Einstein-Langevin equation
(\ref{Einstein-Langevin eq 3})
can be explicitly seen as follows. The non-local terms in that
equation are due to the kernel $H(x,y)$ which is defined in Eq.
(\ref{H}) as the sum of $H_S(x,y)$ and $H_A(x,y)$. Now,
when the points $x$ and $y$ are spacelike
separated, $\hat{\phi}_{n}(x)$ and $\hat{\phi}_{n}(y)$ commute and,
thus, $G_n^+(x,y) \!=\!
i G\!_{\scriptscriptstyle F_{\scriptstyle \hspace{0.1ex}  n}}
 \hspace{-0.2ex}(x,y) \!=\!
(1/2) \langle 0| \, \{ \hat{\phi}_{n}(x) , \hat{\phi}_{n}(y) \} \,
|0 \rangle$, which is real. Hence, from the above expressions, we
have that $H_{\scriptscriptstyle \! {\rm A}_{\scriptstyle
n}}^{abcd}(x,y) \!=\! H_{\scriptscriptstyle \! {\rm
S}_{\scriptstyle n}}^{abcd}(x,y) \!=\! 0$, and thus
$H_n^{abcd}(x,y)=0$. This fact is expected since, from the
causality of the expectation value of the stress-energy operator
\cite{Wal77}, we know that the non-local dependence on the metric
perturbation in the Einstein-Langevin equation, see Eq.
(\ref{2.11}), must be causal. See Ref.~\cite{HuVer03a} for an
alternative proof of the causal nature of the Einstein-Langevin
equation.

\section{Noise Kernel and Point-Separation}
\label{sec4}

In this section we explore further the properties  of the noise
kernel and the stress energy bi-tensor. Similar to what was done
for the stress energy tensor it is desirable to relate the noise
kernel defined at separated points to the Green function of a
quantum field. We pointed out earlier \cite{stogra} that field
quantities defined at two separated points may possess important
information which could be the starting point for probes into
possible extended structures of spacetime. Of more practical
concern is how one can define a finite quantity at one point or in
some small region around it from the noise kernel defined at two
separated points. When we refer to, say, the fluctuations of
energy density in ordinary (point-wise) quantum field theory, we
are in actuality asking such a question. This is essential for
addressing fundamental issues like a) the validity of
semiclassical gravity \cite{KuoFor93} -- e.g., whether the
fluctuations to mean ratio is a correct criterion
\cite{HuPhi00,PhiHu00,ForSCG,ForWu,AndMolMot02,AndMolMot03}; b)
Whether the fluctuations in the vacuum energy density which drives
some models of inflationary cosmology violates the positive energy
condition; c) Physical effects of black hole horizon fluctuations
and Hawking radiation backreaction -- to begin with, is the
fluctuations finite or infinite?  d) General relativity as a low
energy effective theory in the geometro-hydrodynamic limit towards a
kinetic theory approach to quantum gravity
\cite{grhydro,stogra,kinQG}.

Thus, for comparison with ordinary phenomena at low energy we need
to find a reasonable prescription for obtaining a finite quantity
of the noise kernel in the limit of ordinary (point-defined)
quantum field theory. Regularization schemes used in obtaining a
finite expression for the stress energy tensor  have been applied
to the noise kernel\footnote{It is well-known that several
regularization methods can work equally well for the removal of
ultraviolet divergences in the stress energy tensor of quantum
fields  in curved spacetime. Their mutual relations are known, and
discrepancies explained. This formal structure of regularization
schemes for quantum fields in curved spacetime should remain
intact when applied to the regularization of the noise kernel in
general curved spacetimes; it is the meaning and relevance of
regularization of the noise kernel which is more of a concern (see
comments below).  Specific considerations will of course enter for
each method. But for the methods employed so far, such as
zeta-function, point separation, dimensional, smeared-field,
applied to simple cases (Casimir, Einstein, thermal fields) there
is no new inconsistency or discrepancy.}. This includes the simple
normal ordering \cite{KuoFor93,WuFor01} and smeared field operator
\cite{PhiHu00} methods applied to the Minkowski and Casimir
spaces, zeta-function \cite{EliEtal94,Kir00,Cam90} for spacetimes
with an Euclidean section, applied to the Casimir effect
\cite{CogGuiEli02} and the Einstein Universe \cite{PhiHu97}, or the covariant
point-separation methods applied to the Minkowski \cite{PhiHu00},
hot flat space and the Schwarzschild spacetime \cite{PhiHu03}.
There are differences and deliberations on
whether it is meaningful to seek a
point-wise expression for the noise kernel, and if so what is the
correct way to proceed -- e.g., regularization by a subtraction
scheme or by integrating over a test-field. Intuitively the smear
field method \cite{PhiHu00} may better preserve the
integrity of the noise kernel as it provides a sampling of the two
point function rather than using
a subtraction scheme which alters its innate
properties by forcing a nonlocal quantity into a local one. More
investigation is needed to clarify these points, which bear on
important issues like the validity of semiclassical gravity. We
shall set a more modest goal here, to derive a general expression
for the noise kernel for quantum fields in an arbitrary curved
spacetime in terms of Green functions and leave the discussion of
point-wise limit to a later date. For this purpose the covariant
point-separation method which highlights the bi-tensor features,
when used not as a regularization scheme, is perhaps closest to
the spirit of stochastic gravity.

The task of finding a general expression of the noise-kernel for
quantum fields in curved spacetimes was carried out by Phillips
and Hu in two papers using the ``modified'' point separation
scheme \cite{Wal75,AdlLieNg77,Wal78}. Their first paper
\cite{PhiHu01} begins with a discussion of the procedures for
dealing with the quantum stress tensor bi-operator at two
separated points, and ends with a general expression of the noise
kernel defined at separated points expressed as products of
covariant derivatives up to the fourth order of the quantum
field's Green function. (The stress tensor involves up to two
covariant derivatives.) This result holds for $x\ne y$ without
recourse to renormalization of the Green function, showing that
$N_{abc'd'}(x,y)$ is always finite for $x\ne y$ (and off the light
cone for massless theories). In particular, for a massless
conformally coupled free scalar field on a four dimensional
manifold they computed the trace of the noise kernel at both
points and found this double trace vanishes identically. This
implies that there is no stochastic correction to the trace
anomaly  for massless conformal fields, in agreement with results
arrived at in Refs. \cite{CalHu94,CamVer96,MarVer99} (see also
section \ref{sec2}). In their second paper \cite{PhiHu03} a
Gaussian approximation for the Green function (which is what
limits the accuracy of the results) is used to derive finite
expressions for two specific classes of spacetimes, ultrastatic
spacetimes, such as the hot flat space, and the conformally-
ultrastatic spacetimes, such as the Schwarzschild spacetime.
Again, the validity of these results may depend on how we view the
relevance and meaning of regularization. We will only report the
result of their first paper here.


\subsection{Point Separation}

The point separation scheme introduced in the 60's by DeWitt
\cite{DeW65}  was brought to more popular use in the 70's in the
context of quantum field theory in curved spacetimes
\cite{DeW75,Chr76,Chr78} as a means for obtaining a finite
quantum stress tensor.  Since the stress-energy tensor is built
from the product of a pair of field operators evaluated at a
single point, it is not well-defined. In this scheme, one
introduces an artificial separation of the single point $x$ to a
pair of closely separated points $x$ and $x'$. The problematic
terms involving field products such as $\hat\phi(x)^2$ becomes
$\hat\phi(x)\hat\phi(x')$, whose expectation value is well
defined. If one is interested in the low energy behavior captured
by the point-defined quantum field theory -- as the effort in the
70's was directed -- one takes the coincidence limit. Once the
divergences present are identified, they may be  removed
(regularization) or moved (by renormalizing the coupling
constants), to produce  a well-defined, finite stress tensor at a
single point.

Thus the first order of business is  the construction of the
stress tensor and then derive the symmetric stress-energy tensor
two point function, the noise kernel, in terms of the Wightman
Green function. In this section we will use the traditional
notation for index tensors in the point-separation context.

\subsubsection{n-tensors and end-point expansions}

An object like the Green function $G(x,y)$ is an example of a
{\em bi-scalar}: it transforms as scalar at both points $x$ and
$y$. We can also define a {\em bi-tensor}\, $T_{a_1\cdots
a_n\,b'_1\cdots b'_m}(x,y)$: upon a coordinate transformation,
this transforms as a rank $n$ tensor at $x$ and a rank $m$ tensor
at $y$. We will extend this up to a {\em quad-tensor}\,
$T_{a_1\cdots a_{n_1}\,b'_1\cdots b'_{n_2}\,
    c''_1\cdots c''_{n_3}\,d'''_1\cdots d'''_{n_4}}$
which has support at four points $x,y,x',y'$, transforming as
rank $n_1,n_2,n_3,n_4$ tensors at each of the four points. This
also sets the notation we will use: unprimed indices referring to
the tangent space constructed above $x$, single primed indices to
$y$, double primed to $x'$ and triple primed to $y'$. For each
point, there is the covariant derivative $\nabla_a$ at that point.
Covariant derivatives at different points commute and the
covariant derivative at, say, point $x'$, does not act on a
bi-tensor defined at, say,  $x$ and $y$:
\begin{equation}
T_{ab';c;d'} = T_{ab';d';c} \quad {\rm and } \quad T_{ab';c''} =
0.
\end{equation}
To simplify notation, henceforth we will eliminate the semicolons
after the first one for multiple covariant derivatives at
multiple points.

Having objects defined at different points, the {\rm coincident
limit} is defined as evaluation ``on the diagonal'', in the sense
of the spacetime support of the function or tensor, and the usual
shorthand $\left[ G(x,y) \right] \equiv G(x,x)$ is used. This
extends to $n$-tensors as
\begin{equation}
\left[ T_{a_1\cdots a_{n_1}\,b'_1\cdots b'_{n_2}\,
    c''_1\cdots c''_{n_3}\,d'''_1\cdots d'''_{n_4}} \right] =
T_{a_1\cdots a_{n_1}\,b_1\cdots b_{n_2}\,
    c_1\cdots c_{n_3}\,d_1\cdots d_{n_4}},
\end{equation}
{\it i.e.}, this becomes a rank $(n_1+n_2+n_3+n_4)$ tensor at $x$.
The multi-variable chain rule relates covariant derivatives
acting at different points, when we are interested in the
coincident limit:
\begin{equation}
\left[ T_{a_1\cdots a_m \,b'_1\cdots b'_n} \right]\!{}_{;c} =
\left[ T_{a_1\cdots a_m \,b'_1\cdots b'_n;c} \right] + \left[
T_{a_1\cdots a_m \,b'_1\cdots b'_n;c'} \right].
\label{ref-Synge's}
\end{equation}
This result is referred to as {\em Synge's theorem} in this
context; we  follow Fulling's \cite{Ful89} discussion.

The bi-tensor of {\em parallel transport}\, $g_a{}^{b'}$ is
defined such that when it acts on a vector $v_{b'}$ at $y$, it
parallel transports the vector along the geodesics connecting $x$
and $y$. This allows us to add vectors and tensors defined at
different points. We cannot directly add a vector $v_a$ at $x$
and vector $w_{a'}$ at $y$. But by using $g_a{}^{b'}$, we can
construct the sum $v^a + g_a{}^{b'} w_{b'}$. We will also need
the obvious property $\left[ g_a{}^{b'} \right] = g_a{}^b$.

The main bi-scalar we need  is the {\em world function}
$\sigma(x,y)$. This is defined as a half of the square of the
geodesic distance between the points $x$ and $y$. It satisfies
the equation
\begin{equation}
\sigma = \frac{1}{2} \sigma^{;p} \sigma_{;p} \label{define-sigma}
\end{equation}
Often in the literature, a covariant derivative is implied when
the world function appears with indices: $\sigma^a \equiv
\sigma^{;a}$, {\it i.e.}taking the covariant derivative at $x$,
while $\sigma^{a'}$ means the covariant derivative at $y$. This
is done since the vector $-\sigma^a$ is the tangent vector to the
geodesic with length equal the distance between $x$ and $y$. As
$\sigma^a$ records information about distance and direction for
the two points  this makes it ideal for constructing a series
expansion of a bi-scalar.  The {\em end point} expansion of a
bi-scalar $S(x,y)$ is of the form
\begin{equation}
S(x,y) = A^{(0)} + \sigma^p A^{(1)}_p + \sigma^p \sigma^q
A^{(2)}_{pq} + \sigma^p \sigma^q  \sigma^r A^{(3)}_{pqr} +
\sigma^p \sigma^q  \sigma^r \sigma^s A^{(4)}_{pqrs} + \cdots
\label{general-endpt-series}
\end{equation}
where, following our convention, the expansion tensors
$A^{(n)}_{a_1\cdots a_n}$ with  unprimed indices have support at
$x$ and hence the name end point expansion. Only the symmetric
part of these tensors  contribute to the expansion. For the
purposes of multiplying series expansions it is convenient to
separate the distance dependence from the direction dependence.
This is done by introducing the unit vector $p^a =
\sigma^a/\sqrt{2\sigma}$. Then the series expansion can be written
\begin{equation}
S(x,y) = A^{(0)} + \sigma^{\frac{1}{2}} A^{(1)} + \sigma  A^{(2)}
+ \sigma^{\frac{3}{2}} A^{(3)} + \sigma^2 A^{(4)} + \cdots
\end{equation}
The expansion scalars are related, via
$A^{(n)} = 2^{n/2} A^{(n)}_{p_1\cdots p_n} p^{p_1}\cdots p^{p_n}$,
to the expansion tensors.

The last object we need  is the {\em VanVleck-Morette}
determinant $D(x,y)$, defined as $D(x,y) \equiv -\det\left(
-\sigma_{;ab'} \right)$. The related bi-scalar
\begin{equation}
{\Delta\!^{1/2}} = \left( \frac{D(x,y)}{\sqrt{g(x)
g(y)}}\right)^\frac{1}{2}
\end{equation}
satisfies the equation
\begin{equation}
{\Delta\!^{1/2}}\left(4-\sigma_{;p}{}^p\right) -
2{\Delta\!^{1/2}}_{\,\,;p}\sigma^{;p} = 0 \label{define-VanD}
\end{equation}
with the boundary condition $\left[{\Delta\!^{1/2}}\right] = 1$.

Further details on these objects and discussions of the
definitions and properties are contained in \cite{Chr76,Chr78}
and \cite{NPsc}. There it is shown how the defining equations for
$\sigma$ and ${\Delta\!^{1/2}}$ are used to determine the
coincident limit expression for the various covariant derivatives
of the world function ($\left[ \sigma_{;a}\right]$, $\left[
\sigma_{;ab}\right]$, {\it etc.}) and how the defining
differential equation for ${\Delta\!^{1/2}}$ can be used to
determine the series expansion of ${\Delta\!^{1/2}}$. We show how
the expansion tensors $A^{(n)}_{a_1\cdots a_n}$ are determined in
terms of the coincident limits of covariant derivatives of the
bi-scalar $S(x,y)$.  (Ref.  \cite{NPsc} details how point
separation can be implemented on the computer to provide easy
access to a wider range of applications involving higher
derivatives of the curvature tensors.)

\subsection{Stress Energy Bi-Tensor Operator and Noise Kernel}

Even though we believe that the point-separated results are more
basic in the sense that it reflects a deeper structure of the
quantum theory of spacetime, we will nevertheless start with
quantities defined at one point because they are what enter in
conventional quantum field theory. We will use point separation
to introduce the bi-quantities. The key issue here is thus the
distinction between point-defined ({\it pt}) and point-separated
({\it bi}) quantities.

For  a free classical scalar field $\phi$ with the action
$S_m[g,\phi]$ defined in Eq. (\ref{2.1}), the classical
stress-energy tensor is
\begin{eqnarray}
T_{ab} &=& \left( 1 - 2\,\xi  \right) \,{\phi {}_;{}_{a}}\,{\phi
{}_;{}_{b}}
 + \left(2\,\xi -{1\over 2} \right) \,{\phi {}_;{}_{p}}
\,{\phi {}^;{}^{p}}\,{g{}_{a}{}_{b}} + 2\xi\,\phi \,
\,\left({\phi {}_;{}_{p}{}^{p} -{\phi {}_;{}_{a}{}_{b}}}
\,{g{}_{a}{}_{b}} \right)  \cr &&+ {{\phi }^2}\,\xi \,
\left({R{}_{a}{}_{b}} - {1\over 2}{ R\,{g{}_{a}{}_{b}}  }
 \right)
 - \frac{1}{2}{{m^2}\,{{\phi }^2}\,{g{}_{a}{}_{b}}},
\label{ref-define-classical-emt}
\end{eqnarray}
which is equivalent to the tensor of Eq. (\ref{2.3}) but written
in a slightly different form for convenience. When we make the
transition to quantum field theory, we promote the field
$\phi(x)$ to a field operator $\hat\phi(x)$. The fundamental
problem of defining a quantum operator for the stress tensor is
immediately visible: the field operator appears quadratically.
Since $\hat\phi(x)$ is an operator-valued distribution, products
at a single point are not well-defined. But if the product is
point separated ($\hat\phi^2(x) \rightarrow
\hat\phi(x)\hat\phi(x')$), they are finite and well-defined.

Let us first seek a point-separated extension of these classical
quantities and then consider the quantum field operators. Point
separation is symmetrically extended to products of covariant
derivatives of the field according to
\begin{eqnarray}
\left({\phi {}_;{}_{a}}\right)\left({\phi {}_;{}_{b}}\right)
&\rightarrow &\frac{1}{2}\left(
g_a{}^{p'}\nabla_{p'}\nabla_{b}+g_b{}^{p'}\nabla_a\nabla_{p'}
\right)\phi(x)\phi(x'),
\nonumber\\
\phi \,\left({\phi {}_;{}_{a}{}_{b}}\right) &\rightarrow&
\frac{1}{2}\left(
\nabla_a\nabla_b+g_a{}^{p'}g_b{}^{q'}\nabla_{p'}\nabla_{q'}
\right)\phi(x)\phi(x').
\nonumber
\end{eqnarray}
The bi-vector of parallel displacement $g_a{}^{a'}(x,x')$ is
included so that we may have objects that are rank 2 tensors at
$x$ and scalars at $x'$.

To carry out  point separation on
(\ref{ref-define-classical-emt}), we first define the
differential operator
\begin{eqnarray}
{\cal T}_{ab} &=&
  \frac{1}{2}\left(1-2\xi\right)
   \left(g_a{}^{a'}\nabla_{a'}\nabla_{b}+g_b{}^{b'}\nabla_a\nabla_{b'}\right)
+ \left(2\xi-\frac{1}{2}\right)
     g_{ab}g^{cd'}\nabla_c\nabla_{d'} \cr
&& - \xi
     \left(\nabla_a\nabla_b+g_a{}^{a'}g_b{}^{b'}\nabla_{a'}\nabla_{b'}\right)
+ \xi g_{ab}
     \left(\nabla_c\nabla^c+\nabla_{c'}\nabla^{c'}\right) \cr
&& +\xi\left(R_{ab} - \frac{1}{2}g_{ab}R\right)
    -\frac{1}{2}m^2 g_{ab}
\label{PSNoise-emt-diffop}
\end{eqnarray}
from which we obtain the classical stress tensor as
\begin{equation}
T_{ab}(x) = \lim_{x' \rightarrow x} {\cal T}_{ab}\phi(x)\phi(x').
\end{equation}
That the classical tensor field no longer appears as a product of
scalar fields at a single point allows a smooth transition to the
quantum tensor field. From the viewpoint of the stress tensor,
the separation of points is an artificial construct so when
promoting the classical field to a quantum one, neither point
should be favored. The product of field configurations is taken
to be the symmetrized operator product, denoted by curly brackets:
\begin{equation}
\phi(x)\phi(y) \rightarrow \frac{1}{2}
 \left\{{\hat\phi(x)},{\hat\phi(y)}\right\}
= \frac{1}{2}\left( {\hat\phi(x)} {\hat\phi(y)} +
                    {\hat\phi(y)} {\hat\phi(x)}
\right)
\end{equation}
With this, the point separated stress energy tensor operator is
defined as
\begin{equation}
\hat T_{ab}(x,x') \equiv \frac{1}{2} {\cal
T}_{ab}\left\{\hat\phi(x),\hat\phi(x')\right\}.
\label{PSNoise-emt-define}
\end{equation}
While the classical stress tensor was defined at the coincidence
limit $x'\rightarrow x$, we cannot attach any physical meaning to
the quantum stress tensor at one point until the issue of
regularization is dealt with, which will happen in the next
section. For now,  we will maintain point separation so as to
have a mathematically meaningful operator.

The expectation value of the point-separated stress tensor can
now be taken. This amounts to replacing the field operators by
their expectation value, which is given by the Hadamard (or
Schwinger) function
\begin{equation}
{G^{(1)}}(x,x') =
     \langle\left\{{\hat\phi(x)},{\hat\phi(x')}\right\}\rangle.
\end{equation}
and the point-separated stress tensor is defined as
\begin{equation}
\langle \hat T_{ab}(x,x') \rangle = \frac{1}{2} {\cal
T}_{ab}{G^{(1)}}(x,x') \label{ref-emt-PSdefine}
\end{equation}
where, since ${\cal T}_{ab}$ is a differential operator, it can
be taken ``outside'' the expectation value. The expectation value
of the point-separated quantum stress tensor for a free, massless
($m=0$)  conformally coupled ($\xi=1/6$) scalar field on a four
dimension spacetime with scalar curvature $R$ is
\begin{eqnarray}
\langle \hat T_{ab}(x,x') \rangle &=&
  \frac{1}{6}\left( {g{}^{p'}{}_{b}}\,{{G^{(1)}}{}_;{}_{p'}{}_{a}}
 + {g{}^{p'}{}_{a}}\,{{G^{(1)}}{}_;{}_{p'}{}_{b}} \right)
 -\frac{1}{12} {g{}^{p'}{}_{q}}\,{{G^{(1)}}{}_;{}_{p'}{}^{q}}
\,{g{}_{a}{}_{b}} \cr && -\frac{1}{12}\left(
{g{}^{p'}{}_{a}}\,{g{}^{q'}{}_{b}}
\,{{G^{(1)}}{}_;{}_{p'}{}_{q'}} + {{G^{(1)}}{}_;{}_{a}{}_{b}}
\right)
\nonumber  \\
&&
 +\frac{1}{12}\left( \left( {{G^{(1)}}{}_;{}_{p'}{}^{p'}}
 + {{G^{(1)}}{}_;{}_{p}{}^{p}} \right) \,{g{}_{a}{}_{b}} \right) \cr
&& +\frac{1}{12} {G^{(1)}}\, \left({R{}_{a}{}_{b}} -{1\over 2}
R\,{g{}_{a}{}_{b}} \right)
\end{eqnarray}

\subsubsection{Finiteness of Noise Kernel}

We now turn our attention to the noise kernel introduced in Eq.
(\ref{2.8}), which  is the symmetrized product of the (mean
subtracted) stress tensor operator:
\begin{eqnarray}
8 N_{ab,c'd'}(x,y) &=& \langle \left\{
        \hat T_{ab}(x)-\langle \hat T_{ab}(x)\rangle,
    \hat T_{c'd'}(y)-\langle \hat T_{c'd'}(y) \rangle
\right\} \rangle \cr &=& \langle \left\{ \hat T_{ab}(x),\hat
T_{c'd'}(y) \right\} \rangle -2 \langle \hat
T_{ab}(x)\rangle\langle \hat T_{c'd'}(y) \rangle
\end{eqnarray}
Since $\hat T_{ab}(x)$ defined at one point can be ill-behaved as
it is generally divergent, one can question the soundness of
these quantities. But as will be shown later, the noise kernel is
finite for $y\neq x$. All field operator products present in the
first expectation value that could be divergent are canceled by
similar products in the second term. We will replace each of the
stress tensor operators in the above expression for the noise
kernel by their point separated versions, effectively separating
the two points $(x,y)$ into the four points $(x,x',y,y')$. This
will allow us to express the noise kernel in terms of a pair of
differential operators acting on a combination of four and two
point functions. Wick's theorem will allow the four point
functions to be re-expressed in terms of two point functions.
{}From this we see that all possible divergences for $y\neq x$ will
cancel. When the coincidence limit is taken divergences do occur.
The above procedure will allow us to isolate the divergences and
obtain a finite result.

Taking the point-separated quantities as more basic, one should
replace each of the stress tensor operators in the above with the
corresponding point separated version (\ref{PSNoise-emt-define}),
with ${\cal T}_{ab}$ acting at $x$ and $x'$ and ${\cal T}_{c'd'}$
acting at $y$ and $y'$. In this framework the noise kernel is
defined as
\begin{equation}
8 N_{ab,c'd'}(x,y) =
   \lim_{x'\rightarrow x}\lim_{y'\rightarrow y}
   {\cal T}_{ab} {\cal T}_{c'd'}\, G(x,x',y,y')
\end{equation}
where the four point function is
\begin{eqnarray}
G(x,x',y,y') &=& \frac{1}{4}\left[
\langle\left\{\left\{{\hat\phi(x)},{\hat\phi(x')}\right\},\left\{{\hat\phi(y)}
,{\hat\phi(y')}\right\}\right\}\rangle \right.
\cr\cr&&\hspace{1cm}\left.
  -2\,\langle\left\{{\hat\phi(x)},{\hat\phi(x')}\right\}\rangle
  \langle\left\{{\hat\phi(y)},{\hat\phi(y')}\right\}\rangle \right].
\label{PSNoise-G4a}
\end{eqnarray}
We assume the pairs $(x,x')$ and $(y,y')$ are each within their
respective Riemann normal coordinate neighborhoods so as to avoid
problems that possible geodesic caustics might be present. When
we later turn our attention to computing the limit $y\rightarrow
x$, after issues of regularization are addressed, we will want to
assume all four points are within the same Riemann normal
coordinate neighborhood.

Wick's theorem, for the case of free fields which we are
considering, gives the simple product four point function in terms
of a sum of products of Wightman functions (we use the shorthand
notation $G_{xy}\equiv G_{+}(x,y) =
\langle{\hat\phi(x)}\,{\hat\phi(y)}\rangle$):
\begin{equation}
\langle{\hat\phi(x)}\,{\hat\phi(y)}\,{\hat\phi(x')}\,{\hat\phi(y')}\rangle
= {G_{xy'}}\,{G_{yx'}} + {G_{xx'}}\,{G_{yy'}} +
{G_{xy}}\,{G_{x'y'}}
\end{equation}
Expanding out the anti-commutators in (\ref{PSNoise-G4a}) and
applying Wick's theorem, the four point function becomes
\begin{equation}
G(x,x',y,y')  = {G_{xy'}}\,{G_{x'y}} + {G_{xy}}\,{G_{x'y'}} +
{G_{yx'}}\,{G_{y'x}} + {G_{yx}} \,{G_{y'x'}}.
\end{equation}
We can now easily see that the noise kernel defined via this
function is indeed well defined for the limit $(x',y')\rightarrow
(x,y)$:
\begin{equation}
G(x,x,y,y) = 2\,\left( {{{G^2_{xy}}}} + {{{G^2_{yx}}}} \right) .
\end{equation}
{}From this we can see that the noise kernel is also well defined
for $y \neq x$; any divergence present in the first expectation
value of (\ref{PSNoise-G4a}) have been cancelled by those present
in the pair of Green functions in the second term, in agreement
with the results of section \ref{sec2}.

\subsubsection{Explicit Form of the Noise Kernel}

We will let the points  separated for a while so we can keep
track of which covariant derivative acts on which arguments of
which Wightman function. As an example (the complete calculation
is quite long), consider the result of the first set of covariant
derivative operators in the differential operator
(\ref{PSNoise-emt-diffop}), from both ${\cal T}_{ab}$ and ${\cal
T}_{c'd'}$, acting on $G(x,x',y,y')$:
\begin{eqnarray}
&&\frac{1}{4}\left(1-2\xi\right)^2
   \left(g_a{}^{p''}\nabla_{p''}\nabla_{b}+
         g_b{}^{p''}\nabla_{p''}\nabla_{a}\right)\cr
&&\hspace{17mm}\times
   \left(g_{c'}{}^{q'''}\nabla_{q'''}\nabla_{d'}
        +g_{d'}{}^{q'''}\nabla_{q'''}\nabla_{c'}\right)
    G(x,x',y,y').
\end{eqnarray}
(Our notation is that $\nabla_a$ acts at $x$, $\nabla_{c'}$ at
$y$, $\nabla_{b''}$ at $x'$ and $\nabla_{d'''}$ at $y'$).
Expanding out the differential operator above, we can determine
which derivatives act on which Wightman function:
\begin{eqnarray}
{{{{\left( 1 - 2\,\xi  \right) }^2}}\over 4} &\times & \left[
    {g{}_{c'}{}^{p'''}}\,{g{}^{q''}{}_{a}}
 \left( {{G_{xy'}}{}_;{}_{b}{}_{p'''}}\,{{G_{x'y}}{}_;{}_{q''}{}_{d'}}
 + {{G_{xy}}{}_;{}_{b}{}_{d'}}\,{{G_{x'y'}}{}_;{}_{q''}{}_{p'''}}
 \right.\right. \cr
&&  \hspace{18mm} + \left. {{G_{yx'}}{}_;{}_{q''}{}_{d'}}
\,{{G_{y'x}}{}_;{}_{b}{}_{p'''}} + {{G_{yx}}{}_;{}_{b}{}_{d'}}
\,{{G_{y'x'}}{}_;{}_{q''}{}_{p'''}} \right) \cr &&+
{g{}_{d'}{}^{p'''}}\,{g{}^{q''}{}_{a}}
 \left( {{G_{xy'}}{}_;{}_{b}{}_{p'''}}\,{{G_{x'y}}{}_;{}_{q''}{}_{c'}}
 + {{G_{xy}}{}_;{}_{b}{}_{c'}}\,{{G_{x'y'}}{}_;{}_{q''}{}_{p'''}} \right. \cr
&&  \hspace{18mm} + \left. {{G_{yx'}}{}_;{}_{q''}{}_{c'}}
\,{{G_{y'x}}{}_;{}_{b}{}_{p'''}} + {{G_{yx}}{}_;{}_{b}{}_{c'}}
\,{{G_{y'x'}}{}_;{}_{q''}{}_{p'''}} \right) \cr &&+
{g{}_{c'}{}^{p'''}}\,{g{}^{q''}{}_{b}}
 \left( {{G_{xy'}}{}_;{}_{a}{}_{p'''}}\,{{G_{x'y}}{}_;{}_{q''}{}_{d'}}
 + {{G_{xy}}{}_;{}_{a}{}_{d'}}\,{{G_{x'y'}}{}_;{}_{q''}{}_{p'''}} \right. \cr
&&  \hspace{18mm} + \left. {{G_{yx'}}{}_;{}_{q''}{}_{d'}}
\,{{G_{y'x}}{}_;{}_{a}{}_{p'''}} + {{G_{yx}}{}_;{}_{a}{}_{d'}}
\,{{G_{y'x'}}{}_;{}_{q''}{}_{p'''}} \right) \cr &&+
{g{}_{d'}{}^{p'''}}\,{g{}^{q''}{}_{b}}
 \left( {{G_{xy'}}{}_;{}_{a}{}_{p'''}}\,{{G_{x'y}}{}_;{}_{q''}{}_{c'}}
 + {{G_{xy}}{}_;{}_{a}{}_{c'}}\,{{G_{x'y'}}{}_;{}_{q''}{}_{p'''}} \right. \cr
&&  \hspace{18mm} +\! \left.\left. {{G_{yx'}}{}_;{}_{q''}{}_{c'}}
{{G_{y'x}}{}_;{}_{a}{}_{p'''}}\! +\! {{G_{yx}}{}_;{}_{a}{}_{c'}}
{{G_{y'x'}}{}_;{}_{q''}{}_{p'''}} \right) \right]\! .
\end{eqnarray}
If we now  let $x'\rightarrow x$ and $y' \rightarrow y$ the
contribution to the noise kernel is (including the factor of
$\frac{1}{8}$ present in the definition of the noise kernel):
\begin{eqnarray}
&&\frac{1}{8}\left\{ {{\left( 1 - 2\,\xi  \right) }^2} \,\left(
{{G_{xy}}{}_;{}_{a}{}_{d'}}\,{{G_{xy}}{}_;{}_{b}{}_{c'}}
 + {{G_{xy}}{}_;{}_{a}{}_{c'}}\,{{G_{xy}}{}_;{}_{b}{}_{d'}}
 \right)  \right. \cr
&&\hspace{20mm} \left. + {{\left( 1 - 2\,\xi  \right) }^2}
\,\left( {{G_{yx}}{}_;{}_{a}{}_{d'}}\,{{G_{yx}}{}_;{}_{b}{}_{c'}}
 + {{G_{yx}}{}_;{}_{a}{}_{c'}}\,{{G_{yx}}{}_;{}_{b}{}_{d'}} \right)
 \right\}.
\end{eqnarray}
That this term can be written as the sum of a part involving
$G_{xy}$ and one involving $G_{yx}$ is a general property of the
entire noise kernel. It thus takes the form
\begin{equation}
N_{abc'd'}(x,y) = N_{abc'd'}\left[ G_{+}(x,y)\right]
                + N_{abc'd'}\left[ G_{+}(y,x)\right].
\end{equation}
We will present the form of the functional $N_{abc'd'}\left[ G
\right]$ shortly. First we note, for $x$ and $y$ time-like
separated, the above split of the noise kernel allows us to
express it in terms of the Feynman (time ordered) Green function
$G_F(x,y)$ and the Dyson (anti-time ordered) Green function
$G_D(x,y)$:
\begin{equation}
N_{abc'd'}(x,y) = N_{abc'd'}\left[ G_F(x,y)\right]
                + N_{abc'd'}\left[ G_D(x,y)\right].
\label{noiker}
\end{equation}
This can be connected  with the zeta function approach
to this problem \cite{PhiHu97} as follows: Recall when the quantum
stress tensor fluctuations determined in the Euclidean section is
analytically continued back to Lorentzian signature ($\tau
\rightarrow i t$), the time ordered product results. On the other
hand, if the continuation is $\tau \rightarrow -i t$, the
anti-time ordered product results. With this in mind, the noise
kernel is seen to be related to the quantum stress tensor
fluctuations derived via the effective action as
\begin{equation}
16 N_{abc'd'} =
   \left.\Delta T^2_{abc'd'}\right|_{t=-i\tau,t'=-i\tau'}
 + \left.\Delta T^2_{abc'd'}\right|_{t= i\tau,t'= i\tau'}.
\end{equation}
The complete form of the functional $N_{abc'd'}\left[ G \right]$
is
\begin{equation}
 N_{abc'd'}\left[ G \right]  =
    \tilde N_{abc'd'}\left[ G \right]
  + g_{ab}   \tilde N_{c'd'}\left[ G \right]
 + g_{c'd'} \tilde N'_{ab}\left[ G \right]
 + g_{ab}g_{c'd'} \tilde N\left[ G \right].
\label{general-noise-kernel}
\end{equation}
with
\begin{eqnarray}
8 \tilde N_{abc'd'} \left[ G \right]\!\!\!\!&=&\!\!\!\!
{{\left(1\!-\!2\xi\right)}^2}\!\left( G{}\!\,_{;}{}_{c'}{}_{b}
     G{}\!\,_{;}{}_{d'}{}_{a}
+
    G{}\!\,_{;}{}_{c'}{}_{a}\,G{}\!\,_{;}{}_{d'}{}_{b} \right)
\nonumber  \\
&&\!\!\!\!\!
+4{{\xi}^2}\left(
G{}\!\,_{;}{}_{c'}{}_{d'}\,G{}\!\,_{;}{}_{a}{}_{b} +
    G\,G{}\!\,_{;}{}_{a}{}_{b}{}_{c'}{}_{d'} \right)  \cr
&& \!\!\!\!\!-\!2\xi\!\left(1\!-\!2\xi\right)\!
  \left( G{}\!\,_{;}{}_{b}G{}\!\,_{;}{}_{c'}{}_{a}{}_{d'} \!+\!
    G{}\!\,_{;}{}_{a}G{}\!\,_{;}{}_{c'}{}_{b}{}_{d'} \!+\!
    G{}\!\,_{;}{}_{d'}G{}\!\,_{;}{}_{a}{}_{b}{}_{c'} \!+\!
    G{}\!\,_{;}{}_{c'}G{}\!\,_{;}{}_{a}{}_{b}{}_{d'}\! \right)  \cr
&&\!\!\!\!\!+\!2\xi\!\left(1\!-\!2\xi\right)\!\left(
G{}\!\,_{;}{}_{a}\,G{}\!\,_{;}{}_{b}\,
     {R{}_{c'}{}_{d'}} + G{}\!\,_{;}{}_{c'}\,G{}\!\,_{;}{}_{d'}\,
     {R{}_{a}{}_{b}} \right)  \cr
&& \!\!\!\!\! -4\!{{\xi}^2}\!\left(
G{}\!\,_{;}{}_{a}{}_{b}\,{R{}_{c'}{}_{d'}} +
    G{}\!\,_{;}{}_{c'}{}_{d'}\,{R{}_{a}{}_{b}} \right)  G
 +  2\,{{\xi}^2}\,{R{}_{c'}{}_{d'}}\,{R{}_{a}{}_{b}} {G^2},
\end{eqnarray}
\begin{eqnarray}
8 \tilde N'_{ab} \left[ G \right]\!\!\!&=&\!\!\!\! 2(1\!-\!2\xi)
\!\left[
   \left(\!2\xi\!-\!{\frac{1}{2}\!}\right)G{}\!\,_{;}{}_{p'}{}_{b}\,
  G{}\!\,_{;}{}^{p'}{}_{a}
 \!+\! \xi\!\left( G{}\!\,_{;}{}_{b}\,G{}\!\,_{;}{}_{p'}{}_{a}{}^{p'} \!+\!
    G{}\!\,_{;}{}_{a}\,G{}\!\,_{;}{}_{p'}{}_{b}{}^{p'} \right)
\right]\cr &&
%
-4\xi \left[
    \left(\!2\xi\!-\!{\frac{1}{2}}\!\right)\,G{}\!\,_{;}{}^{p'}\,
  G{}\!\,_{;}{}_{a}{}_{b}{}_{p'}
  + \xi\,\left( G{}\!\,_{;}{}_{p'}{}^{p'}\,G{}\!\,_{;}{}_{a}{}_{b} +
    G\,G{}\!\,_{;}{}_{a}{}_{b}{}_{p'}{}^{p'} \right)
\right] \cr
%
&& -({m^2}+\xi R')\,\left[(1-2\,\xi)\,G{}\!\,_{;}{}_{a}\,
     G{}\!\,_{;}{}_{b} - 2\,G\,\xi\,G{}\!\,_{;}{}_{a}{}_{b} \right]  \cr
%
&& + 2\xi\,\left[
\left(\!2\xi\!-\!{\frac{1}{2}}\!\right)\,G{}\!\,_{;}{}_{p'}\,
     G{}\!\,_{;}{}^{p'} + 2\,G\,\xi\,G{}\!\,_{;}{}_{p'}{}^{p'} \right] \,
  {R{}_{a}{}_{b}} \cr
%
&& - ({m^2}+\xi R')\,\xi\,{R{}_{a}{}_{b}} {G^2},
\end{eqnarray}
\begin{eqnarray}
8 \tilde N \left[ G \right]\!\!\!\!&=&\!\!\!\!
2{{\left(2\xi\!-\!{\frac{1}{2}}\right)}^2}G{}\!\,_{;}{}_{p'}{}_{q}\,
  G{}\!\,_{;}{}^{p'}{}^{q}
+ 4{{\xi}^2}\left(
G{}\!\,_{;}{}_{p'}{}^{p'}\,G{}\!\,_{;}{}_{q}{}^{q} +
    G\,G{}\!\,_{;}{}_{p}{}^{p}{}_{q'}{}^{q'} \right)  \cr
&&\!\!\!\! + 4\xi\left(2\xi\!-\!{\frac{1}{2}}\right)\,
  \left( G{}\!\,_{;}{}_{p}\,G{}\!\,_{;}{}_{q'}{}^{p}{}^{q'} +
    G{}\!\,_{;}{}^{p'}\,G{}\!\,_{;}{}_{q}{}^{q}{}_{p'} \right)  \cr
&&\!\!\!\! - \left(2\xi\!-\!{\frac{1}{2}}\right)\,
  \left[ \left({m^2}+\xi R\right)\,G{}\!\,_{;}{}_{p'}\,G{}\!\,_{;}{}^{p'} +
    \left({m^2}+\xi R'\right)\,G{}\!\,_{;}{}_{p}\,G{}\!\,^{;}{}^{p} \right]  \cr
&&\!\!\!\! - 2\xi\left[ \left({m^2}+\xi
R\right)\,G{}\!\,_{;}{}_{p'}{}^{p'} +
    \left({m^2}+\xi R'\right)\,G{}\!\,_{;}{}_{p}{}^{p} \right]  G \cr
&& \!\!\!\! +{\frac{1}{2}} \left({m^2}+\xi R\right)\left({m^2}+\xi
R'\right) {G^2}.
\end{eqnarray}

\subsubsection{Trace of the Noise Kernel}

One of the most interesting and surprising results to come  out of
the investigations undertaken in the 1970's of the quantum stress
tensor was the discovery of the trace anomaly
\cite{CapDuf74,Duf75}. When the trace of the stress tensor
$T=g^{ab}T_{ab}$ is evaluated for a field configuration that
satisties the field equation (\ref{2.2}) the trace is seen to
vanish for  massless conformally coupled fields. When this
analysis is carried over to the renormalized expectation value of
the quantum stress tensor, the trace no longer vanishes. Wald
\cite{Wal78} showed this was due to the failure of the
renormalized Hadamard function $G_{\rm ren}(x,x')$ to be symmetric
in $x$ and $x'$, implying it does not necessarily satisfy the
field equation (\ref{2.2}) in the variable $x'$. The definition of
$G_{\rm ren}(x,x')$ in the context of point separation will come
next.)

With this in  mind, we can now determine the noise associated
with the trace. Taking the trace at both points $x$ and $y$ of
the noise kernel functional (\ref{noiker}):
\begin{eqnarray}
N\left[ G \right]\!\!\!\!&=&\!\!\!\!g^{ab}\,g^{c'd'}\, N_{abc'd'}\left[ G
\right] \cr \!\!\!\!&=&\!\!\!\! - 3\,G\,\xi
    \left[
        \left({m^2} + {1\over 2} \xi R \right) \,{G{}_;{}_{p'}{}^{p'}}
      + \left({m^2} + {1\over 2} \xi R'\right) \,{G{}_;{}_{p}{}^{p}}
    \right] \cr
&&
  \!\!\!\! + {{9\,{{\xi }^2}}\over 2}
       \left[
           {G{}_;{}_{p'}{}^{p'}}\,{G{}_;{}_{p}{}^{p}}
          + G\,{G{}_;{}_{p}{}^{p}{}_{p'}{}^{p'}}
       \right]
\nonumber  \\
&&
  \!\!\!\!  +\left({m^2}\! + \!{1\over 2} \xi R \right)  \,
     \left({m^2} + {1\over 2} \xi R'\right)   G^2
\cr &&\!\!\!\!+ 3\! \left( {1\over 6} \!-\! \xi  \right)
     \! \left[
        3 \!{{\left( {1\over 6} \!-\! \xi  \right) }}
              {G{}_;{}_{p'}{}_{p}}\,{G{}_;{}^{p'}{}^{p}}
        \!-\!3\xi\!
           \left(\!
                {G{}_;{}_{p}}\,{G{}_;{}_{p'}{}^{p}{}^{p'}}
             \!+ \! {G{}_;{}_{p'}}\,{G{}_;{}_{p}{}^{p}{}^{p'}}
          \! \right)
      \right.\cr
&&\!\!\!\!\left.\hspace{10mm}+
          \left({m^2}\! +\! {1\over 2} \xi R \right)
{G{}_;{}_{p'}}{G{}_;{}^{p'}}
       \! +\! \left({m^2} \!+ \!{1\over 2} \xi R'\right) {G{}_;{}_{p}}{G{}^;{}^{p}}
\right]\! .
\end{eqnarray}
For the massless conformal case, this reduces to
\begin{equation}
N\left[ G \right] = \frac{1}{144}\left \{ R R' G^2 - 6G\left(R
\Box' + R' \Box\right) G
  + 18\left[ \left(\Box G\right)\left(\Box' G\right)+ \Box' \Box
  G\right] \right\}\! ,
\end{equation}
which holds for any function $G(x,y)$. For  $G$ being the Green
function, it satisfies the field equation (\ref{2.2}):
\begin{equation}
\Box G = (m^2 + \xi R) G .
\end{equation}
We will only assume the Green function satisfies the field
equation in its first variable. Using the fact $\Box' R=0$
(because the covariant derivatives act at a different point than
at which $R$ is supported), it follows that
\begin{equation}
\Box' \Box  G = (m^2 + \xi R)\Box' G.
\end{equation}
With these results, the noise kernel trace becomes
\begin{eqnarray}
N\left[ G \right]\!\!\! &=& \!\!\!\frac{1}{2} \left[
      {m^2}\,\left( 1 - 3\,\xi  \right)
    + 3\,R\,\left( {1\over 6} - \xi  \right) \,\xi
\right] \cr &&\hspace{18mm}\times
  \left[
       {G^2}\,\left( 2\,{m^2} + {R'} \,\xi  \right)
            + \left( 1 - 6\,\xi  \right) \,{G{}_;{}_{p'}}\,{G{}_;{}^{p'}}
            - 6\,G\,\xi \,{G{}_;{}_{p'}{}^{p'}}
  \right] \cr
&&+ \frac{1}{2} \left( {1\over 6} - \xi  \right)  \left[
3\,\left( 2\,{m^2} + {R'}\,\xi  \right)
\,{G{}_;{}_{p}}\,{G{}^;{}^{p}}
  - 18\,\xi \,{G{}_;{}_{p}}\,{G{}_;{}_{p'}{}^{p}{}^{p'}}
\right.\cr&&\hspace{18mm}\left.
  + 18\,\left( {1\over 6} - \xi  \right) \,
         {G{}_;{}_{p'}{}_{p}}\,{G{}_;{}^{p'}{}^{p}} \right],
\end{eqnarray}
which vanishes for the massless conformal case.  We have thus
shown, based solely on the definition of the point separated noise
kernel, there is no noise associated with the trace anomaly. This
result obtained in Ref. \cite{PhiHu03} is completely general since
it is assumed that the Green function is only satisfying the field
equations in its first variable; an alternative proof of this
result was given in Ref. \cite{MarVer99}. This condition holds not
just for the classical field case, but also for the regularized
quantum case, where one does not expect the Green function to
satisfy the field equation in both variables. One can see this
result from the simple observation used in section \ref{sec2}:
since the trace anomaly is known to be locally determined and
quantum state independent, whereas the noise present in the
quantum field is non-local, it is hard to find a noise associated
with it. This general result is in agreement with previous
findings \cite{CalHu94,HuSin95,CamVer96}, derived from the
Feynman-Vernon influence functional formalism
\cite{FeyVer63,FeyHib65} for some particular cases.

\section{Metric fluctuations in Minkowski spacetime}
\label{sec:flucminspa}

Although the Minkowski vacuum is an eigenstate of the total
four-momentum operator of a field in Minkowski spacetime, it is
not an eigenstate of the stress-energy operator. Hence, even for
those solutions of semiclassical gravity such as the Minkowski
metric, for which the expectation value of the stress-energy
operator can always be chosen to be zero, the fluctuations of this
operator are non-vanishing. This fact leads to consider the
stochastic metric perturbations induced by these fluctuations.

Here we derive the Einstein-Langevin equation for the metric
perturbations in a Minkowski background. We solve this equation
for the linearized Einstein tensor and compute the associated
two-point correlation functions, as well as, the
two-point correlation functions for the metric
perturbations. Even though, in
this case, we expect to have negligibly small values for these
correlation functions for points separated by lengths larger than
the Planck length, there are several reasons why it is worth
carrying out this calculation.

On the one hand, these are the first
back-reaction solutions of
the full Einstein-Langevin
equation.
There are analogous solutions to a ``reduced'' version
of this equation inspired in a ``mini-superspace'' model
\cite{CamVer97,CalCamVer97}, and there is also a previous
attempt to obtain a solution to
the Einstein-Langevin equation in Ref.~\cite{CamVer96},
but, there, the non-local terms in the
Einstein-Langevin equation were neglected.

On the other hand, the results of this calculation, which confirm our
expectations that gravitational fluctuations are negligible at length
scales larger than the Planck length,
but also predict that the fluctuations are strongly
suppressed on small scales, can be considered a first test
of stochastic semiclassical gravity.
In addition, these results reveal an important connection between
stochastic gravity and the large $N$ expansion of quantum gravity.
We can also extract
conclusions on the possible qualitative behavior of the solutions to
the Einstein-Langevin equation. Thus, it is interesting to
note that the correlation functions at short
scales are characterized by
correlation lengths of the order of the Planck length;
furthermore, such correlation lengths enter in a non-analytic
way in the correlation functions.

We advise the reader that his section is rather technical since it
deals with an explicit non trivial back reaction computation in
stochastic gravity. We tried to make it reasonable self-contained
and detailed,
however a more detailed exposition can be found in
Ref.~\cite{MarVer00}.


\subsection{Perturbations around Minkowski spacetime}


The Minkowski metric $\eta_{ab}$, in a manifold ${\cal M}$ which
is topologically ${\rm I\hspace{-0.4 ex}R}^{4}$, and the usual
Minkowski vacuum, denoted as $|0 \rangle$, are the class of
simplest solutions to the semiclassical Einstein equation
(\ref{2.5}), the so called trivial solutions of semiclassical
gravity \cite{FlaWal96}. They constitute the ground state of
semiclassical gravity. In fact, we can always choose a
renormalization scheme in which the renormalized expectation value
$\langle 0|\, \hat{T}_{R}^{ab}\,[\eta] |0 \rangle =0$. Thus,
Minkowski spacetime $({\rm I\hspace{-0.4 ex}R}^{4},\eta_{ab})$ and
the vacuum state $|0 \rangle$ are a solution to the semiclassical
Einstein equation with renormalized cosmological constant
$\Lambda\!=\!0$. The fact that the vacuum expectation value of the
renormalized stress-energy operator in Minkowski spacetime should
vanish was originally proposed by Wald \cite{Wal77} and it may be
understood as a renormalization convention
\cite{Ful89,GriMamMos94}. Note that other possible solutions of
semiclassical gravity with zero vacuum expectation value of the
stress-energy tensor are the exact gravitational plane waves,
since they are known to be vacuum solutions of Einstein equations
which induce neither particle creation nor vacuum polarization
\cite{Gib75,Des75,GarVer91}.

As we have already mentioned the
vacuum $|0 \rangle$ is an eigenstate of the total four-momentum
operator in Minkowski spacetime, but
not an eigenstate of $\hat{T}^{R}_{ab}[\eta]$. Hence, even in
the Minkowski background, there are quantum
fluctuations in the stress-energy tensor and, as a result,
the noise kernel does not vanish.
This fact leads to consider the stochastic corrections
to this class of trivial solutions of semiclassical
gravity.
Since, in this case, the Wightman and Feynman functions
(\ref{Wightman and Feynman functions}), their values in the two-point
coincidence limit, and the products of derivatives of two of such
functions appearing in expressions (\ref{Wightman expression 2}) and
(\ref{Feynman expression 2})
are known in dimensional regularization,
we can compute the Einstein-Langevin
equation using the methods outlined in sections \ref{sec2}
and \ref{sec3}.

To perform explicit calculations it is convenient to work in a global
inertial coordinate system $\{ x^\mu\}$ and in the associated basis,
in which the components of the flat metric are simply
$\eta_{\mu\nu}={\rm diag}(-1,1,\dots,1)$.
In Minkowski spacetime, the components of the classical stress-energy
tensor (\ref{2.3}) reduce to
\be
T^{\mu\nu}[\eta,\phi]=\partial^{\mu}\phi
\partial^{\nu} \phi - {1\over 2}\, \eta^{\mu\nu} \hspace{0.2ex}
\partial^{\rho}\phi \partial_{\rho} \phi
-{1\over 2}\, \eta^{\mu\nu}\hspace{0.2ex} m^2 \phi^2
+\xi \left( \eta^{\mu\nu} \Box
-\partial^{\mu} \partial^{\nu} \right) \phi^2,
\label{flat class s-t}
\ee
where $\Box \!\equiv\! \partial_{\mu}\partial^{\mu}$, and the formal
expression for the
components of the corresponding ``operator''
in dimensional regularization, see Eq. (\ref{regul s-t 2}), is
\be
\hat{T}_{n}^{\mu\nu}[\eta] = {1\over 2} \{
     \partial^{\mu}\hat{\phi}_{n} ,
     \partial^{\nu}\hat{\phi}_{n} \}
     + {\cal D}^{\mu\nu} \hat{\phi}_{n}^2,
\label{flat regul s-t}
\ee
where ${\cal D}^{\mu\nu}$ is the differential operator
(\ref{diff operator}), with $g_{\mu\nu}=\eta_{\mu\nu}$,
$R_{\mu\nu}=0$, and $\nabla_\mu=\partial_\mu$.
The field
$\hat{\phi}_{n}(x)$ is the field operator in the Heisenberg
representation in
a $n$-dimensional Minkowski spacetime, which satisfies the
Klein-Gordon equation (\ref{2.2}).
We use here a stress-energy tensor which differs from the
canonical one, which corresponds to $\xi=0$, both tensors,
however, define the same total momentum.

The Wightman and Feynman functions
(\ref{Wightman and Feynman functions}) when
$g_{\mu\nu}=\eta_{\mu\nu}$, are well
known:
\be
G_n^+(x,y) = i \hspace{0.2ex}\Delta_n^+(x-y),
\ \ \ \
G\!_{\scriptscriptstyle F_{\scriptstyle \hspace{0.1ex}  n}}
 \hspace{-0.2ex}(x,y)
  =
  \Delta_{\scriptscriptstyle F_{\scriptstyle \hspace{0.1ex} n}}
  \hspace{-0.2ex}(x-y),
\label{flat Wightman and Feynman functions}
\ee
with
\bea
&&\Delta_n^+(x)=-2 \pi i \int \! {d^n k \over (2\pi)^n} \,
e^{i kx}\, \delta (k^2+m^2) \,\theta (k^0),
\nn   \\
&&\Delta_{\scriptscriptstyle F_{\scriptstyle \hspace{0.1ex} n}}
  \hspace{-0.2ex}(x)=- \int \! {d^n k \over (2\pi)^n} \,
{e^{i kx}  \over k^2+m^2-i \epsilon} ,
\hspace{5ex} \epsilon \!\rightarrow \! 0^+,
\label{flat propagators}
\eea
where
$k^2 \equiv \eta_{\mu\nu} k^{\mu} k^{\nu}$ and
$k x \equiv \eta_{\mu\nu} k^{\mu} x^{\nu}$.
Note that the derivatives of these functions satisfy
$\partial_{\mu}^{x}\Delta_n^+(x-y)
= \partial_{\mu}\Delta_n^+(x-y)$ and
$\partial_{\mu}^{y}\Delta_n^+(x-y)=
 - \partial_{\mu}\Delta_n^+(x-y)$,
and similarly for the Feynman propagator
$\Delta_{\scriptscriptstyle F_{\scriptstyle \hspace{0.1ex} n}}
 \hspace{-0.2ex}(x-y)$.

To write down the semiclassical Einstein equation
(\ref{2.5}) in $n$-dimensions for this case, we need to
compute the vacuum expectation value of the
stress-energy operator components
(\ref{flat regul s-t}). Since, from
(\ref{flat Wightman and Feynman functions}), we have that
$\langle 0 |\hat{\phi}_{n}^2(x)|0 \rangle=
i\Delta_{\scriptscriptstyle F_{\scriptstyle \hspace{0.1ex} n}}
\hspace{-0.2ex}(0)
=i \Delta_n^+(0)$, which is a constant (independent
of $x$), we have simply
\begin{equation}
\langle 0 | \hat{T}_{n}^{\mu\nu} [\eta]|0 \rangle =
-i \int {d^n k \over (2\pi)^n} \,
{k^{\mu} k^{\nu} \over k^2+m^2-i \epsilon}
= {\eta^{\mu\nu} \over 2} \left( {m^2 \over 4 \pi} \right)^{\! n/2}
\! \Gamma \!\left(- {n \over 2}\right),
\label{vev}
\end{equation}
where the integrals in dimensional regularization have been computed
in the standard way (see Ref. \cite{MarVer00})
and where $\Gamma (z)$ is the
Euler's gamma function. The semiclassical Einstein equation
(\ref{2.5}) in $n$-dimensions before renormalization
reduces now to
\be
{\Lambda_{B} \over 8 \pi G_{B}}\, \eta^{\mu\nu}
= \mu^{-(n-4)}
\langle 0 | \hat{T}_{n}^{\mu\nu}[\eta]|0 \rangle  .
\label{flat semiclassical eq}
\ee
This equation, thus, simply sets the value
of the bare coupling constant
$\Lambda_{B}/G_{B}$.
Note, from (\ref{vev}), that in order to have
$\langle 0|\, \hat{T}_{R}^{\mu\nu}\, |0 \rangle [\eta]\!=\! 0$,
the renormalized and regularized stress-energy tensor
``operator'' for a scalar field in Minkowski spacetime,
see Eq. (\ref{2.4}),
has to be defined as
\be
\hat{T}_{R}^{\mu\nu}[\eta] =
\mu^{-(n-4)}\, \hat{T}_{n}^{\mu\nu}[\eta]
-{ \eta^{\mu\nu} \over 2} \, {m^4 \over (4\pi)^2}
\left( {m^2 \over 4 \pi \mu^2}
\right)^{\!_{\scriptstyle n-4 \over 2}}
\! \Gamma \!\left(- {n \over 2}\right),
\label{flat renorm s-t operator}
\ee
which corresponds to a renormalization of the cosmological constant
\be
{\Lambda_{B} \over G_{B}}={\Lambda \over G}
-{2 \over \pi} \, {m^4 \over n \hspace{0.2ex}(n\!-\!2)}
\: \kappa_n
+O(n-4),
\label{cosmological ct renorm 2}
\ee
where
\be
\kappa_n \equiv {1 \over (n\!-\!4)}
\left({e^\gamma m^2 \over 4 \pi \mu^2} \right)
^{\!_{\scriptstyle n-4 \over 2}}=
{1 \over n\!-\!4}
+{1\over 2}\,
\ln \!\left({e^\gamma m^2 \over 4 \pi \mu^2} \right)+O (n-4),
\label{kappa}
\ee
being $\gamma$ the Euler's constant. In the case of a
massless scalar field, $m^2\!=\!0$, one simply has
$\Lambda_{B} / G_{B}=\Lambda / G$. Introducing this renormalized
coupling constant into Eq.~(\ref{flat semiclassical eq}), we can
take the limit $n \!\rightarrow \! 4$.
We find that,
for $({\rm I\hspace{-0.4 ex}R}^{4}, \eta_{ab},|0 \rangle )$ to
satisfy the semiclassical Einstein equation,
we must take $\Lambda\!=\!0$.

We can now write down the
Einstein-Langevin equations for the components
$h_{\mu\nu}$ of the stochastic metric perturbation
in dimensional regularization.
In our case, using $\langle 0 |\hat{\phi}_{n}^2(x)|0 \rangle=
i\Delta_{\scriptscriptstyle F_{\scriptstyle \hspace{0.1ex} n}}
\hspace{-0.2ex}(0)$
and the explicit expression of
Eq. (\ref{Einstein-Langevin eq 3})
we obtain
\bea
&&\!\!\!\!\!{1\over 8 \pi G_{B}}\Biggl[
G^{{\scriptscriptstyle (1)}\hspace{0.1ex} \mu\nu} +
\Lambda_{B} \left( h^{\mu\nu}
\!-\!{1\over 2}\, \eta^{\mu\nu} h \right)
\Biggr](x) -
{4\over 3}\, \alpha_{B} D^{{\scriptscriptstyle
(1)}\hspace{0.1ex} \mu\nu}(x)
-2\beta_{B}B^{{\scriptscriptstyle (1)}\hspace{0.1ex} \mu\nu}(x)
\nn   \\
&&\!\!\!\!\!-\! \xi G^{{\scriptscriptstyle (1)}\hspace{0.1ex} \mu\nu}\!(x)
\mu^{-(n-4)}\! i\Delta_{\scriptscriptstyle F_{\scriptstyle
\hspace{0.1ex} n}} \hspace{-0.2ex}(0)\! +\! {1\over 2} \!\int\!\! d^ny
 \mu^{-(n-4)}\! H_n^{\mu\nu\alpha\beta}\!(x,y)h_{\alpha\beta}(y)
\!=\! \xi^{\mu\nu}\!(x)\!. \label{flat Einstein-Langevin eq}
\eea
The
indices in $h_{\mu\nu}$ are raised with the Minkowski metric and
$h \equiv h_{\rho}^{\rho}$, and here a superindex ${\scriptstyle
(1)}$  denotes the components of a tensor linearized around the
flat metric. Note that in $n$-dimensions the
two-point correlation functions for the
field $\xi^{\mu\nu}$ is written as \be \langle
\xi^{\mu\nu}(x)\xi^{\alpha\beta}(y) \rangle_s =\mu^{-2
\hspace{0.2ex} (n-4)}  N_n^{\mu\nu\alpha\beta}(x,y),
\label{correlator} \ee

Explicit expressions for $D^{{\scriptscriptstyle
(1)}\hspace{0.1ex} \mu\nu}$ and $B^{{\scriptscriptstyle
(1)}\hspace{0.1ex} \mu\nu}$ are given by
\be
D^{{\scriptscriptstyle (1)}\hspace{0.1ex} \mu\nu}(x)= {1 \over
2}\, {\cal F}^{\mu\nu\alpha\beta}_{x} \, h_{\alpha\beta}(x),
\hspace{7.2ex} B^{{\scriptscriptstyle (1)}\hspace{0.1ex}
\mu\nu}(x)= 2  {\cal F}^{\mu\nu}_{x} {\cal F}^{\alpha\beta}_{x}
h_{\alpha\beta}(x),
\label{D, B tensors}
\ee
with the differential operators ${\cal F}^{\mu\nu}_{x} \equiv
\eta^{\mu\nu} \Box_x -\partial^\mu_{x} \partial^\nu_{x}$ and
${\cal F}^{\mu\nu\alpha\beta}_{x} \equiv 3 {\cal
F}^{\mu(\alpha}_{x} {\cal F}^{\beta)\nu}_{x} -{\cal
F}^{\mu\nu}_{x} {\cal F}^{\alpha\beta}_{x}$.

\subsection{The kernels in the Minkowski background}

Since the two kernels (\ref{finite kernels}) are free of
ultraviolet divergences in the limit $n\!\rightarrow \! 4$, we
can deal directly with the $F^{\mu\nu\alpha\beta}(x-y)\equiv
\lim_{n \rightarrow 4} \mu^{-2 \hspace{0.2ex} (n-4)} \,
F^{\mu\nu\alpha\beta}_n$ in Eq. (\ref{bitensor F}). The kernels $
N^{\mu\nu\alpha\beta}(x,y) ={\rm Re}\, F^{\mu\nu\alpha\beta}(x-y)$
and $ H_{\scriptscriptstyle \!{\rm A}}^{\mu\nu\alpha\beta}(x,y) =
{\rm Im}\, F^{\mu\nu\alpha\beta}(x-y)$ are actually the components
of the ``physical'' noise and dissipation kernels that will appear
in the Einstein-Langevin equations once the renormalization
procedure has been carried out. The bi-tensor
$F^{\mu\nu\alpha\beta}$ can be expressed in terms of the Wightman
function in four spacetime dimensions, according to (\ref{Wightman
expression 2}). The different terms in this kernel can be easily
computed using the integrals \be I(p) \equiv \int\! {d^4 k \over
(2\pi)^4} \: \delta (k^2+m^2) \,\theta (-k^0)  \, \delta
[(k-p)^2+m^2]\,\theta (k^0-p^0), \label{integrals} \ee and
$I^{\mu_1 \dots \mu_r}(p)$ which are defined as the previous one
by inserting the momenta $k^{\mu_1}\dots k^{\mu_r}$ with $r \!=\!
1, 2, 3 ,4$ in the integrand. All these integral can be expressed
in terms of $I(p)$; see Ref. \cite{MarVer00} for the explicit
expressions. It is convenient to separate  $I(p)$ in its even and
odd parts with respect to the variables $p^{\mu}$ as \be
I(p)=I_{\scriptscriptstyle {\rm S}}(p) +I_{\scriptscriptstyle {\rm
A}}(p), \label{I} \ee where $I_{\scriptscriptstyle {\rm S}}(-p)=
I_{\scriptscriptstyle {\rm S}}(p)$ and $I_{\scriptscriptstyle {\rm
A}}(-p)= -I_{\scriptscriptstyle {\rm A}}(p)$. These two functions
are explicitly given by \bea &&I_{\scriptscriptstyle {\rm
S}}(p)={1 \over 8 \, (2 \pi)^3} \; \theta (-p^2-4m^2) \, \sqrt{1+4
\,{m^2 \over p^2} },
\nn  \\
&&I_{\scriptscriptstyle {\rm A}}(p)={-1 \over 8 \, (2 \pi)^3} \;
{\rm sign}\,p^0 \;
\theta (-p^2-4m^2) \, \sqrt{1+4 \,{m^2 \over p^2} }.
\label{S and A parts of I}
\eea
After some manipulations, we find
\bea
F^{\mu\nu\alpha\beta}(x)&\!\!\!\!=\!\!\!\!& {\pi^2 \over 45}\,
 {\cal F}^{\mu\nu\alpha\beta }_{x}
\int\! {d^4 p \over (2\pi)^4} \,
e^{-i px}\hspace{0.1ex}
\left(1+4 \,{m^2 \over p^2} \right)^2 I(p)
\nn   \\
&&+\,{8 \pi^2 \over 9 } \, {\cal F}^{\mu\nu}_{x}{\cal
F}^{\alpha\beta}_{x} \int\! {d^4 p \over (2\pi)^4} \, e^{-i
px}\hspace{0.1ex} \left(3 \hspace{0.3ex}\Delta \xi+{m^2 \over p^2}
\right)^2 I(p), \label{M 3} \eea where $\Delta \xi \equiv \xi -
1/6$. The real and imaginary parts of the last expression, which
yield the noise and dissipation kernels, are easily recognized as
the terms containing $I_{\scriptscriptstyle {\rm S}}(p)$ and
$I_{\scriptscriptstyle {\rm A}}(p)$, respectively. To write them
explicitly, it is useful to introduce the new kernels
\bea
&&N_{\rm A}(x;m^2) \equiv {1 \over 480 \pi} \int\! {d^4 p \over
(2\pi)^4} \, e^{i px}\, \theta (-p^2-4m^2) \, \sqrt{1+4 \,{m^2
\over p^2} } \left(1+4 \,{m^2 \over p^2} \right)^2,
\nn \\
&&N_{\rm B}(x;m^2,\Delta \xi) \equiv {1 \over 72 \pi} \int\! {d^4
p \over (2\pi)^4} \, e^{i px}\, \theta (-p^2-4m^2)
\nonumber\\
&&\hspace{55mm}\times
 \sqrt{1+4
\,{m^2 \over p^2} } \left(3 \hspace{0.3ex}\Delta \xi+{m^2 \over
p^2} \right)^2,
\nn \\
&&D_{\rm A}(x;m^2) \equiv {-i \over 480 \pi} \int\! {d^4 p \over
(2\pi)^4} \, e^{i px}\, {\rm sign}\,p^0 \; \theta (-p^2-4m^2)
\nonumber\\
&&\hspace{55mm}\times \sqrt{1+4 \,{m^2 \over p^2} } \left(1+4
\,{m^2 \over p^2} \right)^2,
\nn \\
&&D_{\rm B}(x;m^2,\Delta \xi) \equiv {-i \over 72 \pi} \int\! {d^4
p \over (2\pi)^4} \, e^{i px}\, {\rm sign}\,p^0 \; \theta
(-p^2-4m^2)
\nonumber\\
&&\hspace{55mm}\times \sqrt{1+4 \,{m^2 \over p^2} } \left(3
\hspace{0.3ex}\Delta \xi+{m^2 \over p^2} \right)^{\!2}\!\!\!\!,
\label{N and D kernels} \eea and we finally get \bea
\!\!\!\!\!&&\!\!\!\!\!N^{\mu\nu\alpha\beta}(x,y)= {1 \over 6}
{\cal F}^{\mu\nu\alpha\beta}_{x} N_{\rm A}(x\!-\!y;m^2) +{\cal
F}^{\mu\nu}_{x}{\cal F}^{\alpha\beta}_{x} N_{\rm
B}(x\!-\!y;m^2,\Delta \xi),
\nn \\
\!\!\!\!\!&&\!\!\!\!\!H_{\scriptscriptstyle \!{\rm
A}}^{\mu\nu\alpha\beta}(x,y)= {1 \over 6}
 {\cal F}^{\mu\nu\alpha\beta}_{x}
D_{\rm A}(x\!-\!y;m^2) +{\cal F}^{\mu\nu}_{x}{\cal
F}^{\alpha\beta}_{x} D_{\rm B}(x\!-\!y;m^2,\Delta \xi) .
\label{noise and dissipation kernels 2} \eea Notice that the noise
and dissipation kernels defined in (\ref{N and D kernels}) are
actually real because, for the noise kernels, only the $\cos px$
terms of the exponentials $e^{i px}$ contribute to the integrals,
and, for the dissipation kernels, the only contribution of such
exponentials comes from the $i \sin px$ terms.

The evaluation of the kernel $H_{\scriptscriptstyle \!{\rm
S}_{\scriptstyle n}} ^{\mu\nu\alpha\beta}(x,y)$ is a more involved
task. Since this kernel contains divergences in the limit
$n\!\rightarrow \! 4$, we use dimensional regularization. Using
Eq.~(\ref{Feynman expression 2}), this kernel can be written in
terms of the Feynman propagator (\ref{flat propagators}) as \be
\mu^{-(n-4)} H_{\scriptscriptstyle \!{\rm S}_{\scriptstyle n}}
^{\mu\nu\alpha\beta}(x,y)= {\rm Im}\, K^{\mu\nu\alpha\beta}(x-y),
\label{kernel H_S} \ee where
\bea \!\!\!\!\!\!\! &&\!\!\!\!\!\!\!
K^{\mu\nu\alpha\beta}(x) \equiv - \mu^{-(n-4)} \biggl\{ 2
\partial^\mu \partial^{( \alpha} \Delta_{\scriptscriptstyle
F_{\scriptstyle \hspace{0.1ex} n}}
   \hspace{-0.2ex}(x) \,
\partial^{\beta )} \partial^\nu
\Delta_{\scriptscriptstyle F_{\scriptstyle \hspace{0.1ex} n}}
   \hspace{-0.2ex}(x)
\nonumber\\
\!\!\!&&\!\!\!\!\!\! +2 {\cal D}^{\mu\nu} \!\left( \partial^\alpha
\Delta_{\scriptscriptstyle F_{\scriptstyle \hspace{0.1ex} n}}
   \hspace{-0.2ex}(x)
\partial^\beta
\Delta_{\scriptscriptstyle F_{\scriptstyle \hspace{0.1ex} n}}
   \hspace{-0.2ex}(x)  \right)
+ 2 {\cal D}^{\alpha\beta}  \Bigl( \partial^\mu
\Delta_{\scriptscriptstyle F_{\scriptstyle \hspace{0.1ex} n}}
   \hspace{-0.2ex}(x) \,
\partial^\nu
\Delta_{\scriptscriptstyle F_{\scriptstyle \hspace{0.1ex} n}}
   \hspace{-0.2ex}(x) \Bigr)
\nonumber\\
\!\!\!&&\!\!\!\!\!\! +2 {\cal D}^{\mu\nu} {\cal D}^{\alpha\beta}
\!\left( \Delta_{\scriptscriptstyle F_{\scriptstyle \hspace{0.1ex}
n}}^2
   \hspace{-0.2ex}(x) \right)
\!+\!\biggl[ \eta^{\mu\nu} \partial^{( \alpha}
\Delta_{\scriptscriptstyle F_{\scriptstyle \hspace{0.1ex} n}}
   \hspace{-0.2ex}(x)
\partial^{\beta )}
\!+ \!\eta^{\alpha\beta} \partial^{( \mu}
 \Delta_{\scriptscriptstyle F_{\scriptstyle \hspace{0.1ex} n}}
   \hspace{-0.2ex}(x)
\partial^{\nu )}
\nn   \\
\!\!\!&&\!\!\!\!\!\! +\! \Delta_{\scriptscriptstyle
F_{\scriptstyle \hspace{0.1ex} n}}
   \hspace{-0.2ex}(0) \! \left( \eta^{\mu\nu}
{\cal D}^{\alpha\beta}\!+\! \eta^{\alpha\beta} {\cal D}^{\mu\nu}
\right) \!+\!{1 \over 4}\eta^{\mu\nu}\! \eta^{\alpha\beta}
\!\left( \Delta_{\scriptscriptstyle F_{\scriptstyle \hspace{0.1ex}
n}}
   \hspace{-0.2ex}(x) \Box
\!-\!m^2\!
\Delta_{\scriptscriptstyle F_{\scriptstyle \hspace{0.1ex} n}}
   \hspace{-0.2ex}(0)\!  \right) \!\biggr] \!\delta^n (x)\!
\biggr\}\!. \label{K} \eea Let us define the integrals \be J_n(p)
\equiv \mu^{-(n-4)} \!\int\! {d^n k \over (2\pi)^n} \: {1 \over
(k^2+m^2-i \epsilon) \, [(k-p)^2+m^2-i \epsilon] },
\label{integrals in n dim} \ee and $J_n^{\mu_1 \dots \mu_r}(p)$
obtained by inserting the momenta $k^{\mu_1}\dots k^{\mu_r}$ into
the previous integral, together with \be I_{0_{\scriptstyle n}}
\equiv \mu^{-(n-4)} \!\int\! {d^n k \over (2\pi)^n} \: {1 \over
(k^2+m^2-i \epsilon) }, \label{constant integrals in n dim} \ee
and $I_{0_{\scriptstyle n}}^{\mu_1 \dots \mu_r}$ which are also
obtained by inserting momenta in the integrand. Then, the
different terms in Eq.~(\ref{K}) can be computed. These integrals
are explicitly given in Ref. \cite{MarVer00}. It is found that
$I_{0_{\scriptstyle n}}^{\mu}=0$ and the remaining integrals can
be written in terms of $I_{0_{\scriptstyle n}}$ and $J_n(p)$. It
is useful to introduce the projector $P^{\mu\nu}$ orthogonal to
$p^\mu$ and the tensor $P^{\mu\nu\alpha\beta}$ as \be p^2
P^{\mu\nu} \!\equiv \! \eta^{\mu\nu} p^2- p^\mu p^\nu, \ \ \ \ \ \
P^{\mu\nu\alpha\beta}\equiv
3P^{\mu(\alpha}P^{\beta)\nu}-P^{\mu\nu}P^{\alpha\beta},
\label{projector} \ee then the action of the operator ${\cal
F}^{\mu\nu}_{x}$ is simply written as ${\cal F}^{\mu\nu}_{x}
\int\! d^n p \, e^{i p x}\, f(p) = - \!\int\! d^n p \, e^{i p x}\,
f(p) \, p^2 P^{\mu\nu}$ where $f(p)$ is an arbitrary function of
$p^\mu$.

Finally after a rather long but straightforward calculation,
and after expanding around $n\!=\!4$, we get,
\bea
&&\!\!\!K^{\mu\nu\alpha\beta}(x)={i \over (4\pi)^2}
\Biggl\{ \kappa_n \left[ {1 \over 90}
{\cal F}^{\mu\nu\alpha\beta}_{x} \delta^n (x)
+4  \Delta \xi^2
{\cal F}^{\mu\nu}_{x}{\cal F}^{\alpha\beta}_{x}
 \delta^n (x)
\right.
\nn  \\
&&
+{2 \over 3}{m^2 \over (n\!-\!2)} \:
\bigr( \eta^{\mu\nu} \eta^{\alpha\beta} \Box_x
-\eta^{\mu (\alpha } \eta^{\beta )\nu} \Box_x
+\eta^{\mu (\alpha } \partial^{\beta )}_x \partial^\nu_x
+\eta^{\nu (\alpha } \partial^{\beta )}_x \partial^\mu_x
\nonumber\\
&&
-\eta^{\mu\nu} \partial^\alpha_x \partial^\beta_x
-\eta^{\alpha\beta} \partial^\mu_x \partial^\nu_x \bigl)
\delta^n (x)
+ {4 \hspace{0.2ex} m^4 \over n (n\!-\!2)} \:
(2 \hspace{0.2ex}\eta^{\mu (\alpha } \eta^{\beta )\nu}
\!- \eta^{\mu\nu} \eta^{\alpha\beta}) \, \delta^n (x)
\biggr]
\nonumber\\
&&
+{1 \over 180}
 {\cal F}^{\mu\nu\alpha\beta}_{x}
  \!\int \! {d^n p \over (2\pi)^n}
e^{i p x} \left(1+4 \,{m^2 \over p^2} \right)^2 \! \bar\phi (p^2)
\nn  \\
&&
 +{2 \over 9}  \, {\cal F}^{\mu\nu}_{x}{\cal
F}^{\alpha\beta}_{x} \!\int \! {d^n p \over (2\pi)^n} \, e^{i p x}
\left(3 \hspace{0.2ex}\Delta \xi+{m^2 \over p^2} \right)^2 \!
\bar\phi (p^2)
\nn  \\
&&
- \left[ {4 \over 675} \,
 {\cal F}^{\mu\nu\alpha\beta}_{x}
+{1 \over 270} \, (60 \hspace{0.1ex}\xi \!-\! 11) \,
{\cal F}^{\mu\nu}_{x}{\cal F}^{\alpha\beta}_{x}
\right] \delta^n (x)
\nn  \\
&&
- m^2 \left[ {2 \over 135} \,
 {\cal F}^{\mu\nu\alpha\beta}_{x}
+{1 \over 27} \, {\cal F}^{\mu\nu}_{x}{\cal F}^{\alpha\beta}_{x}
\right] \Delta_n(x) \Biggr\}+ O(n-4), \label{result for K} \eea
where $\kappa_n$ has been defined in (\ref{kappa}), and $\bar\phi
(p^2)$ and $\Delta_n(x)$ are given by
\be
\bar\phi (p^2)
\!\equiv\! \int_0^1 \!d\alpha \: \ln \biggl(\!1\!+\!{p^2
\over m^2} \alpha (1\!-\!\alpha)\!-\!i \epsilon \!\biggr)\! =\! -i \pi
\theta (-p^2\!-\!4m^2)\sqrt{1\!+\!4 {m^2 \over p^2} } \!+\!\varphi
(p^2\!)\!,
\label{phi}
\ee
\be
\Delta_n(x)\!\equiv\!\int\! {d^n p \over (2\pi)^n}
\: e^{i p x}\; {1 \over p^2},
\label{Delta_n}
\ee
where $ \varphi
(p^2) \equiv  \int_0^1 d\alpha \: \ln | 1+{p^2 \over m^2} \,
\alpha (1\!-\!\alpha)| $.
The imaginary part of (\ref{result for K}) gives the kernel components
$\mu^{-(n-4)}
H_{\scriptscriptstyle \!{\rm S}_{\scriptstyle n}}
^{\mu\nu\alpha\beta}(x,y)$, according to  (\ref{kernel H_S}).
It can be easily obtained multiplying
this expression by $-i$ and retaining only the real part,
$\varphi (p^2)$, of the function $\bar\phi (p^2)$.

\subsection{The Einstein-Langevin equation}

With the previous results for the kernels we can
now write the $n$-dimensional Einstein-Langevin equation
(\ref{flat Einstein-Langevin eq}),
previous to the renormalization.
Taking also into account
Eqs.~(\ref{vev}) and (\ref{flat semiclassical eq}),
we may finally write:
\bea
&&{1\over 8 \pi G_{B}}
G^{{\scriptscriptstyle (1)}\hspace{0.1ex} \mu\nu}(x)
-{4\over 3}\, \alpha_{B} D^{{\scriptscriptstyle
(1)}\hspace{0.1ex} \mu\nu}(x)
-2\beta_{B}B^{{\scriptscriptstyle (1)}\hspace{0.1ex} \mu\nu}(x)
\nonumber\\
&&
+{\kappa_n \over (4\pi)^2} \, \Biggl[
-4 \hspace{0.2ex}\Delta \xi \, {m^2 \over (n\!-\!2)} \,
G^{{\scriptscriptstyle (1)}\hspace{0.1ex} \mu\nu}
+{1 \over 90} \,
D^{{\scriptscriptstyle (1)}\hspace{0.1ex} \mu\nu}
 \Delta \xi^2
B^{{\scriptscriptstyle (1)}\hspace{0.1ex} \mu\nu}
\Biggr]
\hspace{-0.2ex} (x)
\nonumber\\
&&
+{1 \over 2880 \pi^2} \, \Biggl\{
-{16 \over 15} \,
D^{{\scriptscriptstyle (1)}\hspace{0.1ex} \mu\nu}(x)
+\left({1 \over 6}-\! 10\hspace{0.2ex} \Delta \xi \right) \!
B^{{\scriptscriptstyle (1)}\hspace{0.1ex} \mu\nu}(x)
\nn  \\
&&
+ \int\! d^n y \!
\int\! {d^n p \over (2\pi)^n} \, e^{i p (x-y)} \,
\varphi (p^2) \,
\Biggl[\left(1+4 {m^2 \over p^2} \right)^2 \!
D^{{\scriptscriptstyle (1)} \mu\nu}(y)
\nonumber\\
&&
\hspace{18mm}+10 \!
\left(3 \hspace{0.2ex}\Delta \xi+{m^2 \over p^2} \right)^2 \!
B^{{\scriptscriptstyle (1)}\hspace{0.1ex} \mu\nu}(y)
\Biggr]
\nn  \\
&& -\, {m^2 \over 3} \!\int\! d^n y \, \Delta_n(x\!-\!y) \, \Bigl(
8 D^{{\scriptscriptstyle (1)}\hspace{0.1ex} \mu\nu} + 5
B^{{\scriptscriptstyle (1)}\hspace{0.1ex} \mu\nu}
\Bigr)\hspace{-0.2ex} (y) \Biggr\}
\nonumber\\
&&
+{1\over 2} \!\int\! d^ny \,
\mu^{-(n-4)} H_{\scriptscriptstyle \!{\rm A}_{\scriptstyle n}}
^{\mu\nu\alpha\beta}(x,y)\, h_{\alpha\beta}(y) +O(n\!-\!4)
= \xi^{\mu\nu}(x). \label{flat Einstein-Langevin eq 2} \eea
Notice that the terms containing the bare cosmological constant
have cancelled. These equations can now be renormalized, that is,
we can now write the bare coupling constants as renormalized
coupling constants plus some suitably chosen counterterms and take
the limit $n\!\rightarrow \! 4$. In order to carry out such a
procedure, it is convenient to distinguish between massive and
massless scalar fields. The details of the calculation can be
found in Ref.~\cite{MarVer00}.

It is convenient to introduce the two new kernels
\bea &&H_{\rm
A}(x;m^2) \equiv {1 \over 480 \pi^2}\! \int\! {d^4 p \over
(2\pi)^4} \, e^{i px}\, \Biggl\{ \! \left(1+4 \,{m^2 \over p^2}
\right)^{\! 2}
\nonumber\\
&&
\ \ \ \times\Biggl[ - i \pi {\rm sign}p^0 \theta
(-p^2\!-\!4m^2)  \sqrt{1+4 {m^2 \over p^2} }
+\varphi(p^2) \Biggr]
-{8 \over 3} {m^2 \over p^2} \Biggr\},
\nn \\
&&H_{\rm B}(x;m^2,\Delta \xi) \equiv {1 \over 72 \pi^2}\! \int\!
{d^4 p \over (2\pi)^4}  e^{i px} \Biggl\{ \! \left(3
\Delta \xi\!+\!{m^2 \over p^2} \right)^{\! 2}
\nonumber\\
&&
\ \ \ \times\Biggl[\! -
i \pi{\rm sign}\,p^0 \theta (-p^2\!-\!4m^2)  \sqrt{1\!+\!4
{m^2 \over p^2} }
\! +\! \varphi(p^2) \Biggr]\! -{1 \over 6} \! {m^2
\over p^2} \!\Biggr\},
\label{H kernels}
\eea
where $\varphi(p^2)$
is given by the restriction to $n=4$ of expression (\ref{phi}).
The renormalized coupling constants $1/G$, $\alpha$ and $\beta$
are easily computed as it was done in Eq. (\ref{cosmological ct
renorm 2}). Substituting  their expressions into Eq.~(\ref{flat
Einstein-Langevin eq 2}), we can take the limit $n\!\rightarrow \!
4$, using the fact that, for $n=4$, $D^{{\scriptscriptstyle
(1)}\hspace{0.1ex} \mu\nu}(x)= (3/ 2) \, A^{{\scriptscriptstyle
(1)}\hspace{0.1ex} \mu\nu}(x)$, we obtain the corresponding
semiclassical Einstein-Langevin equation.

For the massless case one needs the limit $m \! \rightarrow \! 0$
of equation (\ref{flat Einstein-Langevin eq 2}). In this case it is
convenient to separate $\kappa_n$ in (\ref{kappa}) as
$\kappa_n=\tilde{\kappa}_n +{1 \over 2}\ln (m^2/\mu
^2)+O(n\!-\!4)$, where \be \tilde{\kappa}_n \equiv {1 \over
(n\!-\!4)} \left({e^\gamma \over 4 \pi} \right) ^{\!_{\scriptstyle
n-4 \over 2}}= {1 \over n\!-\!4} +{1\over 2}\, \ln
\!\left({e^\gamma \over 4 \pi } \right)+O (n-4), \label{kappa
tilde} \ee and use that, from Eq.~(\ref{phi}), we have \be
\lim_{m^2 \rightarrow 0} \left[ \varphi (p^2)+\ln (m^2/\mu ^2)
\right]=-2+\ln \left| \hspace{0.2ex} {p^2 \over \mu^2}
\hspace{0.2ex}\right|. \label{massless limit} \ee The coupling
constants are then easily renormalized. We note that in the
massless limit, the Newtonian gravitational constant is not
renormalized and, in the conformal coupling case, $\Delta \xi=0$,
we have that $\beta_{B}\!=\! \beta$. Note also that, by making $m
\!=\!0$ in (\ref{N and D kernels}), the noise and dissipation
kernels can be written as \bea &&N_{\rm A}(x;m^2\!=\!0)=N(x),
\hspace{7ex} N_{\rm B}(x;m^2\!=\!0,\Delta \xi) =60 \hspace{0.2ex}
\Delta \xi^2 \hspace{0.2ex}  N(x),
\nn \\
&&D_{\rm A}(x;m^2\!=\!0)=D(x), \hspace{7ex} D_{\rm
B}(x;m^2\!=\!0,\Delta \xi) =60 \hspace{0.2ex} \Delta \xi^2
\hspace{0.2ex} D(x), \label{massless N and D kernels}
\eea
where
\be N(x) \!\equiv\! {1 \over 480 \pi} \!\int \! {d^4 p \over (2\pi)^4}
\! e^{i px}\! \theta (-p^2)\!,\ \
D(x) \!\equiv\! {-i \over
480 \pi} \!\int \! {d^4 p \over (2\pi)^4} \! e^{i px}\! {\rm
sign}\,p^0  \theta (-p^2)\!.
\label{N and D}
\ee
It is also
convenient to introduce the new kernel \bea H(x;\mu^2) &\equiv &
{1 \over 480 \pi^2} \!\int \! {d^4 p \over (2\pi)^4} \, e^{i px}
\left[ \ln \left| \hspace{0.2ex} {p^2 \over \mu^2}
\hspace{0.2ex}\right| - i \pi \, {\rm sign}\,p^0 \; \theta (-p^2)
\right]
\nn  \\
&=& {1 \over 480 \pi^2} \lim_{\epsilon \rightarrow 0^+} \!\int \!
{d^4 p \over (2\pi)^4} \, e^{i px} \, \ln\! \left( {-(p^0+i
\epsilon)^2+p^i p_i \over \mu^2} \right). \label{Hnew} \eea This
kernel is real and can be written as the sum of an even part and
an odd part in the variables $x^\mu$, where the odd part is the
dissipation kernel $D(x)$. The Fourier transforms (\ref{N and D})
and (\ref{Hnew}) can actually be computed and, thus, in this case
we have explicit expressions for the kernels in position space;
see, for instance, Refs.~\cite{Jon66,CamMarVer95,Hor80}.

Finally, the Einstein-Langevin equation for the physical
stochastic perturbations $h_{\mu\nu}$ can be written in both
cases, for $m \!\neq \!0$ and for $m\!=\!0$, as \bea
\!\!\!\!&&\!\!\!\!{1\over 8 \pi G} G^{{\scriptscriptstyle
(1)}\hspace{0.1ex} \mu\nu}(x) \hspace{-0.2ex}-\hspace{-0.2ex} 2
\left(\bar{\alpha} A^{{\scriptscriptstyle (1)}\hspace{0.1ex}
\mu\nu}(x) \hspace{-0.2ex}+\hspace{-0.2ex} \bar{\beta}
B^{{\scriptscriptstyle (1)}\hspace{0.1ex} \mu\nu}(x) \right)\!
\nn \\
&& \ \ \ + {1\over 4}\int\! d^4y \left[ H_{\rm A}(x\!-\!y)
A^{{\scriptscriptstyle (1)}\hspace{0.1ex} \mu\nu}(y)
\hspace{-0.2ex}+\hspace{-0.2ex} H_{\rm B}(x\!-\!y)
B^{{\scriptscriptstyle (1)}\hspace{0.1ex} \mu\nu}(y) \right] =
\xi^{\mu\nu}(x), \label{unified Einstein-Langevin eq} \eea where
in terms of the renormalized constants $\alpha$ and $\beta$ the
new constants are $\bar{\alpha}=\alpha+(3600\pi^2)^{-1}$ and
$\bar{\beta}=\beta-(1/12-5\Delta\xi)(2880\pi^2)^{-1}$. The kernels
$H_{\rm A}(x)$ and $H_{\rm B}(x)$ are given by Eqs. (\ref{H
kernels}) when $m\neq 0$, and $H_{\rm A}(x)=H(x;\mu^2)$, $H_{\rm
B}(x)=60 \hspace{0.2ex} \Delta \xi^2 H(x;\mu^2)$ when $m \!=\! 0$.
In the massless case, we can use the arbitrariness of the mass
scale $\mu$ to eliminate one of the parameters $\bar{\alpha}$ or
$\bar{\beta}$. The components of the Gaussian stochastic source
$\xi^{\mu\nu}$ have zero mean value and their two-point
correlation functions are given by $
\langle\xi^{\mu\nu}(x)\xi^{\alpha\beta}(y) \rangle_s
=N^{\mu\nu\alpha\beta}(x,y)$, where the noise kernel is given in
(\ref{noise and dissipation kernels 2}), which in the massless
case reduces to (\ref{massless N and D kernels}).

It is interesting to consider the massless conformally coupled
scalar field, {\it i.e.}, the case $\Delta \xi\!=\!0$, of
particular interest because of its similarities with the
electromagnetic field, and also because of its interest in
cosmology: massive fields become conformally invariant when their
masses are negligible compared to the spacetime curvature. We have
already mentioned that for a conformally coupled, field, the
stochastic source tensor must be traceless (up to first order in
perturbation theory around semiclassical gravity), in the sense
that the stochastic variable $\xi^\mu_\mu \!\equiv
\!\eta_{\mu\nu}\xi^{\mu\nu}$ behaves deterministically as a
vanishing scalar field. This can be directly checked by noticing,
from Eqs.~(\ref{noise and dissipation kernels 2}) and
(\ref{massless N and D kernels}), that, when $\Delta \xi\!=\!0$,
one has $\langle\xi^\mu_\mu(x)\xi^{\alpha\beta}(y) \rangle_s =0$,
since ${\cal F}^\mu_\mu\!=\! 3 \hspace{0.2ex}\Box $ and ${\cal
F}^{\mu \alpha}{\cal F}^\beta_\mu \!=\! \Box {\cal
F}^{\alpha\beta}$. The Einstein-Langevin equations for this
particular case (and generalized to a spatially flat
Robertson-Walker background) were first obtained in
Ref.~\cite{CamVer96}, where the coupling constant $\beta$ was
fixed to be  zero. See also Ref.~\cite{HuVer03a} for a discussion
of this result and its connection to the problem of structure
formation in the trace anomaly driven inflation
\cite{Sta80,Vil85,HawHerRea01}.

Note that the expectation value of the renormalized stress-energy
tensor for a scalar field can be obtained by comparing
Eq.~(\ref{unified Einstein-Langevin eq}) with the
Einstein-Langevin equation (\ref{2.11}), its explicit expression
is given in Ref.~\cite{MarVer00}. The results agree with the
general form found by Horowitz \cite{Hor80,Hor81} using an
axiomatic approach and coincides with that given in
Ref.~\cite{FlaWal96}. The particular cases of conformal coupling,
$\Delta \xi \!=\!0$, and minimal coupling, $\Delta \xi \!=\!-1/6$,
are also in agreement with the results for this cases given in
Refs.~\cite{Hor80,Hor81,Sta81,CamVer94,Jor87}, modulo local terms
proportional to $A^{{\scriptscriptstyle (1)}\hspace{0.1ex}
\mu\nu}$ and $B^{{\scriptscriptstyle (1)}\hspace{0.1ex} \mu\nu}$
due to different choices of the renormalization scheme. For the
case of a massive minimally coupled scalar field, $\Delta \xi
\!=\!-1/6$, our result is equivalent to that of
Ref.~\cite{TicFla98}.


\subsection{Correlation functions for gravitational
perturbations}


Here we solve the Einstein-Langevin equations (\ref{unified
Einstein-Langevin eq}) for the components $G^{{\scriptscriptstyle
(1)}\hspace{0.1ex} \mu\nu}$ of the linearized Einstein tensor.
Then we use these solutions to compute the corresponding two-point
correlation functions, which give a measure of the gravitational
fluctuations predicted by the stochastic semiclassical theory of
gravity in the present case. Since the linearized Einstein tensor
is invariant under gauge transformations of the metric
perturbations, these two-point correlation functions are also
gauge invariant. Once we have computed the two-point correlation
functions for the linearized Einstein tensor, we find the
solutions for the metric perturbations and compute the associated
two-point correlation functions. The procedure used to solve the
Einstein-Langevin equation is similar to the one used by Horowitz
\cite{Hor80}, see also Ref.~\cite{FlaWal96}, to analyze the
stability of Minkowski spacetime in semiclassical gravity.

We first note that the tensors $A^{{\scriptscriptstyle
(1)}\hspace{0.1ex} \mu\nu}$ and $B^{{\scriptscriptstyle
(1)}\hspace{0.1ex} \mu\nu}$ can be written in terms of
$G^{{\scriptscriptstyle (1)}\hspace{0.1ex} \mu\nu}$ as \be
A^{{\scriptscriptstyle (1)}\hspace{0.1ex} \mu\nu} = {2 \over 3} \,
({\cal F}^{\mu\nu} G^{{\scriptscriptstyle
(1)}}\mbox{}^{\alpha}_\alpha -{\cal F}^{\alpha}_\alpha
G^{{\scriptscriptstyle (1)}\hspace{0.1ex} \mu\nu}), \hspace{10ex}
B^{{\scriptscriptstyle (1)}\hspace{0.1ex} \mu\nu} = 2
\hspace{0.2ex} {\cal F}^{\mu\nu} G^{{\scriptscriptstyle
(1)}}\mbox{}^{\alpha}_\alpha, \label{A and B} \ee where we have
used that $3 \hspace{0.2ex}\Box={\cal F}^{\alpha}_\alpha$.
Therefore, the Einstein-Langevin equation (\ref{unified
Einstein-Langevin eq})  can be seen as a linear
integro-differential stochastic equation for the components
$G^{{\scriptscriptstyle (1)}\hspace{0.1ex} \mu\nu}$. In order to
find solutions to Eq. (\ref{unified Einstein-Langevin eq}), it is
convenient to Fourier transform. With the convention $\tilde
f(p)=\int d^4x e^{-ipx}f(x)$ for a given field $f(x)$, one finds,
from (\ref{A and B}),
\begin{eqnarray}
&&
\tilde{A}^{{\scriptscriptstyle
(1)}\mu\nu}(p)\!=\! 2 p^2
\tilde{G}^{{\scriptscriptstyle (1)}\mu\nu}(p)\! -\! {2
\over 3} \! p^2\! P^{\mu\nu} \tilde{G}^{{\scriptscriptstyle
(1)}}\mbox{}^{\alpha}_\alpha(p),
\nonumber\\
&&
\tilde{B}^{{\scriptscriptstyle (1)}\mu\nu}(p)\!=\! -2
p^2 \!P^{\mu\nu}\! \tilde{G}^{{\scriptscriptstyle
(1)}}\mbox{}^{\alpha}_\alpha(p).
\end{eqnarray}

The Fourier transform of the
Einstein-Langevin Eq.~(\ref{unified Einstein-Langevin eq}) now reads
\be F^{\mu\nu}_{\hspace{2ex}\alpha\beta}(p) \,
\tilde{G}^{{\scriptscriptstyle (1)}\hspace{0.1ex} \alpha\beta}(p)=
8 \pi G \, \tilde{\xi}^{\mu\nu}(p),
\label{Fourier transf of E-L
eq}
\ee
where
\be F^{\mu\nu}_{\hspace{2ex} \alpha\beta}(p) \equiv
F_1(p) \, \delta^\mu_{( \alpha} \delta^\nu_{\beta )}+ F_2(p) \,
p^2 P^{\mu\nu} \eta_{\alpha\beta},
\label{F def}
\ee
with
\begin{eqnarray}
&&F_1(p) \equiv 1+16 \pi G \, p^2 \left[ {1\over 4}\tilde{H}_{\rm A}(p)-2
\bar{\alpha}\right],
\nonumber\\
&&
F_2(p) \equiv -{16 \over 3} \,
\pi G \left[ {1\over 4}\tilde{H}_{\rm A}(p)+{3\over 4} \tilde{H}_{\rm B}(p) -2
\bar{\alpha}-6 \bar{\beta}\right].
\label{F_1 and F_2}
\end{eqnarray}
In the
Fourier transformed Einstein-Langevin Eq.~(\ref{Fourier transf of
E-L eq}), $\tilde{\xi}^{\mu\nu}(p)$, the Fourier transform of
$\xi^{\mu\nu}(x)$, is a Gaussian stochastic source of zero average
and
\be \langle \tilde{\xi}^{\mu\nu}(p)
\tilde{\xi}^{\alpha\beta}(p^\prime) \rangle_s = (2 \pi)^4 \,
\delta^4(p+p^\prime) \, \tilde{N}^{\mu\nu\alpha\beta}(p),
\label{Fourier transf of corr funct}
\ee
where we have introduced
the Fourier transform of the noise kernel. The explicit expression
for $\tilde{N}^{\mu\nu\alpha\beta}(p)$ is found from (\ref{noise
and dissipation kernels 2}) and (\ref{N and D kernels}) to be
\bea
\tilde{N}^{\mu\nu\alpha\beta}(p)
\!\!\!\! &=& \!\!\!\! {1 \over 720 \pi}\theta
(-p^2\!-\!4m^2) \, \sqrt{1+4 \,{m^2 \over p^2} }
\left[ {1 \over 4} \left(1+4 \,{m^2 \over p^2} \right)^{\!2}
(p^2)^2  P^{\mu\nu\alpha\beta} \right.
\nn  \\
&&\hspace{25mm} \left. + 10 \! \left(3 \hspace{0.2ex}\Delta
\xi+{m^2 \over p^2} \right)^{\!2}\! (p^2)^2 P^{\mu\nu}
P^{\alpha\beta} \right], \label{Fourier transf of noise 2} \eea
which in the massless case reduces to \be \lim_{m \rightarrow 0}\!
\tilde{N}^{\mu\nu\alpha\beta}(p)= {1 \over 480 \pi} \: \theta
(-p^2) \left[ {1 \over 6} \, (p^2)^2 \, P^{\mu\nu\alpha\beta} +60
\hspace{0.2ex} \Delta \xi^2 (p^2)^2 P^{\mu\nu} P^{\alpha\beta}
\right]. \label{Fourier transf of massless noise} \ee

\subsubsection{Correlation functions for the linearized
Einstein tensor}


In general, we can write
$G^{{\scriptscriptstyle (1)}\hspace{0.1ex} \mu\nu}=
\langle G^{{\scriptscriptstyle (1)}\hspace{0.1ex} \mu\nu} \rangle_s
+G_{\rm f}^{{\scriptscriptstyle (1)}\hspace{0.1ex} \mu\nu}$,
where
$G_{\rm f}^{{\scriptscriptstyle (1)}\hspace{0.1ex} \mu\nu}$
is a solution to Eqs.~(\ref{unified Einstein-Langevin eq})
with zero average,
or (\ref{Fourier transf of E-L eq}) in the Fourier transformed version.
The averages
$\langle G^{{\scriptscriptstyle (1)}\hspace{0.1ex} \mu\nu} \rangle_s$
must be a solution of the linearized semiclassical Einstein equations
obtained by averaging Eqs.~(\ref{unified Einstein-Langevin eq}),
or (\ref{Fourier transf of E-L eq}).
Solutions to these equations (specially in the massless case,
$m \!=\! 0$) have been studied by several authors
\cite{Hor80,Hor81,HorWal78,Ran81,Ran82,Sue89a,Sue89b,HarHor81,Sim91,%
Jor87,FlaWal96}, particularly in connection with the problem of
the stability of the ground state of semiclassical gravity. The
two-point correlation functions for the linearized Einstein tensor
are defined by
\begin{eqnarray}
{\cal G}^{\mu\nu\alpha\beta}(x,x^{\prime}) \!\!\! &\equiv &\!\!\!
\langle G^{{\scriptscriptstyle (1)}\hspace{0.1ex} \mu\nu}(x)
G^{{\scriptscriptstyle (1)}\hspace{0.1ex} \alpha\beta}(x^{\prime})
\rangle_s
-\langle G^{{\scriptscriptstyle (1)}\hspace{0.1ex} \mu\nu}(x)
\rangle_s
\langle G^{{\scriptscriptstyle (1)}\hspace{0.1ex} \alpha\beta}
(x^{\prime})\rangle_s
\nonumber\\
 &= &\!\!\!
\langle G_{\rm f}^{{\scriptscriptstyle (1)}\hspace{0.1ex} \mu\nu}(x)
G_{\rm f}^{{\scriptscriptstyle (1)}\hspace{0.1ex} \alpha\beta}
(x^{\prime})\rangle_s.
\label{two-p corr funct}
\end{eqnarray}

Now we shall seek the family of solutions to the Einstein-Langevin
equation which can be written as a linear functional of the
stochastic source and whose Fourier transform,
$\tilde{G}^{{\scriptscriptstyle (1)}\hspace{0.1ex} \mu\nu}(p)$,
depends locally on $\tilde{\xi}^{\alpha\beta}(p)$. Each of such
solutions is a Gaussian stochastic field and, thus, it can be
completely characterized by the averages $\langle
G^{{\scriptscriptstyle (1)}\hspace{0.1ex} \mu\nu} \rangle_s$ and
the two-point correlation functions (\ref{two-p corr funct}). For
such a family of solutions, $\tilde{G}_{\rm
f}^{{\scriptscriptstyle (1)}\hspace{0.1ex} \mu\nu}(p)$ is the most
general solution to Eq.~(\ref{Fourier transf of E-L eq}) which is
linear, homogeneous and local in $\tilde{\xi}^{\alpha\beta}(p)$.
It can be written as \be \tilde{G}_{\rm f}^{{\scriptscriptstyle
(1)}\hspace{0.1ex} \mu\nu}(p) = 8 \pi G \,
D^{\mu\nu}_{\hspace{2ex} \alpha\beta}(p) \,
\tilde{\xi}^{\alpha\beta}(p), \label{G_f} \ee where
$D^{\mu\nu}_{\hspace{2ex} \alpha\beta}(p)$ are the components of a
Lorentz invariant tensor field distribution in Minkowski spacetime
(by ``Lorentz invariant'' we mean invariant under the
transformations of the orthochronous Lorentz subgroup; see
Ref.~\cite{Hor80} for more details on the definition and
properties of these tensor distributions), symmetric under the
interchanges $\alpha \! \leftrightarrow \!\beta$ and $\mu \!
\leftrightarrow \!\nu$, which is the most general solution of \be
F^{\mu\nu}_{\hspace{2ex} \rho\sigma}(p) \,
D^{\rho\sigma}_{\hspace{2ex} \alpha\beta}(p)= \delta^\mu_{(
\alpha} \delta^\nu_{\beta )}. \label{eq for D} \ee In addition, we
must impose the conservation condition: $p_\nu \tilde{G}_{\rm
f}^{{\scriptscriptstyle (1)}\hspace{0.1ex} \mu\nu}(p) = 0$, where
this zero must be understood as a stochastic variable which
behaves deterministically as a zero vector field. We can write
$D^{\mu\nu}_{\hspace{2ex} \alpha\beta}(p)= D^{\mu\nu}_{p
\hspace{1.2ex} \alpha\beta}(p)+ D^{\mu\nu}_{h \hspace{1.2ex}
\alpha\beta}(p)$, where $D^{\mu\nu}_{p \hspace{1.2ex}
\alpha\beta}(p)$ is a particular solution to Eq.~(\ref{eq for D})
and $D^{\mu\nu}_{h \hspace{1.2ex} \alpha\beta}(p)$ is the most
general solution to the homogeneous equation. Consequently, see
Eq.~(\ref{G_f}), we can write $\tilde{G}_{\rm
f}^{{\scriptscriptstyle (1)}\hspace{0.1ex} \mu\nu}(p)
=\tilde{G}_p^{{\scriptscriptstyle (1)}\hspace{0.1ex} \mu\nu}(p)+
\tilde{G}_h^{{\scriptscriptstyle (1)}\hspace{0.1ex} \mu\nu}(p)$.
To find the particular solution, we try an ansatz of the form \be
D^{\mu\nu}_{p \hspace{1.2ex} \alpha\beta}(p)= d_1(p) \,
\delta^\mu_{( \alpha} \delta^\nu_{\beta )} + d_2(p) \, p^2
P^{\mu\nu} \eta_{\alpha\beta}. \label{ansatz for D} \ee
Substituting this ansatz into Eqs.~(\ref{eq for D}), it is easy to
see that it solves these equations if we take \be d_1(p)=\left[ {1
\over F_1(p)} \right]_r, \hspace{7ex} d_2(p)= - \left[
{F_2(p)\over F_1(p) F_3(p)} \right]_r, \label{d's} \ee with \be
F_3(p) \equiv F_1(p) + 3 p^2 F_2(p)= 1-48 \pi G \, p^2 \left[
{1\over 4}\tilde{H}_{\rm B}(p)-2 \bar{\beta}\right], \label{F_3}
\ee and where the notation $[ \;\; ]_r$ means that the zeros of
the denominators are regulated with appropriate prescriptions in
such a way that $d_1(p)$ and $d_2(p)$ are well defined Lorentz
invariant scalar distributions. This yields a particular solution
to the Einstein-Langevin equations: \be
\tilde{G}_p^{{\scriptscriptstyle (1)}\hspace{0.1ex} \mu\nu}(p) = 8
\pi G \, D^{\mu\nu}_{p \hspace{1.2ex} \alpha\beta}(p) \,
\tilde{\xi}^{\alpha\beta}(p), \label{solution} \ee which, since
the stochastic source is conserved, satisfies the conservation
condition. Note that, in the case of a massless scalar field,
$m\!=\!0$, the above solution has a functional form analogous to
that of the solutions of linearized semiclassical gravity found in
the Appendix of Ref.~\cite{FlaWal96}. Notice also that, for a
massless conformally coupled field, $m\!=\!0$ and $\Delta
\xi\!=\!0$, the second term on the right hand side of
Eq.~(\ref{ansatz for D}) will not contribute in the correlation
functions (\ref{two-p corr funct}), since in this case the
stochastic source is traceless.

A detailed analysis given in Ref.~\cite{MarVer00}
concludes that the homogeneous solution
$\tilde{G}_h^{{\scriptscriptstyle (1)}\hspace{0.1ex} \mu\nu}(p)$ gives
no contribution to the correlation functions
(\ref{two-p corr funct}). Consequently
${\cal G}^{\mu\nu\alpha\beta}(x,x^{\prime}) \!=\!
\langle G_p^{{\scriptscriptstyle (1)}\hspace{0.1ex} \mu\nu}(x)
G_p^{{\scriptscriptstyle (1)}\hspace{0.1ex} \alpha\beta}
(x^{\prime})\rangle_s$, where
$G_p^{{\scriptscriptstyle (1)}\hspace{0.1ex} \mu\nu}(x)$ is the
inverse Fourier transform of (\ref{solution}),
and the correlation functions (\ref{two-p corr funct}) are
\be
\langle \tilde{G}_p^{{\scriptscriptstyle
(1)}\hspace{0.1ex} \mu\nu}\!(p)  \tilde{G}_p^{{\scriptscriptstyle
(1)}\hspace{0.1ex} \alpha\beta}\! (p^\prime) \rangle_s\! = \! 64 \!(2
\pi)^6 \! G^2 \! \delta^4\!(p\!+\!p^\prime) \! D^{\mu\nu}_{p
\hspace{1.2ex}\rho\sigma}(p) \! D^{\alpha\beta}_{p \hspace{1.2ex}
\lambda\gamma}(-p) \! \tilde{N}^{\rho\sigma\lambda\gamma}\!(p).
\ee
It is easy to see from the above analysis that the prescriptions
$[ \;\; ]_r$ in the factors $D_p$ are irrelevant in the last
expression and, thus, they can be suppressed. Taking into account
that $F_l(-p) \!=\! F^{\displaystyle \ast}_l(p)$, with $l \!=\!
1,2,3$, we get from Eqs.~(\ref{ansatz for D}) and (\ref{d's}) \bea
\!\!\!\! &&\!\!\!\!\langle \tilde{G}_p^{{\scriptscriptstyle (1)}
\hspace{0.1ex}
\mu\nu}(p) \, \tilde{G}_p^{{\scriptscriptstyle (1)}\hspace{0.1ex}
\alpha\beta} (p^\prime) \rangle_s =
\nonumber\\
\!\!\!\!\!\! && \!\!\!\!\!\!
64
(2 \pi)^6 \, G^2 \: {\delta^4(p+p^\prime) \over \left|
\hspace{0.1ex} F_1(p) \hspace{0.1ex}\right|^2 } \left[
\tilde{N}^{\mu\nu\alpha\beta}(p) - {F_2(p) \over F_3(p)} \: p^2
P^{\mu\nu} \hspace{0.2ex}
\tilde{N}^{\alpha\beta\rho}_{\hspace{3.3ex} \rho}(p)
\right.     \nn  \\
\!\! \!\!\! && \!\!\!\!\!
\left. -
{F_2^{\displaystyle \ast}(p) \over F_3^{\displaystyle \ast}(p)}
\: p^2 P^{\alpha\beta} \hspace{0.2ex}
\tilde{N}^{\mu\nu\rho}_{\hspace{3.3ex} \rho}(p)
\! +\! { \left| F_2(p) \right|^2
\over \left| F_3(p) \right|^2 }
p^2 P^{\mu\nu}  p^2 P^{\alpha\beta}
\tilde{N}
^{\rho \hspace{0.9ex} \sigma}_{\hspace{1ex} \rho \hspace{1.1ex}
\sigma} (p) \right]\!.   \nn  \\
\mbox{} \eea This last expression is well defined as a
bi-distribution and can be easily evaluated using
Eq.~(\ref{Fourier transf of noise 2}). The final explicit result
for the Fourier transformed correlation function for the Einstein
tensor is thus \bea
 \!\!\!\!\! && \!\!\!\!\!
\langle \tilde{G}_p^{{\scriptscriptstyle
(1)}\hspace{0.1ex} \mu\nu}(p) \, \tilde{G}_p^{{\scriptscriptstyle
(1)}\hspace{0.1ex} \alpha\beta} (p^\prime) \rangle_s=
\nonumber\\
 \!\!\!\!\! &=& \!\!\!\!\!{2 \over
45} \!(2 \pi)^5 \, G^2 \:
{\delta^4(p+p^\prime) \over \left| \hspace{0.1ex} F_1(p)
\hspace{0.1ex}\right|^2 } \: \theta (-p^2\!-\!4m^2) \, \sqrt{1+4
\,{m^2 \over p^2} }
\nn   \\
&&  \!\!\!\!  \times \!
\left[{1 \over 4} \left(1+4 \,{m^2 \over p^2} \right)^{\!2} \!
(p^2)^2 \,
 P^{\mu\nu\alpha\beta }  \right.
\nn   \\
&&  \left.
+\! 10 \!
\left(3 \hspace{0.2ex}\Delta \xi\!+\!{m^2 \over p^2} \right)^{\!2} \!
(p^2)^2 P^{\mu\nu} P^{\alpha\beta}
\left| 1\!-\!3 p^2  {F_2(p) \over F_3(p)} \right|^2
\right]\!\!.
\label{Fourier tr corr funct}
\eea

To obtain the correlation functions in coordinate space,
Eq.~(\ref{two-p corr funct}), we take the inverse Fourier
transform. The final result is: \be {\cal
G}^{\mu\nu\alpha\beta}(x,x^{\prime})= {\pi \over 45} \, G^2
\,{\cal F}^{\mu\nu\alpha\beta}_{x} \, {\cal G}_{\rm A}
(x-x^{\prime})+ {8 \pi \over 9} \, G^2 \, {\cal F}^{\mu\nu}_{x}
{\cal F}^{\alpha\beta}_{x} \, {\cal G}_{\rm B} (x-x^{\prime}),
\label{corr funct} \ee with \bea
\!\!\!\!\!\!&&\!\!\!\!\!\!\tilde{{\cal G}}_{\rm A}(p) \!\equiv\!
\theta (-p^2-4m^2) \, \sqrt{1+4 \,{m^2 \over p^2} } \left(1+4
\,{m^2 \over p^2} \right)^{\!2} \! {1 \over \left| \hspace{0.1ex}
F_1(p) \hspace{0.1ex}\right|^2 }  ,
\nn   \\
\!\!\!\!\!\!&&\!\!\!\!\!\!\tilde{{\cal G}}_{\rm B}(p)\! \equiv\!
\theta (-p^2\!-\!4m^2)  \sqrt{1\!+\!4{m^2 \over p^2} }\! \left(\!3
\Delta \xi\!+\!{m^2 \over p^2}\!\right)^{\!2} \!\! {1 \over \left|
F_1(p) \right|^2 }\! \left| 1\!-\! 3 p^2 {F_2(p) \over F_3(p)}
\right|^2\!\!, \label{distri} \eea where $F_l(p)$, $l=1,2,3$, are
given in (\ref{F_1 and F_2}) and (\ref{F_3}). Notice that, for a
massless field ($m \!=\! 0$), we have \bea &&F_1(p)= 1+ 4 \pi G
\hspace{0.2ex} p^2 \hspace{0.2ex}
          \tilde{H}(p;\bar{\mu}^2),
\nn  \\
&&F_2(p)= - {16 \over 3} \, \pi G \left[
(1 +180 \hspace{0.2ex}\Delta \xi^2 ) \, {1\over 4}\tilde{H}(p;\bar{\mu}^2)
-6 \Upsilon \right],
\nn  \\
&&F_3(p)= 1- 48 \pi G \hspace{0.2ex} p^2
\left[ 15 \hspace{0.2ex}\Delta \xi^2 \hspace{0.2ex}
\tilde{H}(p;\bar{\mu}^2) -2 \Upsilon \right],
\eea
with $\bar{\mu} \equiv \mu\, \exp (1920 \pi^2 \bar{\alpha})$
and
$\Upsilon \equiv \bar{\beta}
 -60 \hspace{0.2ex}\Delta \xi^2 \hspace{0.2ex}\bar{\alpha}$, and where
$\tilde{H}(p;\mu^2)$ is the Fourier transform of
$H(x;\mu^2)$ given in (\ref{Hnew}).


\subsubsection{Correlation functions for the metric
perturbations}


Starting from the solutions found for the linearized Einstein tensor,
which are characterized by the two-point correlation functions
(\ref{corr funct}) [or, in terms of Fourier transforms,
(\ref{Fourier tr corr funct})], we can now solve the equations for the
metric perturbations. Working in the harmonic gauge,
$\partial_{\nu} \bar{h}^{\mu\nu} \!=\! 0$ (this zero must be
understood in a statistical sense) where
$\bar{h}_{\mu\nu} \!\equiv \! h_{\mu\nu}
\!-\! (1/2)\hspace{0.2ex} \eta_{\mu\nu} \hspace{0.2ex}h_\alpha^\alpha$,
the equations for the metric perturbations in terms of
the Einstein tensor are
\be
\Box \bar{h}^{\mu\nu}(x) \!=\! -2
G^{{\scriptscriptstyle (1)}\hspace{0.1ex} \mu\nu}(x),
\label{metric and G}
\ee
or, in terms of
Fourier transforms,
$p^2
\tilde{\bar{h}}^{\mbox{}_{\mbox{}_{\mbox{}_{\mbox{}_{\mbox{}
_{\scriptstyle \mu\nu}}}}}}\hspace{-0.5ex} (p)
\!=\! 2 \tilde{G}^{{\scriptscriptstyle (1)}\hspace{0.1ex} \mu\nu}(p)$.
Similarly to the analysis of  the equation
for the Einstein tensor, we can write
$\bar{h}^{\mu\nu} \!=\! \langle \bar{h}^{\mu\nu} \rangle_s
\!+\! \bar{h}^{\mu\nu}_{\rm f}$, where $\bar{h}^{\mu\nu}_{\rm f}$ is a
solution to these equations with zero average, and the two-point
correlation functions are defined by
\begin{eqnarray}
{\cal H}^{\mu\nu\alpha\beta}(x,x^{\prime})\!\!\!\! &\equiv &\!\!\!\!
\langle \bar{h}^{\mu\nu}(x) \bar{h}^{\alpha\beta}(x^{\prime})
\rangle_s - \langle \bar{h}^{\mu\nu}(x) \rangle_s
\langle \bar{h}^{\alpha\beta}(x^{\prime}) \rangle_s
\nonumber\\
\!\!\!\! &=&\!\!\!\!
\langle \bar{h}^{\mu\nu}_{\rm f}(x)
\bar{h}^{\alpha\beta}_{\rm f}(x^{\prime}) \rangle_s.
\label{co fu}
\end{eqnarray}

We seek solutions of
the Fourier transform of Eq.~(\ref{metric and G})
of the form
$\tilde{\bar{h}}^{\mbox{}_{\mbox{}_{\mbox{}_{\mbox{}_{\mbox{}
_{\scriptstyle \mu\nu}}}}}}_{\rm f}\hspace{-0.5ex} (p)
\!=\! 2 D(p)
\tilde{G}^{{\scriptscriptstyle (1)}\hspace{0.1ex} \mu\nu}_{\rm f}(p)$,
where $D(p)$ is a Lorentz invariant scalar distribution in Minkowski
spacetime, which is the most general solution of $p^2 D(p) \!=\! 1$.
Note that, since the linearized Einstein tensor is conserved,
solutions of this form automatically satisfy the harmonic
gauge condition. As in the previous subsection,
we can write $D(p) \!=\! [ 1/p^2 ]_r
\!+\! D_h(p)$, where $D_h(p)$ is the most general solution to the
associated homogeneous equation and, correspondingly, we have
$\tilde{\bar{h}}^{\mbox{}_{\mbox{}_{\mbox{}_{\mbox{}_{\mbox{}
_{\scriptstyle \mu\nu}}}}}}_{\rm f}\hspace{-0.5ex} (p)
\!=\!
\tilde{\bar{h}}^{\mbox{}_{\mbox{}_{\mbox{}_{\mbox{}_{\mbox{}
_{\scriptstyle \mu\nu}}}}}}_p\hspace{-0.5ex} (p) +
 \tilde{\bar{h}}^{\mbox{}_{\mbox{}_{\mbox{}_{\mbox{}_{\mbox{}
_{\scriptstyle \mu\nu}}}}}}_h \hspace{-0.5ex} (p)$. However, since
$D_h(p)$ has support on the set of points for which $p^2 \!=\! 0$,
it is easy to see from Eq.~(\ref{Fourier tr corr funct}) (from the
factor $\theta (-p^2-4 m^2)$) that $\langle
\tilde{\bar{h}}^{\mbox{}_{\mbox{}_{\mbox{}_{\mbox{}_{\mbox{}
_{\scriptstyle \mu\nu}}}}}}_h \hspace{-0.5ex} (p)
\tilde{G}^{{\scriptscriptstyle (1)}\hspace{0.1ex} \alpha\beta}
_{\rm f}(p^{\prime}) \rangle_s \!=\! 0$ and, thus, the two-point
correlation functions (\ref{co fu}) can be computed from $\langle
\tilde{\bar{h}}^{\mbox{}_{\mbox{}_{\mbox{}_{\mbox{}_{\mbox{}
_{\scriptstyle \mu\nu}}}}}}_{\rm f}\hspace{-0.5ex} (p)
\tilde{\bar{h}}^{\mbox{}_{\mbox{}_{\mbox{}_{\mbox{}_{\mbox{}
_{\scriptstyle \alpha\beta}}}}}}_{\rm
f}\hspace{-0.5ex}(p^{\prime}) \rangle_s \!=\! \langle
\tilde{\bar{h}}^{\mbox{}_{\mbox{}_{\mbox{}_{\mbox{}_{\mbox{}
_{\scriptstyle \mu\nu}}}}}}_p\hspace{-0.5ex} (p)
\tilde{\bar{h}}^{\mbox{}_{\mbox{}_{\mbox{}_{\mbox{}_{\mbox{}
_{\scriptstyle \alpha\beta}}}}}}_p\hspace{-0.5ex}(p^{\prime})
\rangle_s$. {}From Eq.~(\ref{Fourier tr corr funct}) and due to
the factor $\theta (-p^2-4 m^2)$, it is also easy to see that the
prescription $[ \;\; ]_r$ is irrelevant in this correlation
function and we obtain \be \langle
\tilde{\bar{h}}^{\mbox{}_{\mbox{}_{\mbox{}_{\mbox{}_{\mbox{}
_{\scriptstyle \mu\nu}}}}}}_p\hspace{-0.5ex} (p)
\tilde{\bar{h}}^{\mbox{}_{\mbox{}_{\mbox{}_{\mbox{}_{\mbox{}
_{\scriptstyle \alpha\beta}}}}}}_p\hspace{-0.5ex}(p^{\prime})
\rangle_s = {4 \over (p^2)^2} \, \langle
\tilde{G}_p^{{\scriptscriptstyle (1)}\hspace{0.1ex} \mu\nu}(p) \,
\tilde{G}_p^{{\scriptscriptstyle (1)}\hspace{0.1ex} \alpha\beta}
(p^\prime) \rangle_s, \ee where $\langle
\tilde{G}_p^{{\scriptscriptstyle (1)}\hspace{0.1ex} \mu\nu}(p) \,
\tilde{G}_p^{{\scriptscriptstyle (1)}\hspace{0.1ex} \alpha\beta}
(p^\prime) \rangle_s$ is given by  Eq.~(\ref{Fourier tr corr
funct}). The right hand side of this equation is a well defined
bi-distribution, at least for $m \!\neq \! 0$ (the $\theta$
function provides the suitable cutoff). In the massless field
case, since the noise kernel is obtained as the limit $m
\!\rightarrow \!0$ of the noise kernel for a massive field, it
seems that the natural prescription to avoid the divergences on
the lightcone $p^2 \!=\! 0$ is a Hadamard finite part, see
Refs.~\cite{Sch57,Zem87} for its definition. Taking this
prescription, we also get a well defined bi-distribution for the
massless limit of the last expression.

The final result for the two-point correlation function
for the field $\bar h^{\mu\nu}$ is:
\be
{\cal H}^{\mu\nu\alpha\beta}(x,x^{\prime})=
{4 \pi \over 45} \, G^2 \,{\cal F}^{\mu\nu\alpha\beta}_{x} \,
{\cal H}_{\rm A} (x-x^{\prime})+
{32 \pi \over 9} \, G^2 \,
{\cal F}^{\mu\nu}_{x} {\cal F}^{\alpha\beta}_{x} \,
{\cal H}_{\rm B} (x-x^{\prime}),
\label{corr funct 2}
\ee
where $\tilde{{\cal H}}_{\rm A}(p) \!\equiv \!
[1/(p^2)^2]\, \tilde{{\cal G}}_{\rm A}(p)$ and
$\tilde{{\cal H}}_{\rm B}(p) \!\equiv \!
[1/(p^2)^2]\, \tilde{{\cal G}}_{\rm B}(p)$, with
$\tilde{{\cal G}}_{\rm A}(p)$ and
$\tilde{{\cal G}}_{\rm B}(p)$ given by (\ref{distri}).
The two-point correlation functions for the metric perturbations
can be easily obtained using $h_{\mu\nu} \!=\!
\bar{h}_{\mu\nu}
\!-\! (1/2) \hspace{0.2ex}\eta_{\mu\nu}
\hspace{0.2ex}\bar{h}^{\alpha}_{\alpha}$.


\subsubsection{Conformally coupled field}


For a conformally coupled field, {\it i.e.}, when $m = 0$ and
$\Delta \xi=0$, the previous correlation functions are greatly
simplified and can be approximated explicitly in terms of analytic
functions. The detailed results are given in Ref. \cite{MarVer00},
here we outline the main features.

When $m=0$ and $\Delta \xi=0$ we have
${\cal G}_{\rm B} (x) \!=\!0$ and
$
\tilde{{\cal G}}_{\rm A}(p)= \theta(-p^2)
\left|\hspace{0.2ex} F_1(p) \hspace{0.2ex} \right|^{-2}$.
Thus the two-point correlations functions for
the Einstein tensor is
\be
{\cal G}^{\mu\nu\alpha\beta}(x,x^{\prime})=
{\pi \over 45} \, G^2 \,{\cal F}^{\mu\nu\alpha\beta}_{x} \,
\int {d^4p\over (2\pi)^4}\frac
{e^{ip(x-x^\prime)}\,\theta(-p^2)}
{| 1+4\pi Gp^2\tilde H(p;\bar\mu^2)|^2},
\label{corr funct conf}
\ee
where $\tilde H(p,\mu^2)=(480\pi^2)^{-1}\ln
[-((p^0+i\epsilon)^2+p^ip_i)/\mu^2]$, see Eq. (\ref{Hnew}).

To estimate this integral
let us consider spacelike separated points
$(x-x^{\prime})^\mu=(0,\!{\bf x}-{\bf x}^\prime)$, and define
${\bf y}={\bf x}-{\bf x}^\prime$. We may now
formally change the momentum variable $p^\mu$
by the dimensionless vector $s^\mu$: $p^\mu=s^\mu/|{\bf y}|$, then
the previous integral denominator is
$|1+16\pi (L_P/|{\bf y}|)^2s^2\tilde H(s)|^2$,
where we have introduced the Planck length $L_P=\sqrt{G}$.
It is clear that we can consider two regimes: (a) when
$L_P \ll |{\bf y}|$, and (b) when $|{\bf y}|\sim L_P$.
In  case (a) the correlation function, for the $0000$
component, say,
will be of the order
$$
{\cal G}^{0000}({\bf y})\sim {L_P^4\over|{\bf y}|^8}.
$$
In case (b) when the denominator has zeros
a detailed calculation  carried out in Ref. \cite{MarVer00}
shows that:
$$
{\cal G}^{0000}({\bf y})\sim e^{-|{\bf y}|/L_P}\left(
{L_P\over |{\bf y}|^5}+\dots +{1\over L_P^2|{\bf y}|^2}\right)
$$
which indicates an exponential decay at distances around
the Planck scale. Thus short scale fluctuations are
strongly suppressed.

For the two-point metric correlation the results
are similar. In this case we have
\be
{\cal H}^{\mu\nu\alpha\beta}(x,x^{\prime})=
{4\pi \over 45} \, G^2 \,{\cal F}^{\mu\nu\alpha\beta}_{x} \,
\int {d^4p\over (2\pi)^4}\frac
{e^{ip(x-x^\prime )}\theta(-p^2)}
{(p^2)^2| 1+4\pi Gp^2\tilde H(p;\bar\mu^2)|^2}.
\label{corr funct conf metric}
\ee
The integrand has the same behavior of the correlation function
of Eq. (\ref{corr funct conf})
thus matter fields tends to suppress the short scale metric
perturbations.
In this case we find, as for the correlation of
the Einstein tensor,
that for case (a) above we have,
$$
{\cal H}^{0000}({\bf y})\sim {L_P^4\over|{\bf y}|^4},
$$
and for case (b) we have
$$
{\cal H}^{0000}({\bf y})\sim e^{-|{\bf y}|/L_P}\left(
{L_P\over |{\bf y}|}+\dots \right).
$$

It is interesting to write expression (\ref{corr funct conf
metric}) in an alternative way. If we use the dimensionless tensor
$P^{\mu\nu\alpha\beta}$ introduced in Eq.~(\ref{projector}), which
accounts for the effect of the operator ${\cal
F}^{\mu\nu\alpha\beta}_{\,x}$, we can write \be {\cal
H}^{\mu\nu\alpha\beta}(x,x^{\prime})= {4\pi \over 45} \, G^2 \,
\int {d^4p\over (2\pi)^4}\,\frac
{e^{ip(x-x^\prime)}\,P^{\mu\nu\alpha\beta}\,\theta(-p^2)} {|
1+4\pi Gp^2\tilde H(p;\bar\mu^2)|^2}. \label{corr funct conf
metric2} \ee This expression allows a direct comparison with the
graviton propagator for linearized quantum gravity in the $1/N$
expansion found by Tomboulis \cite{Tom77}. One can see that the
imaginary part of the graviton propagator leads, in fact, to
Eq.~(\ref{corr funct conf metric2}). In Ref.~\cite{RouVer03b} it
is shown that, in fact, the two-point correlation functions for
the metric perturbations derived from the Einstein-Langevin
equation are equivalent to the symmetrized quantum two-point
correlation functions for the metric fluctuations in the large $N$
expansion of quantum gravity interacting with $N$ matter fields.


\subsection{Discussion}


The main results of this section are the
correlation functions  (\ref{corr funct})
and (\ref{corr funct 2}). In the case of a conformal field, the
correlation functions of the linearized Einstein
tensor have been explicitly estimated.
{}From the exponential factors $e^{-|{\bf y}|/L_P}$ in these results
for scales near the Planck length,
we see that the correlation functions of the linearized Einstein
tensor have the  Planck length as the correlation length.
A similar behavior is found for the
correlation functions of the metric perturbations.
Since these fluctuations are induced by the matter fluctuations
we infer that the effect of the matter fields is to suppress the
fluctuations of the metric at very small scales.
On the other hand,
at scales much larger than the Planck length
the induced metric fluctuations are small
compared with the free graviton propagator which goes like
$L_P^2/|{\bf y}|^2$, since the action for the free
graviton goes like $S_h\sim\int d^4 x\,L_P^{-2}h\Box h$

For background solutions of semiclassical gravity with other
scales present apart from the Planck scales (for instance, for
matter fields in a thermal state), stress-energy fluctuations may
be important at larger scales. For such backgrounds, stochastic
semiclassical gravity might predict correlation functions with
characteristic correlation lengths larger than the Planck scales.
It seems quite plausible, nevertheless, that these correlation
functions would remain non-analytic in their characteristic
correlation lengths. This would imply that these correlation
functions could not be obtained from a calculation involving a
perturbative expansion in the characteristic correlation lengths.
In particular, if these correlation lengths are proportional to
the Planck constant $\hbar$, the gravitational correlation
functions could not be obtained from an expansion in $\hbar$.
Hence, stochastic semiclassical gravity might predict a behavior
for gravitational correlation functions different from that of the
analogous functions in perturbative quantum gravity
\cite{Don94a,Don94b,Don96a,Don96b}. This is not necessarily
inconsistent with having neglected action terms of higher order in
$\hbar$ when considering semiclassical gravity as an effective
theory \cite{FlaWal96}. It is, in fact, consistent with the closed
connection of stochastic gravity with the large $N$ expansion of
quantum gravity interacting with $N$ matter fields.

\section{Structure formation}
\label{sec:strfor}

Cosmological structure formation is a key problem in modern
cosmology \cite{KolTur90,Pad93} and inflation offers a natural
solution to this problem. If an inflationary period is present,
the initial seeds for the generation of the primordial
inhomogeneities that lead to the large scale structure have their
source in the quantum fluctuations of the inflaton field, the
field which is generally  responsible for driving inflation.
Stochastic gravity provides a sound and natural formalism for the
derivation of the cosmological perturbations generated during
inflation.

In Ref.~\cite{RouVer03a} it was shown that the correlation
functions that follow from the Einstein-Langevin equation which
emerges in the framework of stochastic gravity coincide with that
obtained with the usual quantization procedures \cite{MukFelBra92}
when both the metric perturbations and the inflaton fluctuations
are both linearized. Stochastic gravity, however, can naturally
deal with the fluctuations of the inflaton field even beyond the
linear approximation.

Here we will illustrate the equivalence with the usual formalism,
based on the quantization of the linear cosmological and inflaton
perturbations, with one of the simplest chaotic inflationary
models in which the background spacetime is a quasi de Sitter
universe \cite{RouVer00,RouVer03a}.


\subsection{The model}


In this chaotic inflationary model \cite{Lin90} the inflaton field
$\phi$ of mass $m$ is described by the following Lagrangian
density,
\begin{equation} {\cal
L}(\phi)={1\over 2}g^{ab}\nabla_a\phi \nabla_b\phi + {1\over 2}m^2\phi^2.
\label{1.14}
\end{equation}
The conditions for the existence of an inflationary period, which
is characterized by an accelerated cosmological expansion, is that
the value of the field over a region with the typical size of the
Hubble radius is higher than the Planck mass $m_P$. This is
because in order to solve the cosmological horizon and flatness
problem more than 60 e-folds of expansion are needed, to achieve
this the scalar field should begin with a value higher than
$3m_P$. The inflaton mass is small: as we will see, the large
scale anisotropies measured in the cosmic background radiation
\cite{Smo92} restrict the inflaton mass to be of the order of
$10^{-6}m_P$. We will not discuss the naturalness of this
inflationary model and we will simply assume that if one such
region is found (inside a much larger universe) it will inflate to
become our observable universe.

We want to study the metric perturbations produced by the
stress-energy tensor
fluctuations of the inflaton field on the homogeneous background of a flat
Friedmann-Robertson-Walker model, described by the cosmological scale
factor $a(\eta)$, where $\eta$ is the conformal time, which is driven by
the homogeneous inflaton field $\phi(\eta)=\langle\hat\phi\rangle$. Thus we
write the inflaton field in the following form
\begin{equation}
\hat\phi=\phi(\eta)+ \hat\varphi (x),
\label{1.15}
\end{equation}
where $\hat\varphi (x)$ corresponds to a free massive quantum scalar field
with zero expectation value on the homogeneous background metric:
$\langle\hat\varphi\rangle=0$.
We will restrict ourselves to scalar-type metric perturbations because
these are the ones that couple to the inflaton fluctuations in the linear
theory. We note that this is not so if we were to consider inflaton
fluctuations beyond the linear approximation, then tensorial and vectorial
metric perturbations would also be driven. The perturbed metric $\tilde
g_{ab}=g_{ab}+h_{ab}$ can be written in the longitudinal gauge as,
\begin{equation}
ds^2=a^2(\eta)[-(1+2\Phi(x))d\eta^2+(1-2\Psi(x))\delta_{ij}dx^idx^j],
\label{1.16} \end{equation} where the scalar metric perturbations
$\Phi(x)$ and $\Psi(x)$ correspond to Bardeen's gauge invariant
variables \cite{Bar80}.


\subsection{Einstein-Langevin equation for scalar metric perturbations}


The Einstein-Langevin
equation as described in  section \ref{sec2} is gauge invariant, and thus
we can work in a desired gauge and then extract the gauge invariant
quantities. The Einstein-Langevin
equation (\ref{2.11}) reads now:
\begin{equation}
G^{(0)}_{ab}-8\pi G\langle\hat T^{(0)}_{ab}\rangle+
G^{(1)}_{ab}(h)-8\pi G\langle\hat T^{(1)}_{ab}(h)\rangle=
 8\pi G\xi_{ab},
\label{1.17}
\end{equation}
where the two first terms cancel, that is
$G^{(0)}_{ab}-8\pi G\langle\hat T^{(0)}_{ab}\rangle=0$,
as the background metric
satisfies the semiclassical Einstein equations. Here the subscripts $(0)$
and $(1)$ refer to functions in the background
metric $g_{ab}$ and linear
in the metric perturbation $h_{ab}$, respectively. The stress
tensor operator $\hat T_{ab}$ for
the minimally coupled inflaton field in the perturbed metric is:
\begin{equation}
\hat T_{ab}=
\tilde\nabla_{a}\hat\phi \tilde\nabla_{b}\hat\phi+{1\over2}\tilde
g_{ab} (\tilde\nabla_{c} \hat\phi \tilde\nabla^{c}\hat\phi+
m^2\hat\phi^2).
\label{1.18}
\end{equation}

Using the decomposition of the scalar field into its
homogeneous and inhomogeneous part, see  Eq.~(\ref{1.15}), and the metric
$\tilde g_{ab}$ into its homogeneous background $g_{ab}$ and
its perturbation $h_{ab}$,
the renormalized expectation value for the stress-energy
tensor operator can be written as
\begin{equation}
\langle \hat T^R_{ab}[\tilde g]\rangle=
\langle \hat T_{ab}[\tilde g]\rangle_{\phi\phi}+
\langle \hat T_{ab}[\tilde g]\rangle_{\phi\varphi}+
\langle \hat T^R_{ab}[\tilde g]\rangle_{\varphi\varphi},
\label{1.19}
\end{equation}
where the subindices indicate the degree of dependence
on the homogeneous field $\phi$ and its perturbation $\varphi$.
The first term in this equation depends only on the homogeneous field
and it is given by the classical expression.
The second term is proportional to
$\langle\hat\varphi[\tilde g]\rangle$ which is not zero because the field
dynamics is considered on the perturbed spacetime, {\it i.e.}, this term
includes the coupling of the field with $h_{ab}$ and
may be obtained from the expectation value of the linearized
Klein-Gordon equation,
\begin{equation}
\left( \Box_{g+h}-m^2\right)\hat\varphi =0.
\label{1.19a}
\end{equation}
The last term in Eq.~(\ref{1.19}) corresponds to the
expectation value to the stress tensor for a free scalar field on the
spacetime of the perturbed metric.

After using the previous decomposition, the noise kernel
$N_{abcd}[g;x,y)$ defined in Eq.~(\ref{2.8}) can be written as
\begin{eqnarray}
\langle \{\hat t_{ab}[g;x),\hat t_{cd}[g;y)\}\rangle
\!\!\! &=&\!\!\!
\langle \{\hat t_{ab}[g;x),\hat t_{cd}[g;y)\}
\rangle_{(\phi\varphi)^2}
\nonumber\\
&&\!\!\! +
\langle \{\hat t_{ab}[g;x),\hat t_{cd}[g;y)\}
\rangle_{(\varphi\varphi)^2},
\label{1.20}
\end{eqnarray}
where we have used the fact that $\langle\hat\varphi\rangle=0
=\langle\hat\varphi\hat\varphi\hat\varphi\rangle$ for Gaussian states on
the background geometry. We consider the vacuum state to be the
Euclidean vacuum which is preferred in the de Sitter
background, and this state is Gaussian. In the above equation the first
term is quadratic in $\hat\varphi$ whereas  the second one is quartic,
both contributions to the noise kernel are separately conserved since
both $\phi(\eta)$ and $\hat\varphi$ satisfy the Klein-Gordon field
equations on the background spacetime. Consequently, the two terms can be
considered separately. On the other hand if one treats $\hat \varphi$
as a small perturbation the second term in (\ref{1.20}) is of lower order
than the first and may be consistently neglected, this corresponds to
neglecting the last term of Eq.~(\ref{1.19}). The stress tensor fluctuations
due to a term of that kind were considered in ref. \cite{RouVer99}.

We can now write down the Einstein-Langevin equations (\ref{1.17})
to linear order in the inflaton fluctuations. It is easy to check
\cite{RouVer03a} that the {\it space-space} components coming from
the stress tensor expectation value terms and the stochastic
tensor are diagonal, i.e. $\langle\hat T_{ij}\rangle=0= \xi_{ij}$
for $i\not= j$. This, in turn, implies that the two functions
characterizing the scalar metric perturbations are equal:
$\Phi=\Psi$ in agreement with ref. \cite{MukFelBra92}. The
equation for $\Phi$ can be obtained from the $0i$-component of the
Einstein-Langevin equation, which in Fourier space reads
\begin{equation} 2ik_i({\cal H}\Phi_k+\Phi'_k)= 8\pi
G(\xi_{0i})_k, \label{1.21}
\end{equation}
where $k_i$ is the comoving momentum component associated to the
comoving coordinate $x^i$, and we have used the definition
$\Phi_k(\eta)= \int d^3 x \exp(-i\vec k\cdot\vec x)\Phi(\eta,\vec
x)$. Here primes denote derivatives with respect to the conformal
time $\eta$ and ${\cal H}=a'/a$. A nonlocal term of dissipative
character which comes from the second term in Eq.~(\ref{1.19})
should also appear on the left hand side of Eq.~(\ref{1.21}), but
we have neglected it to simplify the forthcoming expressions. Its
inclusion does not change the large scale spectrum in an essential
way \cite{RouVer03a}.  Note, however, that the equivalence of the
stochastic approach to linear order in $\hat\varphi$ and the usual
linear cosmological perturbations approach is independent of that
approximation \cite{RouVer03a}. To solve Eq.~(\ref{1.21}), whose
left-hand side comes from the linearized Einstein tensor for the
perturbed metric \cite{MukFelBra92}, we need the retarded
propagator for the gravitational potential $\Phi_k$,
\begin{equation} G_k(\eta,\eta')= -i {4\pi\over k_i m_P^2}\left(
\theta(\eta-\eta') {a(\eta')\over a(\eta)}+f(\eta,\eta')\right),
\label{1.22} \end{equation} where $f$ is a
homogeneous solution of Eq.~(\ref{1.21}) related to the initial conditions
chosen and $m_P^2=1/G$. For instance, if we take
$f(\eta,\eta')=-\theta(\eta_0-\eta')a(\eta')/a(\eta)$ the solution would
correspond to ``turning on" the stochastic source at $\eta_0$.
With the solution of the Einstein-Langevin equation (\ref{1.21}) for the
scalar metric perturbations we are in the position to compute the
two-point correlation functions for these perturbations.


\subsection{Correlation functions for scalar metric perturbations}


The two-point
correlation function for the scalar metric perturbations induced by the
inflaton fluctuations is thus given by
\begin{eqnarray}
\!\!\! && \!\!\!\langle\Phi_k(\eta)\Phi_{k'}(\eta')\rangle_s=
(2\pi)^2\delta(\vec
k+\vec k')
\nonumber\\
&& \times\int^\eta \!d\eta_1\int^{\eta'}\!d\eta_2 G_k(\eta,\eta_1)
G_{k'}(\eta',\eta_2)
\langle(\xi_{0i})_k(\eta_1)(\xi_{0i})_{k'}(\eta_2)\rangle_s .
\label{1.23}
\end{eqnarray}
Here two-point correlation function for the stochastic
source, which is connected to
the stress-energy tensor fluctuations through the noise kernel is given by,
\begin{eqnarray}
\langle (\xi_{0i})_k(\eta_1)(\xi_{0i})_{-k}(\eta_2)\rangle_s
\!\!\! &= &\!\!\!{1\over2}
\langle\{(\hat t_{0i})_k(\eta_1),(\hat
t_{0i})_{-k}(\eta_2)\}\rangle_{\phi\varphi}
\nonumber\\
\!\!\! &=&\!\!\! {1\over2}
k_ik_i\phi'(\eta_1)\phi'(\eta_2)G_k^{(1)}(\eta_1,\eta_2),
\label{1.24}
\end{eqnarray}
where $G_k^{(1)}(\eta_1,\eta_2)=\langle\{\hat\varphi_k(\eta_1),
\hat\varphi_{-k}(\eta_2)\}\rangle$ is the $k$-mode Hadamard function for a
free minimally coupled scalar field in the
appropriate vacuum state on the Friedmann-Robertson-Walker background.

In practice, to make the explicit computation of the Hadamard
function we will assume that the field state is in the
Euclidean vacuum and the background spacetime is de Sitter.
Furthermore we will compute the Hadamard function for
a massless field, and will make
a perturbative expansion in terms of the dimensionless parameter
$m/m_P$. Thus we consider
$$\bar
G_k^{(1)}(\eta_1,\eta_2)=
\langle 0|\{\hat y_k(\eta_1),\hat y_{-k}(\eta_2)\}|0\rangle=
2{\cal R}\left(u_k(\eta_1)u_k^*(\eta_2)\right),$$
with $\hat y_k(\eta)= a(\eta)\hat\varphi_k(\eta)=
\hat a_k u_k(\eta)+\hat a_{-k}^\dagger u_{-k}^*(\eta)$ and where
$$u_k=(2k)^{-1/2}e^{ik\eta}(1-i/\eta),$$
are the positive
frequency $k$-mode for a massless minimally coupled scalar field
on a de Sitter background, which define the
Euclidean vacuum state: $\hat a_k|0\rangle=0$ \cite{BirDav82}.

The assumption of a massless field for the computation of the
Hadamard function is made because massless modes in de Sitter are
much simpler to deal with than massive modes. We can see that this
is, however, a reasonable approximation as follows. For a given
mode the $m=0$ approximation is reasonable when its wavelength
$\lambda$ is shorter that the Compton wavelength, $\lambda_c=1/m$.
In our case we have a very small mass $m$ and the horizon size
$H^{-1}$, where $H$ is the Hubble constant $H=\dot a/a$ (here
$a(t)$ with $t$ the physical time $dt=ad\eta$) satisfies that
$H^{-1}<\lambda_c$. Thus, for modes inside the horizon
$\lambda<\lambda_c$ and $m=0$ is a reasonable approximation.
Outside the horizon massive modes decay in amplitude as $\sim \exp
(-m^2 t/H)$ whereas massless modes remain constant, thus when
modes leave the horizon the approximation will eventually break
down. However, we only need to ensure that the approximation is
still valid after $60$ e-folds, {\it i.e.} $Ht\sim 60$, but this
is the case since $60\; m^2< H^2$ given that $m\sim 10^{-6}m_P$,
and $m\ll H$ as in most inflationary models \cite{KolTur90,Pad93}.

The background geometry is not exactly that of de Sitter
spacetime, for which $a(\eta)=-(H\eta)^{-1}$ with $-\infty <\eta<
0$. One can expand in terms of the ``slow-roll" parameters and
assume that to first order $\dot\phi(t)\simeq m_P^2(m/m_P)$, where
$t$ is the physical time. The correlation function for the metric
perturbation (\ref{1.23}) can then be easily computed; see
Ref.~\cite{RouVer00,RouVer03a} for details. The final result,
however, is very weakly dependent on the initial conditions as one
may understand from the fact that the accelerated expansion of de
quasi-de Sitter spacetime during inflation erases the information
about the initial conditions. Thus one may take the initial time
to be $\eta_0=-\infty$ and obtain to lowest order in $m/m_P$ the
expression
\begin{equation}
\langle\Phi_k(\eta)\Phi_{k'}(\eta')\rangle_s\simeq
8\pi^2\left( {m\over m_P}\right)^2 k^{-3}(2\pi)^3\delta(\vec k+\vec k')
\cos k(\eta-\eta').
\label{1.25}
\end{equation}

{}From this result two main conclusions are derived. First, the
prediction of an almost Harrison-Zel'dovich scale-invariant
spectrum for large scales, i.e. small values of $k$. Second, since
the correlation function is of order of $(m/m_P)^2$ a severe bound
to the mass $m$ is imposed by the gravitational fluctuations
derived from the small values of the Cosmic Microwave Background
(CMB) anisotropies detected by COBE. This bound is of the order of
$(m/m_P)\sim 10^{-6}$ \cite{Smo92,MukFelBra92}.

We should now comment on some differences with those works in
Ref.~\cite{CalHu95,Mat97a,Mat97b,CalGon97} which used a
self-interacting scalar field or a scalar field interacting
nonlinearly with other fields. In those works an important
relaxation of the ratio $m/m_p$ was found. The long wavelength
modes of the inflaton field  were regarded as an open system in an
environment made out of the shorter wavelength modes. Then,
Langevin type equations were used to compute the correlations of
the long wavelength modes driven by the fluctuations of the
shorter wavelength modes. In order to get a significant relaxation
on the above ratio, however, one had to assume that the
correlations of the free long wavelength modes, which correspond
to the dispersion of the system initial state, had to be very
small. Otherwise they dominate by several orders of magnitude
those fluctuations that come from the noise of the environment.
This would require a great amount of fine-tuning for the initial
quantum state of each mode \cite{RouVer03a}. We should remark that
in the model discussed here there is no environment for the
inflaton fluctuations. The inflaton fluctuations, however, are
responsible for the noise that induce the metric perturbations.


\subsection{Discussion}


One important advantage of the Einstein-Langevin approach to the
gravitational fluctuations in inflaton over the approach based on
the quantization of the linear perturbations of both the metric
and the inflaton field \cite{MukFelBra92}, is that an exact
treatment of the inflaton quantum fluctuations is possible. This
leads to corrections to the almost scale invariant spectrum for
scalar metric perturbations at large scales, and has implications
for the spectrum of the cosmic microwave background anisotropies.
However, in the standard inflationary models these corrections are
subdominant. Furthermore when the full non linear effect of the
quantum field is considered, tensorial metric perturbations are
also induced by the inflaton fluctuations. An estimation of this
effect, presumably subdominant over the free tensorial
fluctuations, has not been performed.

We should remark that although the gravitational fluctuations are
here assumed to be classical, the correlation functions obtained
correspond to the expectation values of the symmetrized quantum
metric perturbations \cite{CalRouVer03,RouVer03a}. This means that
even in the absence of decoherence the fluctuations predicted by
the Einstein-Langevin equation, whose solutions do not describe
the actual dynamics of the gravitational field any longer, still
give the correct symmetrized quantum two-point functions.

Another important advantage of the stochastic gravity approach is
that one may also compute the gravitational fluctuations in
inflationary models which are not driven by an inflaton field,
such as Starobinsky inflation which is driven by the trace anomaly
due to conformally coupled quantum fields. In fact, Einstein
semiclassical equation (\ref{2.5}) for a massless quantum field
which is conformally coupled to the gravitational field admits an
inflationary solution which is almost de Sitter initially and ends
up in a matter-dominated like regime \cite{Sta80,Vil85}. In these
models the standard approach based on the quantization of the
gravitational and the matter fields to linear order cannot be
used. This is because the calculation of the metric perturbations
correspond to having only the last term in the noise kernel in
Eq.~(\ref{1.20}), since there is no homogeneous field $\phi(\eta)$
as the expectation value $\langle\hat \phi\rangle=0$, and
linearization becomes trivial.

In the trace anomaly induced inflation framework Hawking et al.
\cite{HawHerRea01} were able to compute the two-point quantum
correlation function for scalar and tensorial metric perturbations
in a spatially closed de Sitter universe, making use of the
anti-de Sitter conformal field theory correspondence. They find
that short scale metric perturbations are strongly suppressed by
the conformal matter fields. This is similar to what we obtained
in section \ref{sec:flucminspa} for the induced metric
fluctuations in Minkowski spacetime. In the stochastic gravity
context, the noise kernel in a spatially closed de Sitter
background was derived in Ref.~\cite{RouVer99}. But in a spatially
flat arbitrary Friedmann-Robertson-Walker model the
Einstein-Langevin equation describing the metric perturbations was
first obtained in Ref.~\cite{CamVer96}, see also \cite{HuVer03a}.
The two-point correlation functions for the metric perturbations
can be derived from its solutions, but this is work still in
progress.

\section{Black Hole Backreaction}
\label{sec:bhbkrn}


As another illustration of the application of stochastic gravity
we consider fluctuations and backreaction in black hole
spacetimes. The celebrated Hawking effect of particle creation
from black holes is constructed from a quantum field theory in
curved spacetime framework. The oft-mentioned `black hole
evaporation' referring to the reduction of the mass of a black
hole due to particle creation must entail backreaction
considerations. Backreaction of Hawking radiation
\cite{HajIsr80,Bar81,Yor83,Yor85,Yor86,HocKep93,HocKepYor93,AndEtal94}
could alter the evolution of the background spacetime and change
the nature of its end state, more drastically so for Planck size
black holes. Because of the higher symmetry in cosmological
spacetimes, backreaction studies of processes therein have
progressed further than the corresponding black hole problems,
which to a large degree is still concerned with finding the right
approximations for the regularized energy momentum tensor
\cite{JenMcLOtt95,ParPir94,Mas95,AndHisSam93,AndHisSam95,AndHisLor95,HisLarAnd97}
 for even the
simplest spacetimes such as the spherically symmetric family
including the important Schwarzschild metric
(for a summary of the cosmological
backreaction problem treated in the stochastic gravity theory, see
\cite{HuVer03a}).
The latest important
work is that of Hiscock, Larson and Anderson \cite{HisLarAnd97} on
backreaction in the interior of a black hole, where one can find a
concise summary of earlier work. To name a few of the important
landmarks in this endeavor (this is adopted from
\cite{HisLarAnd97}), Howard and Candelas \cite{HowCan84,How84}
have computed the stress-energy of a conformally invariant scalar
field in the Schwarzschild geometry. Jensen and Ottewill
\cite{JenOtt89} have computed the vacuum stress-energy of a
massless vector field in Schwarzschild. Approximation methods have
been developed by Page, Brown, and Ottewill
\cite{Pag82,BroOtt85,BroOttPag86} for conformally invariant fields
in Schwarzschild spacetime, Frolov and Zel'nikov \cite{FroZel87}
for conformally invariant fields in a general static spacetime,
Anderson, Hiscock and Samuel \cite{AndHisSam93,AndHisSam95} for
massless arbitrarily coupled scalar fields in a general static
spherically symmetric spacetime. Furthermore the DeWitt-Schwinger
approximation has been derived by Frolov and
Zel'nikov \cite{FroZel82,FroZel84} for massive fields in Kerr
spacetime, Anderson Hiscock and Samuel
\cite{AndHisSam93,AndHisSam95} for a general (arbitrary curvature
coupling and mass) scalar field in a general static spherically
symmetric spacetime and have applied their method to the
Reissner-Nordstr\"{o}m geometry \cite{AndHisLor95}. Though arduous
and demanding, the effort continues on because of the importance
of backreaction effects of Hawking radiation in black holes. They
are expected to address some of the most basic issues such as
black hole thermodynamics
\cite{Isr75,Par75,Wal75,Bek73,Bek94,BekMuk95,Sor98,Jac99,Wal02,Wal01,%
SusUgl94,KabSheStr95,StrVaf96,Hor96,HorPol97,MalStrWit97,Mal98}
and the black hole end-state and information loss puzzles
\cite{Pag94}.

Here we wish to address the black hole backreaction problem with
new insights provided by stochastic semiclassical gravity. (For
the latest developments see reviews, e.g.,
\cite{Banff,stogra,HVErice,HuVer03a}). It is not our intention to
seek better approximations for the regularized energy momentum
tensor, but to point out new ingredients lacking in the existing
framework based on semiclassical gravity. In particular one needs
to consider both the dissipation and the fluctuations aspects in
the back reaction of particle creation or vacuum polarization.

In a short note \cite{Vishu} Raval, Sinha and one of us discussed
the formulation of the problem in this new light, commented on
some shortcomings of existing works, and sketched the strategy
\cite{SinRavHu03} behind
the stochastic gravity theory approach to
the black hole fluctuations and backreaction problem. Here we
follow their treatment with focus  on the class of quasi-static
black holes.

{}From the new perspective provided by statistical field theory
and stochastic gravity, it is not difficult to postulate that
backreaction effect is the manifestation of a
fluctuation-dissipation relation
\cite{Ein05,Ein06,Nyq28,CalWel51,CalGre52,Web56}. This was first
conjectured by Candelas and Sciama
\cite{CanSci77,Sci79,SciCanDeu81} for a dynamic Kerr black hole
emitting Hawking radiation, and Mottola \cite{Mottola} for a
static black hole (in a box) in quasi-equilibrium with its
radiation via linear response theory
\cite{Kub57,BerCal59,Kub66,LanLifPit80,KubTodHas85}. While the
fluctuation-dissipation relation in a linear response theory
captures the response of the system (e.g., dissipation of the
black hole) to the environment (in these cases the matter field)
linear response theory (in the way it is commonly presented in
statistical thermodynamics)  cannot provide a full description of
self-consistent backreaction on at least two counts: First,
because it is usually based on the assumption of a specified
background spacetime (static in this case) and state (thermal) of
the matter field(s)(e.g., \cite{Mottola}). The spacetime and the
state of matter should be determined in a self-consistent manner
by their dynamics and mutual influence. Second, the fluctuation
part represented by the noise kernel is amiss (e.g.,
\cite{AndMolMot02,AndMolMot03}) This is also a problem in the
fluctuation-dissipation relation proposed by Candelas and Sciama
\cite{CanSci77,Sci79,SciCanDeu81} (see below). As will be shown in
an explicit example later, the back reaction is intrinsically a
dynamic process. The Einstein-Langevin equation in stochastic
gravity overcomes both of these deficiencies.

For Candelas and Sciama \cite{CanSci77,Sci79,SciCanDeu81},
the classical formula they
showed relating the dissipation in area linearly to the squared
absolute value of the shear amplitude is suggestive of  a
fluctuation-dissipation relation. When the gravitational
perturbations are quantized (they choose the quantum state to be
the Unruh vacuum) they argue that it approximates a flux of
radiation from the hole at large radii. Thus the dissipation in
area due to the Hawking flux of gravitational radiation is
allegedly related to the quantum fluctuations of gravitons.
The criticism in Ref. \cite{Vishu} is that their's is not an
fluctuation-dissipation relation in the truly statistical
mechanical sense because it does not relate dissipation of a
certain quantity (in this case, horizon area) to the fluctuations
of {\it the same quantity}. To do so would require one to compute
the two point function of the area, which, being a four-point
function of the graviton field, is related to a two-point function
of the stress tensor. The stress tensor is the true ``generalized
force'' acting on the spacetime via the equations of motion, and
the dissipation in the metric must eventually be related to the
fluctuations of this generalized force for the relation to qualify
as an fluctuation-dissipation relation.

{}From this reasoning, we see that the stress energy bi-tensor and
its vacuum expectation value known as the noise kernel, are the
new ingredients in backreaction considerations. But these are
exactly the centerpiece in stochastic gravity. Therefore the
correct framework to address semiclassical backreaction problems
is stochastic gravity theory, where fluctuations and dissipation
are the equally essential components. The noise kernel for quantum
fields in Minkowski and de Sitter spacetime has been carried out
by Martin, Roura and Verdaguer
\cite{MarVer99a,MarVer00,RouVer03a}, for thermal fields in black
hole spacetime and scalar fields in general spacetimes by Campos,
Hu and Phillips \cite{CamHu98,CamHu99,PhiHu01,PhiHu03,PhiHu03b}.
Earlier, for cosmological backreaction problems Hu and Sinha
\cite{HuSin95} derived a generalized expression relating
dissipation (of anisotropy in Bianchi Type I universes) and
fluctuations (measured by particle numbers created in neighboring
histories). This example shows that one can understand the
backreaction of particle creation as a manifestation of a
(generalized) fluctuation-dissipation relation.

As an illustration of the application of stochastic gravity theory
we outline the steps in a black hole backreaction calculation,
focusing on the manageable quasi-static class. We adopt the
Hartle-Hawking picture \cite{HarHaw76} where the black hole is
bathed eternally -- actually in quasi-thermal equilibrium -- in
the Hawking radiance it emits. It is described here by a massless
scalar quantum field at the Hawking temperature. As is
well-known, this quasi-equilibrium condition is possible only if
the black hole is enclosed in a box of size suitably larger than
the event horizon.
We can divide our consideration into the far field case and the
near horizon case. Campos and Hu \cite{CamHu98,CamHu99} have
treated a relativistic thermal plasma in a weak gravitational
field.  Since the far field limit of a Schwarzschild metric is
just the perturbed Minkowski spacetime, one can perform a
perturbation expansion off hot flat space using the thermal Green
functions \cite{GibPer78}. Strictly speaking the location of the
box holding the black hole in equilibrium with its thermal
radiation is as far as one can go, thus the metric may not reach
the perturbed Minkowski form. But one can also put the black hole
and its radiation in an anti-de Sitter space \cite{HawPag83},
which contains such a region. Hot flat space has been studied
before for various purposes. See e.g.,
\cite{GroPerYaf82,Reb91,Reb92,AlmBraFre94,BraFre98}. Campos and Hu
derived a stochastic CTP effective action and from it an equation
of motion, the Einstein Langevin equation, for the dynamical
effect of a scalar quantum field on a background spacetime. To
perform calculations leading to the Einstein-Langevin equation one
needs to begin with a self-consistent solution of the
semiclassical Einstein equation for the thermal field and the
perturbed background spacetime. For a black hole background, a
semiclassical gravity solution is provided by York
\cite{Yor83,Yor85,Yor86}. For a Robertson-Walker background with
thermal fields it is given by Hu \cite{Hu81}. Recently Sinha,
Raval and Hu \cite{SinRavHu03} outlined a strategy for treating the near
horizon case, following the same scheme of Campos and Hu. In both
cases two new terms appear which are absent in semiclassical
gravity considerations: a nonlocal dissipation and a (generally
colored) noise kernel. When one takes the noise average one
recovers York's \cite{Yor83,Yor85,Yor86} semiclassical equations
for radially perturbed quasi-static black holes. For the near
horizon case one cannot obtain the full details yet, because the
Green function for a scalar field in the Schwarzschild metric
comes only in an approximate form (e.g. Page approximation
\cite{Pag82}), which, though reasonably accurate for the stress
tensor, fails poorly for the noise kernel \cite{PhiHu03,PhiHu03b}.
In addition a formula is derived in \cite{SinRavHu03} expressing the CTP
effective action in terms of the Bogolyubov coefficients. Since it
measures not only the number of particles created, but also the
difference of particle creation in alternative histories, this
provides a useful avenue to explore the wider set of issues in
black hole physics related to noise and fluctuations.

Since backreaction calculations in semiclassical gravity has been
under study for a much longer time than in stochastic gravity we
will concentrate on explaining how the new stochastic features
arise from the framework of semiclassical gravity, i.e., noise and
fluctuations and their consequences. Technically the goal is to
obtain an influence action for this model of a black hole coupled
to a scalar field and to derive an Einstein-Langevin equation
from it. As a by-product, from the fluctuation-dissipation
relation, one can derive the vacuum susceptibility function and
the isothermal compressibility function for black holes, two
quantities of fundamental interest in characterizing the
nonequilibrium thermodynamic properties of black holes.

\subsection{The Model}

In this model the black hole spacetime is described by a
spherically symmetric static metric with line element of the
following general form written in advanced time
Eddington-Finkelstein coordinates \be ds^2 =
g_{\mu\nu}dx^{\mu}dx^{\nu} = -e^{2\psi}\left(1 - {2m\over
r}\right)dv^2 + 2 e^{2\psi}dvdr + r^2~d{\Omega}^2,
\label{ssmetric} \te where $\psi = \psi(r)$ and $m = m(r)$ , $ v =
t + r + 2Mln\left({r\over 2M} -1 \right)$ and $d{\Omega}^2$ is the
line element on the two sphere. Hawking radiation is described by
a massless, conformally coupled quantum scalar field $\phi$ with
the classical action \be S_m[\phi, g_{\mu\nu}] = -{1\over 2}\int
d^n x \sqrt{-g}[g^{\mu\nu}\partial_{\mu}\phi
\partial_{\nu}\phi + \xi(n) R{\phi}^2], \label{phiact} \te where
$\xi(n) = {(n-2)\over 4(n-1)}$ ($n$ is the dimension of
spacetime) and $R$ is the curvature scalar of the spacetime it
lives in.

Let us consider linear perturbations $h_{\mu\nu}$ off a background
Schwarzschild metric $g^{(0)}_{\mu\nu}$ \be g_{\mu\nu} =
g^{(0)}_{\mu\nu} + h_{\mu\nu}, \label{linearize} \te with standard
line element \be (ds^2)^0 = \left( 1 - {2M\over r}\right)dv^2 +
2dvdr + r^2d{\Omega}^2. \label{schwarz} \te We look for this class
of perturbed metrics in the form given by (\ref{ssmetric}), (thus
restricting our consideration only to spherically symmetric
perturbations): \be e^\psi \simeq  1+ \epsilon \rho(r)
,\label{rho} \te and \be m \simeq M[ 1 + \epsilon \mu (r)],
\label{mu} \te where ${\epsilon\over \lambda M^2} = {1\over 3}a
T_H^4 ;$ $ a ={{\pi}^2\over 30} ; \lambda = 90(8^4)\pi^2$. $T_H$
is the Hawking temperature. This particular parametrization of the
perturbation is chosen following York's \cite{Yor83,Yor85,Yor86} notation. Thus
the only non-zero components of $h_{\mu\nu}$ are \be h_{vv} =
-\left((1 - {2M\over r})2\epsilon \rho(r) + {2M\epsilon \mu
(r)\over r}\right), \label{hvv} \te and \be h_{vr} = \epsilon\rho
(r) \label{hvr}. \te So this represents a metric with small static
and radial perturbations about a Schwarzschild black hole. The
initial quantum state of the scalar field is taken to be the
Hartle Hawking vacuum, which is essentially a thermal state at the
Hawking temperature and it represents a black hole in (unstable)
thermal equilibrium with its own Hawking radiation. In the far
field limit, the gravitational field is described by a linear
perturbation from Minkowski spacetime. In equilibrium  the thermal
bath can be characterized by a relativistic fluid with a
four-velocity (time-like normalized vector field) $u^\mu$, and
temperature in its own rest frame $\beta^{-1}$.

To facilitate later comparisons with our program we briefly recall
York's work \cite{Yor83,Yor85,Yor86}. See also work by Hochberg
and Kephart \cite{HocKep93} for a massless vector field, Hochberg,
Kephart and York \cite{HocKepYor93} for a massless spinor field,
and Anderson, Hiscock, Whitesell, and York \cite{AndEtal94} for a
quantized massless scalar field with arbitrary coupling to
spacetime curvature. York considered the semiclassical Einstein
equation \be G_{\m\n} (g_{\alpha \beta}) = \k \langle
T_{\m\n}\rangle ,\te with $G_{\mu\nu} \simeq G^{(0)}_{\mu\nu} +
\delta G_{\mu\nu}$ where $G^{(0)}_{\mu\nu}$ is the Einstein tensor
for the background spacetime. The zeroth order solution gives a
background metric in empty space, i.e, the Schwarzschild metric.
$\delta G_{\mu\nu}$ is the linear correction to the Einstein
tensor in the perturbed metric. The semiclassical Einstein
equation  in this approximation therefore reduces to \be \delta
G_{\mu\nu}(g^{(0)}, h) = \kappa \langle T_{\mu\nu} \rangle
.\label{pertscee} \te York solved this equation to first order by
using the expectation value of the energy momentum tensor for a
conformally coupled scalar field in the Hartle-Hawking vacuum in
the unperturbed (Schwarzschild) spacetime on the right hand side
and using (\ref{hvv}) and (\ref{hvr}) to calculate $\delta
G_{\mu\nu}$ on the left hand side. Unfortunately, no exact
analytical expression is available for the $\langle
T_{\mu\nu}\rangle$ in a Schwarzschild metric with the quantum
field in the Hartle-Hawking vacuum that goes on the right hand
side. York therefore uses the approximate expression given by Page
\cite{Pag82} which is known to give excellent agreement with
numerical results. Page's approximate expression for $\langle
T_{\mu\nu}\rangle$ was constructed using a thermal Feynman Green's
function obtained  by a conformal transformation of a WKB
approximated Green's function for an optical Schwarzschild metric.
York then solves the semiclassical Einstein equation
(\ref{pertscee}) self consistently to obtain the corrections to
the background metric induced by the backreaction encoded in the
functions $\mu(r)$ and $\rho(r)$. There was no mention of
fluctuations or its effects. As we shall see, in the language of
the previous section, the semiclassical gravity procedure which
York followed working at the equation of motion level is
equivalent to looking at the noise-averaged backreaction effects.

\subsection{CTP Effective Action for the Black Hole}

We first derive the CTP effective action for the model described
in the previous section. Using the  metric (\ref{schwarz}) (and
neglecting the surface terms that appear in an integration by
parts) we have  the action for the scalar field  written
perturbatively as
\begin{equation}
   S_m[\phi,h_{\mu\nu}]
        \ = \  {1\over 2}\int d^nx{\sqrt{-g^{(0)}}}\ \phi
               \left[ \Box^{(0)} + V^{(1)} + V^{(2)} + \cdots
              \right] \phi,
\label{phipert}
\end{equation}
where the first and second order perturbative operators $V^{(1)}$
and $V^{(2)}$ are given by
\begin{eqnarray}
V^{(1)}  \!\!\!\!\!\! & \equiv  &\!\!\!\!\!\! - {1\over
\sqrt{\!-g^{(0)}}} \left\{
\!\partial_\mu\left(\!\sqrt{\!-g^{(0)}}\bar h^{\mu\nu}\right)
                                \!\partial_\nu
                              \!+\!\bar h^{\mu\nu}\partial_\mu
                              \partial_\nu
                            \!+\!\xi(n) R^{(1)}
                     \right\}\!,
               \nonumber \\
V^{(2)}
  \!\!\!\!\!\!  &  \equiv & \!\!\!\!\!\!- \!{1\over \sqrt{\!-g^{(0)}}}
\left\{ \!\partial_\mu \!\left(\!\sqrt{\!-g^{(0)}} \hat
h^{\mu\nu}\right)
                              \!\partial_\nu
                            +\hat h^{\mu\nu}\partial_\mu
                            \partial_\nu
                          \!-\!\xi(n) ( R^{(2)}
                               \!+\!{1\over 2}hR^{(1)})\!\right\}\!\!.
\end{eqnarray}
In the above expressions, $R^{(k)}$ is the $k$-order term in the
perturbation $h_{\mu\nu}(x)$ of the scalar curvature $R$ and $\bar
h_{\mu\nu}$ and $\hat h_{\mu\nu}$ denote a linear and a quadratic
combination of the perturbation, respectively,
\begin{eqnarray}
   \bar h_{\mu\nu}
        &  \equiv  & h_{\mu\nu} - {1\over 2} h g^{(0)}_{\mu\nu},
                     \nonumber \\
   \hat h_{\mu\nu}
        &  \equiv  & h^{\,\, \alpha}_\mu h_{\alpha\nu}
                      -{1\over 2} h h_{\mu\nu}
                      +{1\over 8} h^2 g^{(0)}_{\mu\nu}
                      -{1\over 4} h_{\alpha\beta}h^{\alpha\beta} g^{(0)}_{\mu\nu}.
   \label{eq:def bar h}
\end{eqnarray}
{}From quantum field theory in curved spacetime considerations
discussed above we take the following action for the gravitational
field:
\begin{eqnarray}
  \!\!\!\! S_g[g_{\mu\nu}]
       \!\!\!\!\! &=& \!\!\!\!\!{1\over {(16 \pi G)^{\frac{n-2}{2}}}}\int d^nx\ \sqrt{-g(x)}R(x)
                +{\alpha\bar\mu^{n-4}\over4(n-4)}
                   \int d^nx\ \sqrt{-g(x)} \nonumber \\
      \!\!\!\!\!  &&\!\!\!\!\!
                   \times\left\{ 3R_{\mu\nu\alpha\beta}(x)
                           R^{\mu\nu\alpha\beta}(x)
                         \!-\!\left[ 1\!-\!360 \left(\xi(n)\! - \!{1\over6}\right)^2
                          \right]R^2(x)
                   \right\}\!.
\end{eqnarray}
The first term is the classical Einstein-Hilbert action and the
second term is the counterterm in four dimensions used  to
renormalize the divergent effective action. In this action
$\ell^2_P = 16\pi G_N$, $\alpha = (2880\pi^2)^{-1}$ and $\bar\mu$
is an arbitrary mass scale.

We are interested in computing the CTP effective action
(\ref{phipert}) for the matter action and when the field $\phi$ is
initially in the Hartle-Hawking vacuum. This is equivalent to
saying that the initial state of the field is described by a
thermal density matrix at a finite temperature $T = T_H$. The CTP
effective action at finite temperature $T \equiv 1/\beta$ for
this model is given by (for details see \cite{CamHu98,CamHu99})
\begin{equation}
   S_{rm eff}^\beta [h^\pm_{\mu\nu}]
        \ = \ S_g[h^+_{\mu\nu}]
             -S_g[h^-_{\mu\nu}]
             -{i\over2}Tr\{ \ln\bar G^\beta_{ab}[h^\pm_{\mu\nu}]\},
   \label{eq:eff act two fields}
\end{equation}
where $\pm$ denote the forward and backward time path of the CTP
formalism and $\bar G^\beta_{ab}[h^\pm_{\mu\nu}]$ is the complete
$2\times 2$ matrix propagator ($a$ and $b$ take $\pm$ values:
$G_{++},G_{+-}$ and $G_{--}$ correspond to the Feynman, Wightman
and Schwinger Greens functions respectively) with thermal boundary
conditions for the differential operator $\sqrt{-g^{(0)}}(\Box +
V^{(1)} + V^{(2)} + \cdots)$. 
The actual form of $\bar G^\beta_{ab}$ cannot be explicitly
given. However, it is easy to obtain a perturbative expansion in
terms of $V^{(k)}_{ab}$, the $k$-order matrix version of the
complete differential operator defined by $V^{(k)}_{\pm\pm}
\equiv \pm V^{(k)}_{\pm}$ and $V^{(k)}_{\pm\mp} \equiv 0$, and
$G^\beta_{ab}$, the thermal matrix propagator for a massless
scalar field in Schwarzschild spacetime . To second order $\bar
G^\beta_{ab}$ reads,
\begin{eqnarray}
   \bar G^\beta_{ab}
        \ = \  G^\beta_{ab}
              -G^\beta_{ac}V^{(1)}_{cd}G^\beta_{db}
              -G^\beta_{ac}V^{(2)}_{cd}G^\beta_{db}
              +G^\beta_{ac}V^{(1)}_{cd}G^\beta_{de}
               V^{(1)}_{ef}G^\beta_{fb}
              +\cdots
\end{eqnarray}
Expanding the logarithm and dropping one term independent of the
perturbation $h^\pm_{\mu\nu}(x)$, the CTP effective action may be
perturbatively written as,
\begin{eqnarray}
  \!\!\!\! S_{\rm eff}^\beta [h^\pm_{\mu\nu}]
      \!\! \!\!\!\!\! & = & \!\!\!\!\!\!\! S_g[h^+_{\mu\nu}] \!- \!S_g[h^-_{\mu\nu}]
                \nonumber \\
       \!\! \!\!\!\!\!& &\!\!\!\!\!\!\! +\!{i\over2}Tr[ V^{(1)}_{+}G^\beta_{++}
                               \!-\!V^{(1)}_{-}G^\beta_{--}
                               \!+\!V^{(2)}_{+}G^\beta_{++}
                               \!-\!V^{(2)}_{-}G^\beta_{--}
                              ]
                \nonumber \\
       \!\! \!\!\!\!\!& & \!\!\!\!\!\!\!-{i\over4}Tr[  V^{(1)}_{+}G^\beta_{++}
                                 V^{(1)}_{+}G^\beta_{++}
                               \!+\! V^{(1)}_{-}G^\beta_{--}
                                 V^{(1)}_{-}G^\beta_{--}
                                 \nonumber\\
      \!\!\!\!\!\!\!&&\ \ \ \ \ \ \ \
                               \!-\!2V^{(1)}_{+}G^\beta_{+-}
                                 V^{(1)}_{-}G^\beta_{-+}
                              ]\!.
   \label{eq:effective action}
\end{eqnarray}
In computing the traces, some terms containing divergences are
canceled using counterterms introduced in the classical
gravitational action after dimensional regularization.

\subsection{Near Flat Case}

At this point we divide our considerations into two cases. In the
far field limit  $h_{\mu\nu}$ represent perturbations about flat
space, i.e., $g^{(0)}_{\mu\nu}= \eta_{\m\n}$. The exact
``unperturbed" thermal propagators for scalar fields are known,
i.e., the Euclidean propagator with periodicity $\beta$. Using
the Fourier transformed  form (those quantities are denoted with a
tilde) of the thermal propagators $\tilde G^\beta_{ab}(k)$ , the
trace terms of the form
$Tr[V^{(1)}_{a}G^\beta_{mn}V^{(1)}_{b}G^\beta_{rs}]$ can be
written as \cite{CamHu98,CamHu99},
\begin{eqnarray}
   \!\!\!\!\!\!&&\!\!\!\!\!Tr[V^{(1)}_{a}G^\beta_{mn}V^{(1)}_{b}G^\beta_{rs}]
       = \int d^nxd^nx'\
               h^a_{\mu\nu}(x)h^b_{\alpha\beta}(x')
               \nonumber\\
               &&\ \ \ \ \
               \times\int {d^nk\over(2\pi)^n}{d^nq\over(2\pi)^n}
               e^{ik\cdot (x-x')}
               \tilde G^\beta_{mn}(k+q)\tilde G^\beta_{rs}(q)
               \kl{T}{}{q,k},
   \label{eq:trace}
\end{eqnarray}
where the tensor $\kl{T}{}{q,k}$ is defined in
\cite{CamHu98,CamHu99} after an  expansion in terms of a basis of
14 tensors \cite{Reb91,Reb92}. In particular, the last trace of
(\ref{eq:effective action}) may be split in two different kernels
$\kl{N}{}{x-x'}$ and $\kl{D}{}{x-x'}$,
\begin{eqnarray}
   \!\!\!\!\!&&\!\!\!\!\!{i\over2}
   Tr[V^{(1)}_{+}G^\beta_{+-}V^{(1)}_{-}G^\beta_{-+}]=
   \nonumber\\
   &&\ \ \
   -\int d^4xd^4x'\
               h^+_{\mu\nu}(x)h^-_{\alpha\beta}(x')
               [   \kl{D}{}{x-x'}
                +i \kl{N}{}{x-x'}
               ].
\end{eqnarray}
One can express the Fourier transforms of these kernels,
respectively, as
\begin{eqnarray}
   \kl{\tilde N}{}{k}
       \!\!\!\!\! & = &\!\!\!\!\! \pi^2\int {d^4q\over(2\pi)^4}\
                  \left\{ \theta(k^o+q^o)\theta(-q^o)
                         +\theta(-k^o-q^o)\theta(q^o)
                         \right.
                         \nonumber\\
                         &&\!\!\hspace{2mm}
                         \left.
                         +n_\beta(|q^o|)+n_\beta(|k^o+q^o|)
                  \right.
                \nonumber \\
        & & \!\!\hspace{2mm}
                  \left. +2n_\beta(|q^o|)n_\beta(|k^o+q^o|)
                  \right\}\delta(q^2)\delta[(k+q)^2]\kl{T}{}{q,k},
   \label{eq:N}
\end{eqnarray}
\begin{eqnarray}
   \kl{\tilde D}{}{k}
     \!\!\!\!\!\!   & = &\!\!\!\!\!\! -i\pi^2\int {d^4q\over(2\pi)^4}\
                  \left\{ \theta(k^o+q^o)\theta(-q^o)
                         -\theta(-k^o-q^o)\theta(q^o)
                         \right.
                         \nonumber\\
                         &&\!\!\hspace{2mm}
                         \left.
                         +sg(k^o+q^o) n_\beta(|q^o|)
                  \right.
                \nonumber \\
        &  & \!\!\hspace{2mm}
                  \left. -sg(q^o)n_\beta(|k^o+q^o|)
                  \right\}\delta(q^2)\delta[(k+q)^2]\kl{T}{}{q,k}.
   \label{eq:D}
\end{eqnarray}
Using the property $\kl{T}{}{q,k} = \kl{T}{}{-q,-k}$, it is easy
to see that the kernel $\kl{N}{}{x-x'}$ is symmetric
and $\kl{D}{}{x-x'}$
antisymmetric in their arguments; that is, $\kl{N}{}{x} =
\kl{N}{}{-x}$ and $\kl{D}{}{x} = -\kl{D}{}{-x}$.

The physical meanings of these kernels can be extracted if we
write the renormalized CTP effective action at finite temperature
(\ref{eq:effective action}) in an influence functional form
\cite{CalLeg83,GraSchIng88,HuPazZha92,HuPazZha93}. N, the
imaginary part of the CTP effective action can be identified with
the noise kernel and D, the antisymmetric piece of the real part,
with the dissipation kernel. Campos and Hu \cite{CamHu98,CamHu99}
have shown that these kernels identified as such indeed satisfy a
thermal fluctuation-dissipation relation.

If we denote the difference and the sum of the perturbations
$h^\pm_{\mu\nu}$ defined along each branch $C_\pm$ of the complex
time path of integration $C$ by $[h_{\mu\nu}] \equiv h^+_{\mu\nu}
- h^-_{\mu\nu}$ and $\{h_{\mu\nu}\} \equiv h^+_{\mu\nu} +
h^-_{\mu\nu}$, respectively, the influence functional form of the
thermal CTP effective action may be written to second order in
$h_{\mu\nu}$ as,
\begin{eqnarray}
   S_{\rm eff}^\beta [h^\pm_{\mu\nu}]
        & \ \simeq \ & {1\over 2(16 \pi G_N)} \int d^4x\ d^4x'\
                       [h_{\mu\nu}](x)\kl{L}{(o)}{x-x'}
                       \{h_{\alpha\beta}\}(x')
                     \nonumber \\
        &            &+{1\over2}\int d^4x\
                       [h_{\mu\nu}](x)T^{\mu\nu}_{(\beta)}
                     \nonumber \\
        &            &+{1\over2}\int d^4x\ d^4x'\
                       [h_{\mu\nu}](x)\kl{H}{}{x-x'}
                       \{h_{\alpha\beta}\}(x')
                     \nonumber \\
        &            &-{1\over2}\int d^4x\ d^4x'\
                       [h_{\mu\nu}](x)\kl{D}{}{x-x'}
                       \{h_{\alpha\beta}\}(x')
                     \nonumber \\
        &            &+{i\over2}\int d^4x\ d^4x'\
                       [h_{\mu\nu}](x)\kl{N}{}{x-x'}
                       [h_{\alpha\beta}](x').
\label{CTPbh}
\end{eqnarray}
The first line is the Einstein-Hilbert action to second order in
the perturbation $h^\pm_{\mu\nu}(x)$. $\kl{L}{(o)}{x}$ is a
symmetric kernel ({\sl i.e.} $\kl{L}{(o)}{x}$ =
$\kl{L}{(o)}{-x}$). In the near flat case its Fourier transform
is given by
\begin{eqnarray}
   \kl{\tilde L}{(o)}{k}
       &= &{1\over4}\left[ - k^2 \kl{T}{1}{q,k}
                              +2k^2 \kl{T}{4}{q,k}
                              \right.
                              \nonumber\\
                              && \ \ \ \
                              \left.
                              + \kl{T}{8}{q,k}
                              -2\kl{T}{13}{q,k}
                       \right].
\end{eqnarray}
The fourteen elements of the tensor basis $\kl{T}{i}{q,k}$
($i=1,\cdots,14$) are defined in \cite{Reb91,Reb92}. The second is
a local term linear in $h^\pm_{\mu\nu}(x)$.  Only when far away
from the hole that it takes the form of the stress tensor of
massless scalar particles at temperature $\beta^{-1}$, which has
the form of a perfect fluid stress-energy tensor
\begin{equation}
   T^{\mu\nu}_{(\beta)}
        \ = \ {\pi^2\over30\beta^4}
              \left[ u^\mu u^\nu + {1\over3}(\eta^{\mu\nu}+u^\mu u^\nu)
              \right],
\end{equation}
where $u^\mu$ is the four-velocity of the plasma
and the factor ${\pi^2\over30\beta^4}$ is the
familiar thermal energy density for massless scalar particles at
temperature $\beta^{-1}$. In the
far field limit, taking into account the four-velocity $u^\mu$ of
the fluid, a manifestly Lorentz-covariant approach to thermal
field theory may be used \cite{Wel82}. However, in order to
simplify the involved tensorial structure we work in the
co-moving coordinate system of the fluid where $u^\mu =
(1,0,0,0)$. In the third line, the Fourier
transform of the symmetric kernel $\kl{H}{}{x}$ can be expressed
as
\begin{eqnarray}
   \kl{\tilde H}{}{k}
     \!\!\!\!   & =  & \!\!\!\!\! -{\alpha k^4\over4}
                   \left\{ {1\over2}\ln {|k^2|\over\mu^2}\kl{Q}{}{k}
                          +{1\over3}\kl{\bar Q}{}{k}
                   \right\}
                \nonumber \\
        &       & \!\!\!\! +{\pi^2\over180\beta^4}
                   \left\{ - \kl{T}{1}{u,k}
                           -2\kl{T}{2}{u,k}
                           \right.
                           \nonumber\\
                           &&\ \ \ \ \ \ \ \ \ \
                           \left.
                           + \kl{T}{4}{u,k}
                           +2\kl{T}{5}{u,k}
                   \right\}
                \nonumber \\
        &       & \!\!\!\! +{\xi\over96\beta^2}
                   \left\{    k^2 \kl{T}{1}{u,k}
                           -2 k^2 \kl{T}{4}{u,k}
                           \right.
                           \nonumber\\
                           &&\ \ \ \ \ \ \ \ \ \
                           \left.
                           -      \kl{T}{8}{u,k}
                           +2     \kl{T}{13}{u,k}
                   \right\}
                \nonumber \\
        &       & \!\!\!\! +\pi\int {d^4q\over(2\pi)^4}\
                   \left\{ \delta(q^2)n_\beta(|q^o|)
                           {\cal P}\left[ {1\over(k+q)^2}
                                   \right]
                                   \right.
                           \nonumber\\
                           &&\ \
                           \left.
                          +\delta[(k\!+\!q)^2]n_\beta(|k^o\!+\!q^o|)
                           {\cal P}\!\left[ {1\over q^2}
                                   \right]\!\right\}
        \!\times\!\kl{T}{}{q,k}\!,
   \label{eq:grav pol tensor}
\end{eqnarray}
where $\mu$ is a simple redefinition of the renormalization
parameter $\bar\mu$ given by $\mu \equiv \bar\mu \exp
({23\over15} + {1\over2}\ln 4\pi - {1\over2}\gamma)$, and the
tensors $\kl{Q}{}{k}$ and $\kl{\bar Q}{}{k}$ are defined,
respectively, by
\begin{eqnarray}
   \kl{Q}{}{k}
       \!\!\!\!\! &=& \!\!\!\!\!{3\over2} \left\{               \kl{T}{1}{q,k}
                                    -{1\over k^2} \kl{T}{8}{q,k}
                                    +{2\over k^4} \kl{T}{12}{q,k}
                            \right\}
                \nonumber \\
       & &\hspace{-20mm}-[1\!-\!360(\xi\!-\!{1\over6})^2]\!
                  \left\{ \!              \kl{T}{4}{q,k}
                          \!+\!{1\over k^4} \kl{T}{12}{q,k}
                          \!-\!{1\over k^2} \kl{T}{13}{q,k}
                  \!\right\}\!,
   \label{eq:Q tensor}
\end{eqnarray}
\begin{eqnarray}
   \kl{\bar Q}{}{k}
    \!\!\!\!\!\!    &=&\!\!\!\!\!  [1+576(\xi-{1\over6})^2-60(\xi-{1\over6})(1-36\xi')]
                  \left\{               \kl{T}{4}{q,k}
                  \right.
                  \nonumber\\
                  &&\ \ \ \ \ \ \ \ \ \ \
                  \left.
                          +{1\over k^4} \kl{T}{12}{q,k}
                          -{1\over k^2} \kl{T}{13}{q,k}
                  \right\}.
\end{eqnarray}
In the above and subsequent equations, we denote the coupling
parameter in four dimensions $\xi(4)$ by $\xi$ and consequently
$\xi'$ means $d\xi(n)/dn$ evaluated at $n=4$. $\kl{\tilde H}{}{k}$
is the complete contribution of a free massless quantum scalar
field to the thermal graviton polarization
tensor\cite{Reb91,Reb92,AlmBraFre94,BraFre98} and it is
responsible for the instabilities found in flat spacetime at
finite temperature
\cite{GroPerYaf82,Reb91,Reb92,AlmBraFre94,BraFre98}. Note that the
addition of the contribution of other kinds of matter fields to
the effective action, even graviton contributions, does not change
the tensor structure of these kernels and only the overall factors
are different to leading order \cite{Reb91,Reb92}.
Eq.~(\ref{eq:grav pol tensor}) reflects the fact that the kernel
$\kl{\tilde H}{}{k}$ has thermal as well as non-thermal
contributions. Note that it reduces to the first term in the zero
temperature limit ($\beta\rightarrow\infty$)
\begin{equation}
   \kl{\tilde H}{}{k}
        \ \simeq \ -{\alpha k^4\over4}
                     \left\{ {1\over2}\ln {|k^2|\over\mu^2}\kl{Q}{}{k}
                            +{1\over3}\kl{\bar Q}{}{k}
                     \right\}.
\end{equation}
and at high temperatures the leading term ($\beta^{-4}$) may be
written as
\begin{equation}
   \kl{\tilde H}{}{k}
        \ \simeq \ {\pi^2\over30\beta^4}
                    \sum^{14}_{i=1}
                    \mbox{\rm H}_i(r) \kl{T}{i}{u,K},
\end{equation}
where we have introduced the dimensionless external momentum
$K^\mu \equiv k^\mu/|\vec{k}| \equiv (r,\hat k)$. The $\mbox{\rm
H}_i(r)$ coefficients were first given in \cite{Reb91,Reb92} and
generalized to the next-to-leading order ($\beta^{-2}$) in
\cite{AlmBraFre94,BraFre98}. (They are given with the MTW sign
convention  in \cite{CamHu98,CamHu99}.)

Finally, as defined above, $\kl{N}{}{x}$ is the noise kernel
representing the random fluctuations of the thermal radiance and
$\kl{D}{}{x}$ is the dissipation kernel, describing the
dissipation of energy of the gravitational field.


\subsection{Near Horizon Case}

In this case, since the perturbation is taken around the
Schwarzschild spacetime, exact expressions for the corresponding
unperturbed propagators $G^\beta_{ab}[h^\pm_{\mu\nu}]$ are not
known. Therefore apart from the approximation of computing the CTP
effective action to certain order in perturbation theory, an
appropriate approximation scheme for the unperturbed Green's
functions is also required. This feature manifested itself in
York's calculation of backreaction as well, where, in writing the
$\langle T_{\mu \nu}\rangle$ on the right hand side of the
semiclassical Einstein equation in the unperturbed Schwarzschild
metric, he had to use an approximate expression for $\langle
T_{\mu\nu} \rangle$ in the Schwarzschild metric given by Page
\cite{Pag82}. The additional complication here is that while to
obtain $\langle T_{\mu\nu}\rangle$ as in York's calculation, the
knowledge of only the thermal Feynman Green's function is
required, to calculate the CTP effective action one needs the
knowledge of the full matrix propagator, which involves the
Feynman, Schwinger and Wightman functions.

It is indeed possible to construct the full thermal matrix
propagator $G^\beta_{ab}[h^\pm_{\mu\nu}]$ based on Page's
approximate Feynman Green's function by using identities relating
the Feynman Green's function with the other Green's functions with
different boundary conditions. One can then proceed to explicitly
compute a CTP effective action and hence the influence functional
based on this approximation. However, we desist from delving into
such a calculation for the following reason. Our main interest in
performing such a calculation is to identify and analyze the noise
term which is the new ingredient in the backreaction. We have
mentioned that the noise term gives a stochastic contribution
$\xi^{\mu\nu}$ to the Einstein-Langevin equation (\ref{2.11}). We
had also stated that this term is related to the variance of
fluctuations in $T_{\mu\nu}$, i.e, schematically, to $\langle
T^2_{\mu\nu}\rangle$. However, a calculation of $\langle
T^2_{\mu\nu}\rangle$ in the Hartle-Hawking state in a
Schwarzschild background using the Page approximation was
performed by Phillips and Hu \cite{PhiHu01,PhiHu03,PhiHu03b} and
it was shown that though the approximation is excellent as far as
$\langle T_{\mu\nu}\rangle$ is concerned, it gives unacceptably
large errors for $\langle T^2_{\mu\nu}\rangle$ at the horizon. In
fact, similar errors will be propagated in the non-local
dissipation term as well, because both terms originate from the
same source, that is, they come from the last trace term in
(\ref{eq:effective action}) which contains terms quadratic in the
Green's function. However, the Influence Functional or CTP
formalism itself does not depend on the nature of the
approximation, so we will attempt to exhibit the general structure
of the calculation without resorting to a specific form for the
Greens function and conjecture on what is to be expected. A more
accurate computation can be performed using this formal structure
once a better approximation becomes available.

The general structure of the CTP effective action arising from
the calculation of the traces in equation (\ref{eq:effective
action}) remains the same. But to write down explicit expressions
for the non-local kernels one requires the input of the explicit
form of $G^\beta_{ab}[h^\pm_{\mu\nu}]$  in the Schwarzschild
metric, which is not available in closed form. We can make some
general observations about the terms in there. The first line
containing L does not have an explicit Fourier representation as
given in the far field case, neither will $T_{(\beta)}^{\mu\nu}$
in the second line representing the zeroth order contribution to
$\langle T_{\mu\nu} \rangle$ have a perfect fluid form. The third
and fourth terms containing the remaining quadratic component of
the real part of the effective action will not have any simple or
even complicated analytic form. The symmetry properties of the
kernels $H^{\mu\nu,\alpha\beta}(x,x')$ and
$D^{\mu\nu,\alpha\beta}(x,x')$ remain intact, i.e., they are
respectively even and odd in $x,x'$. The last term in the CTP
effective action gives the imaginary part of the effective action
and the kernel $N(x,x')$ is symmetric.

Continuing our general observations from this CTP effective
action, using the connection between this thermal CTP effective
action to the influence functional \cite{SuEtal88,CalHu94} via an
equation in the schematic form  (\ref{ctpif}). We see that the
nonlocal imaginary term containing the kernel
$N^{\mu\nu,\alpha\beta}(x,x')$ is responsible for the generation
of the stochastic noise term in the Einstein-Langevin equation and
the real non-local term containing kernel
$D^{\mu\nu,\alpha\beta}(x,x')$ is responsible for the non-local
dissipation term. To derive the Einstein-Langevin equation we
first construct the stochastic effective action (\ref{stochastic
eff action}). We then derive the equation of motion, as shown
earlier in (\ref{eq of motion}), by taking its functional
derivative with respect to $[h_{\mu\nu}]$ and equating it to zero.
With the identification of noise and dissipation kernels, one can
write down a linear, non-local relation of the form, \be N(t-t') =
~\int~d(s -s')K(t-t',s-s')\gamma(s -s') \label{FDR}, \te where
$D(t,t')=-\partial_{t'}\gamma (t,t')$.  This is the general
functional form of a fluctuation-dissipation relation and $K(t,s)$
is called the fluctuation-dissipation kernel
\cite{CalLeg83,GraSchIng88,HuPazZha92,HuPazZha93}. In the present
context this relation depicts the backreaction of thermal Hawking
radiance for a black hole in quasi-equilibrium.


\subsection{Einstein-Langevin equation}

In this section we show how  a semiclassical Einstein-Langevin
equation  can be derived from the previous thermal CTP effective
action. This equation depicts the stochastic evolution of the
perturbations of the black hole under the influence of the
fluctuations of the thermal scalar field.

The influence functional ${\cal F}_{\rm IF} \equiv \exp (iS_{\rm
IF})$ previously introduced in Eq. (\ref{influence functional})
can be written in terms of the the CTP effective action $S_{\rm
eff} ^\beta [h^\pm_{\mu\nu}]$ derived in equation (\ref{CTPbh})
using Eq.~(\ref{ctpif}).  The Einstein-Langevin equation follows
from taking the functional derivative of the stochastic effective
action (\ref{stochastic eff action}) with respect to
$[h_{\mu\nu}](x)$ and imposing $[h_{\mu\nu}](x) = 0$. This leads
to
\begin{eqnarray}
 \! \!\!\!\!&&\!\!\!\!\! {1\over 16\pi G_N}
   \int d^4x'\ \kl{L}{(o)}{x-x'} h_{\alpha\beta}(x')
  +{1\over2}\ T^{\mu\nu}_{(\beta)}
  +\int d^4x'\ \left( \kl{H}{}{x-x'}
   \right.
   \nonumber\\
   &&\ \ \ \ \ \ \ \ \ \ \ \ \ \ \ \
  \left.
                     -\kl{D}{}{x-x'}
               \right) h_{\alpha\beta}(x')
  +\xi^{\mu\nu}(x)
         = 0.
\end{eqnarray}
where
\begin{equation}
   \langle \xi^{\mu\nu}(x) \xi^{\alpha\beta}(x') \rangle_j
        \ = \ \kl{N}{}{x-x'},
   \label{eq:correlation}
\end{equation}
In the far field limit this equation should reduce to that
obtained by Campos and Hu \cite{CamHu98,CamHu99}: For gravitational
perturbations $h^{\mu\nu}$ defined in (\ref{eq:def bar h}) under
the harmonic gauge $\bar h^{\mu\nu}_{\,\,\,\,\, ,\nu} = 0$, their
Einstein-Langevin equation is given by
\begin{eqnarray}
  \!\!\!\!\!&&\!\!\!\!\! \Box\bar h^{\mu\nu}(x)
         + {1 \over 16 \pi G_N^2}
               \left\{ T^{\mu\nu}_{(\beta)}
                      +2P_{\rho\sigma,\alpha\beta}
                       \int d^4x'\ \left( \kl{H}{}{x-x'}
                       \right.
       \right.
   \nonumber\\
   &&\ \ \ \ \ \ \ \ \ \ \ \ \ \ \ \
  \left.
                       \left.
                   -\kl{D}{}{x-x'}
                                   \right)\bar h^{\rho\sigma}(x')
                      +2\xi^{\mu\nu}(x)
               \right\} = 0,
\end{eqnarray}
where the tensor $P_{\rho\sigma,\alpha\beta}$ is given by
\begin{equation}
   P_{\rho\sigma,\alpha\beta}
        \ = \ {1\over2}\left( \eta_{\rho\alpha}\eta_{\sigma\beta}
                             +\eta_{\rho\beta}\eta_{\sigma\alpha}
                             -\eta_{\rho\sigma}\eta_{\alpha\beta}
                       \right).
\end{equation}
The expression for  $P_{\rho\sigma,\alpha\beta}$ in the near
horizon limit of course cannot be expressed in such a simple form.
Note that this differential stochastic equation includes a
non-local term responsible for the dissipation of the
gravitational field and a noise source term which accounts for
the fluctuations of the quantum field . Note also that this
equation in combination with the correlation for the stochastic
variable (\ref{eq:correlation}) determine the two-point
correlation for the stochastic metric fluctuations $\langle \bar
h_{\mu\nu}(x) \bar h_{\alpha\beta}(x') \rangle_\xi$
self-consistently.

As we have seen before and here, the Einstein-Langevin equation is
a dynamical equation governing the dissipative evolution of the
gravitational field under the influence of the fluctuations of
the quantum field, which, in the case of black holes, takes the
form of thermal radiance. From its form we can see that even for
the quasi-static case under study the back reaction of Hawking
radiation on the black hole spacetime has an innate dynamical
nature.

For the far field case making use of the explicit forms available
for the noise and dissipation kernels Campos and Hu \cite{CamHu98,CamHu99}
formally proved the existence of a fluctuation-dissipation
relation at all temperatures between the quantum fluctuations of
the thermal radiance and the dissipation of the gravitational
field. They also showed the formal equivalence of this method with
linear response theory for lowest order perturbations of a
near-equilibrium system, and how the response functions such as
the contribution of the quantum scalar field to the thermal
graviton polarization tensor can be derived. An important quantity
not usually obtained in linear response theory, but of equal
importance, manifest in the CTP stochastic approach is the noise
term arising from the quantum and statistical fluctuations in the
thermal field. The example given in this section shows that the
back reaction is intrinsically a dynamic process described (at
this level of sophistication) by the Einstein-Langevin equation.
By comparison, traditional linear response theory calculations
cannot capture the dynamics as fully and thus cannot provide a
complete description of the backreaction problem.


\subsection{Discussions}
We make a few remarks here and draw some connection with related
work on black hole fluctuations.

\subsubsection{Black hole Backreaction}

As remarked earlier, except for the near-flat case, an analytic
form of the Green function is not available. Even the Page
approximation \cite{Pag82} which gives unexpectedly good results
for the stress energy tensor has been shown to fail in the
fluctuations of the energy density \cite{PhiHu03,PhiHu03b}. Thus
using such an approximation for the noise kernel will give
unreliable results for the Einstein-Langevin equation.  If we
confine ourselves to Page's approximation and derive the equation
of motion without the stochastic term, we expect to recover York's
semiclassical Einstein's equation if one retains only the zeroth
order contribution, i.e, the first two terms in the expression for
the CTP effective action in Eq. (\ref{CTPbh}). Thus, this offers a
new route to arrive at York's semiclassical Einstein's equations.
Not only is it a derivation of York's result from a different
point of view, but it also shows how his result arises as an
appropriate limit of a more complete framework, i.e, it arises
when one averages over the noise. Another point worth noting is
that our treatment  will also yield a non-local dissipation term
arising from the fourth term in equation (\ref{CTPbh}) in the CTP
effective action which is absent in York's treatment. This
difference arises primarily due to the difference in the way
backreaction is treated,  at the level of iterative approximations
on the equation of motion as in York, versus the treatment at the
effective action level as pursued here. In York's treatment, the
Einstein tensor is computed to first order in perturbation theory
, while $\langle T_{\mu\nu}\rangle$ on the right hand side of the
semiclassical Einstein equation is replaced by the zeroth order
term. In the effective action treatment the full effective action
is computed to second order in perturbation, and hence includes
the higher order non-local terms.

The other important conceptual point that comes to light from this
approach is that related to the Fluctuation-Dissipation Relation.
In the quantum Brownian motion analog (e.g.,
\cite{CalLeg83,GraSchIng88,HuPazZha92,HuPazZha93} and references
therein), the dissipation of the energy of the Brownian particle
as it approaches equilibrium and the fluctuations at equilibrium
are connected by the Fluctuation-Dissipation relation. Here the
backreaction of quantum fields on black holes also consists of two
forms -- dissipation and fluctuation or noise -- corresponding to
the real and imaginary parts of the influence functional as
embodied in the dissipation and noise kernels. A
fluctuation-dissipation relation relation has been shown to exist
for the near flat case by Campos and Hu \cite{CamHu98,CamHu99} and
we anticipate that it should also exist between the noise and
dissipation kernels for the general case, as it is a categorical
relation \cite{CalLeg83,GraSchIng88,HuPazZha92,HuPazZha93,Banff}.
Martin and Verdaguer have also proved the existence of a
fluctuation-dissipation relation when the semiclassical background
is a stationary spacetime and the quantum field is in thermal
equilibrium. Their result was then extended to a conformal field
in a conformally stationary background \cite{MarVer99a}. The
existence of a fluctuation-dissipation relation for the black hole
case has been discussed by some authors previously
\cite{CanSci77,Sci79,SciCanDeu81,Mottola}. In \cite{Vishu}, Hu,
Raval and Sinha have described how this approach and results
differ from those of previous authors. The fluctuation-dissipation
relation reveals an interesting connection between black holes
interacting with quantum fields and non-equilibrium statistical
mechanics. Even in its restricted quasi-static form, this relation
will allow us to study \textit{nonequilibrium} thermodynamic
properties of the black hole under the influence of stochastic
fluctuations of the energy momentum tensor dictated by the noise
terms.

There are limitations of a technical nature in the specific
example invoked here. For one we have to confine ourselves to
small perturbations about a background metric. For another, as
mentioned above, there is no
reliable approximation to the Schwarzschild thermal Green's
function to explicitly compute the noise and dissipation kernels.
This limits our ability to present explicit analytical
expressions for these kernels. One can try to improve on Page's
approximation by retaining terms to higher order. A less
ambitious first step could be to confine attention to the
horizon  and using approximations that are restricted to near the
horizon and work out the Influence Functional in this
regime. 

Yet another technical limitation of the specific example is the
following. Though we have allowed for backreaction effects to
modify the initial state in the sense that the temperature of the
Hartle-Hawking state gets affected by the backreaction, we have
essentially confined our analysis to a Hartle-Hawking thermal
state of the field. This analysis does not directly extend to a
more general class of states, for example to the case where the
initial state of the field is in the Unruh vacuum. Thus we will
not be able to comment on issues of the stability of an
\textit{isolated} radiating black hole under the influence of
stochastic fluctuations.

\subsubsection{Metric Fluctuations in Black Holes}

In addition to the work described above by Campos, Hu, Raval and
Sinha \cite{CamHu98,CamHu99,Vishu,SinRavHu03} and earlier work quoted
therein, we mention also some recent work on black hole metric
fluctuations and their effect on Hawking radiation. For example,
Casher et al \cite{CasEtal97} and Sorkin \cite{Sor96,SorSud99}
have concentrated on the issue of fluctuations of the horizon
induced by a fluctuating metric. Casher et al \cite{CasEtal97}
considers the fluctuations of the horizon induced by the
``atmosphere" of high angular momentum particles near the horizon,
while Sorkin \cite{Sor96,SorSud99} calculates fluctuations of the
shape of the horizon induced by the quantum field fluctuations
under a Newtonian approximations. Both group of authors come to
the conclusion that horizon fluctuations become large at scales
much larger than the Planck scale (note Ford and Svaiter
\cite{ForSva97} later presented results contrary to this claim).
However, though these works do deal with backreaction, the
fluctuations considered do not arise as an explicit stochastic
noise term as in our treatment. It may be worthwhile exploring the
horizon fluctuations induced by the stochastic metric in our model
and comparing the conclusions with the above authors. Barrabes et
al \cite{BarFroPar99,BarFroPar00}  have considered the propagation
of null rays and massless fields in a black hole fluctuating
geometry and have shown that the stochastic nature of the metric
leads to a modified dispersion relation and helps to confront the
trans-Planckian frequency problem. However, in this case the
stochastic noise is put in by hand and does not naturally arise
from coarse graining as in the quantum open systems approach.  It
also does not take backreaction into account. It will be
interesting to explore how a stochastic black hole metric, arising
as a solution to the Einstein-Langevin equation, hence fully
incorporating backreaction, would affect the trans-Planckian
problem.

Ford and his collaborators  \cite{ForSva97,ForWu,WuFor99} have
also explored the issue of metric fluctuations in detail and in
particular have studied the fluctuations of the black hole horizon
induced by metric fluctuations. However, the fluctuations they
considered are in the context of a fixed background and do not
relate to the backreaction.

Another work originating from the same vein of stochastic gravity
but not complying with the backreaction spirit is that of Hu and
Shiokawa \cite{HuShi98}, who study effects associated with
electromagnetic wave propagation in a Robertson-Walker universe
and the Schwarzschild spacetime with a small amount of given
metric stochasticity. They find that
time-independent randomness can decrease the total luminosity of
Hawking radiation due to multiple scattering of waves outside the
black hole and gives rise to event horizon fluctuations and
fluctuations in the Hawking temperature. The stochasticity in a
background metric in their work is assumed rather than derived
(from quantum field fluctuations, as in this work) and so is not
in the same spirit of backreaction. But it is interesting to
compare their results with that of backreaction, so one can begin
to get a sense of the different sources of stochasticity and their
weights (see, e.g., \cite{stogra} for a list of possible sources
of stochasticity.)

In a subsequent paper Shiokawa \cite{Shio} showed that the scalar
and spinor waves in a stochastic spacetime behave similarly to
the electrons in a disordered system. Viewing this as a quantum
transport problem, he expressed the conductance and its
fluctuations in terms of a nonlinear sigma model in the closed
time path formalism and showed that the conductance fluctuations
are universal, independent of the volume of the stochastic region
and the amount of stochasticity. This result can have significant
importance in characterizing the mesoscopic behavior of
spacetimes resting between the semiclassical and the quantum
regimes.

\section{Concluding Remarks}

In the first part of this review on the fundamentals of theory we
have given two routes to the establishment of stochastic gravity
and derived a general (finite) expression for the noise kernel. In
the second part we gave three applications, the correlation
functions of gravitons in a perturbed Minkowski metric, structure
formation in stochastic gravity theory and the outline of a
program for the study of black hole fluctuations and
backreaction. A central issue which stochastic gravity can
perhaps best address is the validity of semiclassical gravity as
measured by the fluctuations of stress energy compared to the
mean. We will include a review of this topic in a future update.

There are a number of ongoing research related to the topics
discussed in this review. On the theory side, Roura and Verdaguer
\cite{RouVer03b} has recently showed how stochastic gravity can be
related to the large $N$ limit of quantum metric fluctuations.
Given $N$ free matter fields interacting with the gravitational
field, Hartle and Horowitz \cite{HarHor81} and Tomboulis
\cite{Tom77} have shown that semiclassical gravity can be obtained
as the leading order large $N$ limit (while keeping $N$ times the
gravitational coupling constant fixed). It is of interest to find
out where in this setting can one place the fluctuations of the
quantum fields and the metric fluctuations they induce;
specifically, whether the inclusion of these sources will lead to
an Einstein-Langevin equation
\cite{CalHu94,HuMat95,HuSin95,CamVer96,LomMaz97}, as it was
derived historically in other ways,  as described in the first
part of this review. This is useful because it may provide another
pathway or angle in connecting semiclassical to quantum gravity
(a related idea is
the kinetic theory approach to quantum gravity described in
\cite{kinQG}).

Theoretically, stochastic gravity is at the frontline of the
`bottom-up' approach to quantum gravity
\cite{grhydro,stogra,kinQG}. Structurally, as can be seen from the
issues discussed and the applications given, stochastic gravity
has a very rich constituency because it is based on quantum field
theory and nonequilibrium statistical mechanics in a curved
spacetime context. The open systems concepts and the
closed-time-path / influence functional methods constitute an
extended framework suitable for treating the backreaction and
fluctuations problems of dynamical spacetimes interacting with
quantum fields. We have seen it applied to cosmological
backreaction problems. It can also be applied to treat the
backreaction of Hawking radiation in a fully dynamical black hole
collapse situation, and in so doing enable one to address related
issues such as the black hole end state and information loss
puzzles (e.g., \cite{Pag94,Erice95} and references therein). The
main reason why this program has not progressed as swiftly as
desired is due more to technical rather than programatic
difficulties (such as finding reasonable analytic approximations
for the Green function or numerical evaluation of mode-sums near
the black hole horizon). Finally, the multiplex structure of this
theory could be used to explore new lines of inquiry and launch
new programs of research, such as {\it nonequilibrium} black hole
thermodynamics and statistical mechanics.

\section{Acknowledgements}

The materials presented here
originated from research work of BLH with Antonio Campos,
Nicholas Phillips, Alpan Raval and Sukanya Sinha,  and of EV with
Rosario Martin and Albert Roura. We thank them as well as Daniel
Arteaga, Andrew Matacz,  Tom Shiokawa, and Yuhong Zhang for
fruitful collaboration and their cordial friendship since their
Ph. D. days. We enjoy lively discussions with our friends and
colleagues Esteban Calzetta, Diego Mazzitelli and Juan Pablo Paz
whose work in the early years contributed toward the
establishment of this field. We acknowledge useful discussions
with Paul Anderson, Larry Ford, Ted Jacobson, Renaud Parentani
and Raphael Sorkin. This work is supported in part by NSF grant
PHY98-00967, the MICYT Research Project No. FPA-2001-3598 and
European project HPRN-CT-2000-00131.


\end{document}